\documentclass[a4paper,10pt,oneside,onecolumn,preprint,3p]{elsarticle}
\pdfoutput=1

\usepackage{amsmath}
\usepackage{amssymb}
\usepackage{epsfig}
\usepackage{graphicx}% Include figure files
\usepackage{dcolumn}% Align table columns on decimal point
\usepackage{bm}% bold math
\usepackage{bbm}
\usepackage{color}
\usepackage{hyperref}
\usepackage{enumerate}
\hypersetup{colorlinks=true,linkcolor=red,citecolor=blue,urlcolor=blue}

\def\to{\rightarrow}

\def\p{\partial}

\def\s{\sigma}

\renewcommand{\dag}{^{\dagger}}

\newcommand{\beq}{\begin{equation}}
\newcommand{\eeq}{\end{equation}}
\newcommand{\beqa}{\begin{eqnarray}}
\newcommand{\eeqa}{\end{eqnarray}}
\newcommand{\sgn}{\text{sgn}}
\def\gapp{\lower.35em\hbox{$\stackrel{\textstyle>}{\sim}$}}
\def\lapp{\lower.35em\hbox{$\stackrel{\textstyle<}{\sim}$}}

\journal{Physics Reports}

\begin{document}

\begin{frontmatter}

\title{Novel effects of strains in graphene and other two dimensional materials}

\author[label1,label2]{B. Amorim\corref{cor1}}
\ead{amorim.bac@icmm.csic.es, amorim.bac@gmail.com}
\author[label1]{A. Cortijo}
\author[label3,label4]{F. de Juan}
\author[label5]{ A. G. Grushin}
\author[label1,label6,label7]{F. Guinea}
\author[label1]{A. Guti\'errez-Rubio}
\author[label1,label8]{H. Ochoa}
\author[label1,label7]{V. Parente}
\author[label1]{R. Rold\'an}
\author[label1]{P. San-Jose}
\author[label1]{J. Schiefele}
\author[label9]{M. Sturla}
\author[label1]{M. A. H. Vozmediano}

\cortext[cor1]{Corresponding author}

\address[label1]{Instituto de Ciencia de Materiales de Madrid, CSIC, Cantoblanco, E-28049 Madrid, 
Spain}
\address[label2]{Department of Physics and Center of Physics,
University of Minho, P-4710-057, Braga, Portugal}
\address[label3]{Materials Science Division, Lawrence Berkeley National Laboratories, Berkeley, CA 
94720, USA}
\address[label4]{Department of Physics, University of California, Berkeley, CA 94720, USA}
\address[label5]{Max-Planck-Institut fur Physik komplexer Systeme, 01187 Dresden, Germany}
\address[label6]{School of Physics and Astronomy, University of Manchester, Oxford Road, Manchester 
M13 9PL, UK}
\address[label7]{IMDEA Nanociencia Calle de Faraday, 9, Cantoblanco, 28049, Madrid, Spain}
\address[label8]{Donostia International Physics Center (DIPC), 20018 San Sebasti\'an, Spain}
\address[label9]{IFLP-CONICET. Departamento de F\'isica, Universidad Nacional de La Plata, (1900) 
La Plata, Argentina}

\begin{abstract}
The analysis of the electronic properties of strained or lattice deformed graphene combines ideas 
from classical condensed matter physics, soft matter, and geometrical aspects of quantum field 
theory (QFT) in curved spaces. Recent theoretical and experimental work shows the influence of 
strains in many properties of graphene not considered before, such as electronic transport, 
spin-orbit coupling, the formation of Moir\'e patterns and optics. There is also 
significant evidence of anharmonic effects, which can modify the structural properties of graphene. 
These phenomena are not restricted to graphene, and they are being intensively studied in other 
two dimensional materials, such as the transition metal dichalcogenides. We review here recent 
developments related to the role of strains in the structural and electronic properties of graphene 
and other two dimensional compounds.
\end{abstract}


\end{frontmatter}

\tableofcontents

\section{Introduction}
\label{sec_intro}

Graphene has a number of special properties \cite{Geim09}. 
The unique combination of a two dimensional (2D) structure  and massless electronic carriers leads 
to a variety of phenomena never considered before. These features are amplified by the high 
stiffness of the graphene lattice. The analysis of the electronic properties of strained or lattice 
deformed graphene combines ideas from classical condensed matter physics, soft matter  
\cite{SN_88,NP_87,NPW04}, and geometrical aspects of quantum field theory (QFT) in curved spaces 
\cite{BD82}. 

A vast literature already exists on topics such as ripples, effective gauge fields and "strain 
engineering" in graphene \cite{Paco12} and there are already a number of excellent books and 
reviews describing the advances in the understanding of strain related effects  
\cite{CGetal09,VKG10,K__12}. The understanding of these phenomena has motivated the study of related 
effects in other 2D materials.

Recent theoretical and experimental work shows the influence of strains on many properties of 
graphene not considered before, such as electronic transport, spin-orbit coupling, the formation of 
Moir\'e patterns and optics. There is also significant evidence of anharmonic effects, which 
can modify the structural properties of graphene. These phenomena are not restricted to graphene, 
and they are being intensively studied in other 2D materials, such as the metallic 
dichalcogenides.

We review here recent developments related to the role of strains in the structural and electronic 
properties of graphene and other 
2D compounds. We emphasize how strains influence a 
number of features not considered before. 

The review is divided into five self--contained sections which cover the main recent developments.  
A summary of basic properties, formalism and notation used throughout the review are found in Section 
\ref{sec_geometrical}.  This section  introduces the formalism by which elastic deformations of the 
lattice encoded in the strain tensor of the elasticity theory couple to the electronic degrees of 
freedom described by the Dirac equation in the continuum limit. A symmetry approach is used to 
derive all possible  electron--phonon couplings and their physical content is described.  The 
"elastic" gauge fields and the deformation potential to be used through the review are defined 
there. 
This section also includes a geometrical description based on quantum field theory in curved spaces 
that allows to predict some new effects as the Fermi velocity as a space--dependent tensor. This 
formalism is also used to describe some new developments on topological properties of the 
materials. 

Section \ref{sec_anharmonic} is devoted to the new developments concerning the 
membrane properties of the lattices. Although these aspects attracted attention from the birth of 
graphene \cite{K__12}, recent experiments  \cite{BMF09,YYG11,PAL11} have shown that anharmonic 
effects in graphene and other 2D crystalline membranes  have been seriously underestimated. This 
section describes the present situation concerning the effects of these anharmonicities on the 
elastic and mechanical properties of the membranes. A summary of the classical theory of flat 
membranes (Section~\ref{subsec:membrane_classical}) to fix the notations and definitions, is followed by a 
description of the modifications to the electron--phonon interaction due to defects 
(Section~\ref{subsec:membrane_deffects}) and quantum corrections (Section~\ref{subsec:membrane_quantum}).  The 
modifications to the mechanical and thermodynamic properties are described in Sections 
\ref{sec:membrane_mechanical}, \ref{sec:membrane_thermo}. The section concludes with an overview of 
the recent advances on the influence of topological defects on the membrane properties 
(Section~\ref{sec_defects}).

Section \ref{sec_electronic} deals with the new developments concerning the effect of strain on the 
electronic properties of the samples. Particular attention is put on the analysis of the correlation 
between carrier mobilities and amplitudes of strain distributions demonstrated recently (Sec. 
\ref{Sec:general}). The combined influence of strain and interactions is analysed in Sec. 
\ref{sec_interactions} which addresses in particular the access to topological phases through the 
interplay between strain and electron-electron interactions. 
Interaction induced magnetic or superconducting phenomena  in strained graphene are also described.
The optical properties of graphene and other atomically thin materials promise
a wealth of applications.  In particular, graphene plasmonics in the THz and
mid-infrared bands \cite{GPN12}, as well as graphene based broadband devices
working in the near-infrared and visible part of the spectrum \cite{BSH10}, have
recently attracted attention.  Further, the unusual magneto-optical properties
of graphene, which are determined by the nonlinear Landau level spectrum of
massless Dirac Fermions \cite{CLW11,SYY13}, offer a great potential for the
development of non-reciprocal optical devices. Optical properties are described in 
Sec. \ref{sec:optics}.

Isolated atomic planes can now be reassembled into heterostructures made in a precisely chosen 
sequence \cite{GG_13}. Section \ref{Sec:superlattices} deals with the electronic properties of these 
new compounds that include superlattices of graphene deposited on other 2D crystals like hBN, SiC 
or graphene itself.
The induced strains on the pristine sheet, resulting from a competition between the adhesion potential and the elasticity of the layer, are the factor ultimately determining the electronic and mechanical properties of the sheet.  A general description (\ref{Sec:electronic_properties_undeformed}) is followed by an analysis of the spontaneous deformations arising in the samples and their experimental consequences (\ref{Sec:spontaneous_deformations}).

Section \ref{sec_others} is devoted to the new 2D materials whose synthesis followed 
naturally that of graphene. It includes a bibliographic summary of the on-going research in  
isolated mono- and few-layers of hexagonal boron nitride (hBN), molybdenum disulphide, other 
dichalcogenides, black phosphorus and layered oxides.

There are important aspects of the vast subject of physics of strain that have been left out of the 
review. Some of them have been appropriately described in previous literature \cite{VKG10}; others,  
including  mechanical and numerical approaches to analyse the stability of the lattice \cite{BSW13}, have been omitted for lack of expertise. This review is the collective work of a theory group in 
Madrid that has been active in the field from the beginning and continues working on it.

\section{Geometric aspects: Group symmetry approach}
\label{sec_geometrical}
%\section{Geometric aspects: Group symmetry approach}

The aim of this chapter is to discuss the minimal models for the low energy modes of graphene. Both 
symmetry group and differential geometry arguments are employed in order to construct effective 
actions for the dynamics of graphene electrons and phonon modes, and the coupling between them. The 
standard $\mathbf{k}\cdot\mathbf{p}$ derivation is compared with a quantum field theory approach 
based on the study of the Dirac equation in curved spaces. Finally, we review some topological 
aspects of graphene physics, with special emphasis on topological defects, closely related to the 
structural properties of the graphene lattice.

\subsection{Electrons and phonons in graphene. Symmetry approach}

\label{Sec:electrons_phonons}
The idea of constructing effective actions for physical systems based only on symmetry considerations has been
extensively used both in quantum field theory (QFT) and in condensed matter physics and it lies at the heart of
the Landau Fermi liquid theory of metals. In solid state systems, it is the underlying crystal group what imposes the symmetry constraints on the action describing the dynamics of both electrons and phonons.

\begin{figure}
\begin{centering}
\includegraphics[width=0.3\columnwidth]{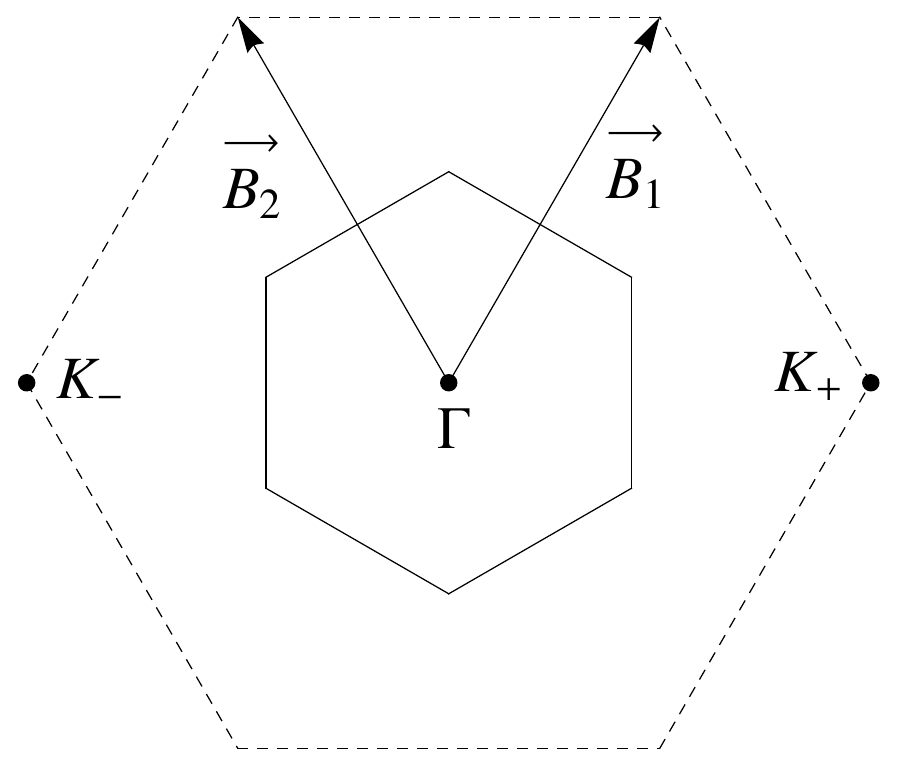}
\par\end{centering}
\caption{\label{Fig:bz}Graphene Brillouin zone corresponding to a real space unit
cell with with 2 (dashed line) and 6 (solid line) atoms in the unit cell.}
\end{figure}

Graphene consists on a single layer of $sp^2$ hybridized carbon atoms forming a honeycomb lattice, which is a triangular Bravais lattice with two  
atoms per unit cell. Three of the four electrons in the outer-shell participate in the strong $\sigma$ bond which keeps them covalently attached 
forming a planar structure with a distance between atoms of $a_{CC}=1.42$ \AA. These $\sigma$ 
electrons are responsible for the structural properties of 
graphene, in particular its stiffness. The remaining electrons occupying the $p_z$ orbital perpendicular to the graphene plane are free to hop between 
neighbouring sites, leading to the $\pi$ bands. In pristine graphene, the Fermi level lies at the 
two inequivalent corners of the hexagonal Brillouin 
zone, $K_{\pm}$ (see Fig. \ref{Fig:bz}).

The three dimensional (3D) point group of the graphene crystal is $D_{6h}=D_6\otimes 
i=C_{6v}\otimes\sigma_h$. We follow the notation of Ref.~\cite{Dresselhaus_book}, 
which can be simplified if we only refer to the operations of the planar group, $C_{6v}$, that contains 12 elements: the identity, five rotations 
along the axis perpendicular to the graphene plane ($z$ axis) and six reflections in planes perpendicular to it. Additionally, we have to take into 
account the reflection operation along the $z$ axis, $\sigma_h$, when dealing with out-of-plane distortions.

Instead of dealing with degenerate states at two inequivalent points
of the Brillouin zone one can enlarge the unit cell in order to
contain six atoms. Therefore, the folded Brillouin zone is three times
smaller and the $K_{\pm}$ points are mapped onto the $\Gamma$ point,
see Fig.~\ref{Fig:bz}. From the point of view of the lattice
symmetries, this means that the two elementary translations
$\left(t_{\vec{a}_1},t_{\vec{a}_2}\right)$ are factorized out of
the translation group and added to the point group, which becomes
$C_{6v}''=C_{6v}+t_{\vec{a}_1}\times C_{6v}+t_{\vec{a}_2}\times
C_{6v}$ \cite{basko}. This allows us to treat all electronic excitations at
the corners of the Brillouin zone on a same footing, including also
possible inter-valley couplings. The Bloch wave function is given by a
6-component vector which represents the amplitude of the $p_z$
orbitals at the 6 atoms of the unit cell. This vector can be reduced
as $A_1+B_2+G'$. The 1-dimensional irreducible representations $A_1$
and $B_2$ correspond to the bonding and anti-bonding states at the
original $\Gamma$ point, whereas $\psi\sim G'$ corresponds to the
Bloch states at the original Brillouin zone corners. Then, in order to
construct the electronic Hamiltonian for quasiparticles around the
$K_{\pm}$ points, we must consider the 16 hermitian operators acting in
the 4-dimensional space defined by the Bloch wave functions
$\psi$. These operators may be classified according to the
transformation rules under the symmetry operations of $C_{6v}''$,
taking into account the algebraical reduction\begin{equation*}
G'\times G'\sim A_1+A_2+B_1+B_2+E_1+E_2+E_1'+E_2'+G'.
\label{Eq:red1}
\end{equation*}
In order to make the discussion more clear we introduce the basis $\psi=\left(\psi_{A+},\psi_{B+},\psi_{A-},\psi_{B-}\right)$, where each entry represents the projection of the Bloch wave function around each valley $K_{\pm}$ on sublattice $A$/$B$. Then, we introduce two inter-commutating Pauli algebras $\sigma_i$ and $\tau_i$ associated to sublattice and valley degrees of freedom respectively. The 16 possible electronic operators are generated by considering the direct products of the elements of these algebras (and the identity). Their symmetry properties are summarized in Tab.~\ref{Tab:opC6v1}. We must take into account also the time reversal operation, which is implemented by the antiunitary operator $\mathcal{T}=\tau_x\mathcal{K}$, where $\mathcal{K}$ represents complex conjugation.

\begin{table}
\begin{centering}
\begin{tabular}{|c|c|}
\hline
$A_1$&$\mathcal{I}$ (+)\\
\hline
$A_2$&$\tau_z\otimes\sigma_z$ (-)\\
\hline
$B_1$&$\tau_z$ (-)\\
\hline
$B_2$&$\sigma_z$ (+)\\
\hline
$E_1$&$\left(\begin{array}{c}
\tau_z\otimes\sigma_x \\
\sigma_y \end{array}\right)$ (-)\\
\hline
$E_2$&$\left(\begin{array}{c}
\sigma_x \\
\tau_z\otimes\sigma_y \end{array}\right)$ (+)\\
\hline
$E_1'$&$\left(\begin{array}{c}
\tau_x\otimes\sigma_x \\
\tau_y\otimes\sigma_x \end{array}\right)$ (+)\\
\hline
$E_2'$&$\left(\begin{array}{c}
-\tau_y\otimes\sigma_y \\
\tau_x\otimes\sigma_y \end{array}\right)$ (-)\\
\hline
$G'$&$\left(\begin{array}{c}
-\tau_y \\
\tau_x\otimes\sigma_z\\
-\tau_y\otimes\sigma_z \\
\tau_x\end{array}\right)$ (+)\\
\hline
\end{tabular}
\caption{Classification of the electronic operators according to the irreducible representations of $C_{6v}''$. The signs $\pm$ denote if the operator is even or odd under time reversal.}
\label{Tab:opC6v1}
\end{centering}
\end{table}

%{\color{red} (Aqui se puede introducir ya el formalismo "relativista" y conectarlo con los operadores de la Tabla I)}

The effective low energy Hamiltonian is constructed as an expansion in powers of the crystalline momentum $\vec{k}=\left(k_x,k_y\right)\sim E_1$ around $K_{\pm}$ points. Up to first order in $\vec{k}$ we have
\begin{align}
H_0=v_F\left(\tau_z\sigma_xk_x+\sigma_yk_y\right),
\label{Eq:effH}
\end{align}
where $v_F$ is the Fermi velocity.
This Dirac Hamiltonian describes the approximately conical dispersion of $\pi$ bands around $K_{\pm}$ points.

Phonon modes of graphene within the tripled unit cell can be also classified according to the irreducible representations of $C_{6v}''$ as shown in Tabs.~\ref{Tab:phononsIN} and \ref{Tab:phononsOUT}. The modes belonging to the valley-diagonal irreducible representations correspond to the acoustic and optical modes at the original $\Gamma$ point, according to which atoms of different sublattices oscillate in-phase or out-of-phase respectively. The modes belonging to the valley off-diagonal irreducible representations correspond to the modes at the $K_{\pm}$ points, in particular, real linear combinations of displacements at both valleys.

\begin{table}
\begin{centering}
\begin{tabular}{|c|c|c|c|c|}
\hline
$E_1$&$E_2$&$E_1'$&$E_2'$&$G'$\\
\hline
\includegraphics[width=0.18\columnwidth]{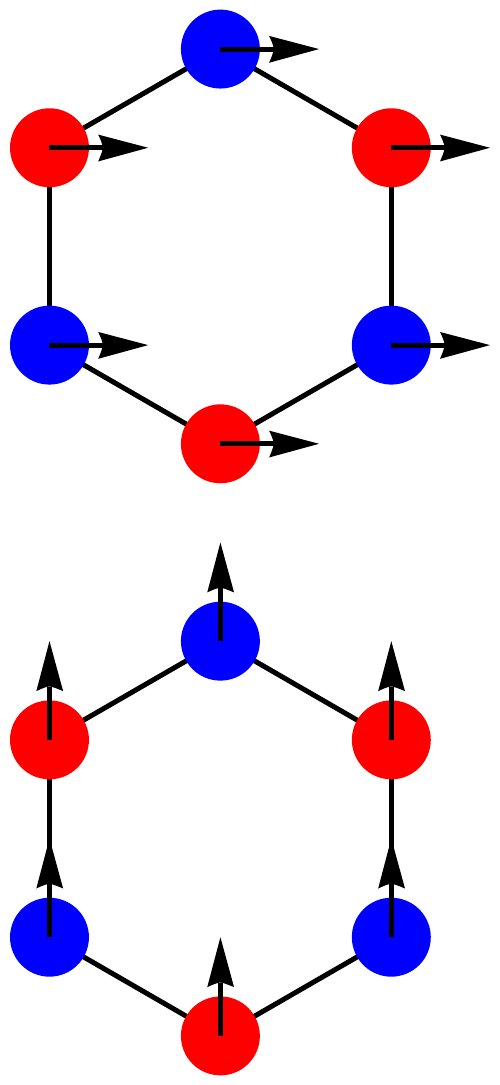}&
\includegraphics[width=0.2\columnwidth]{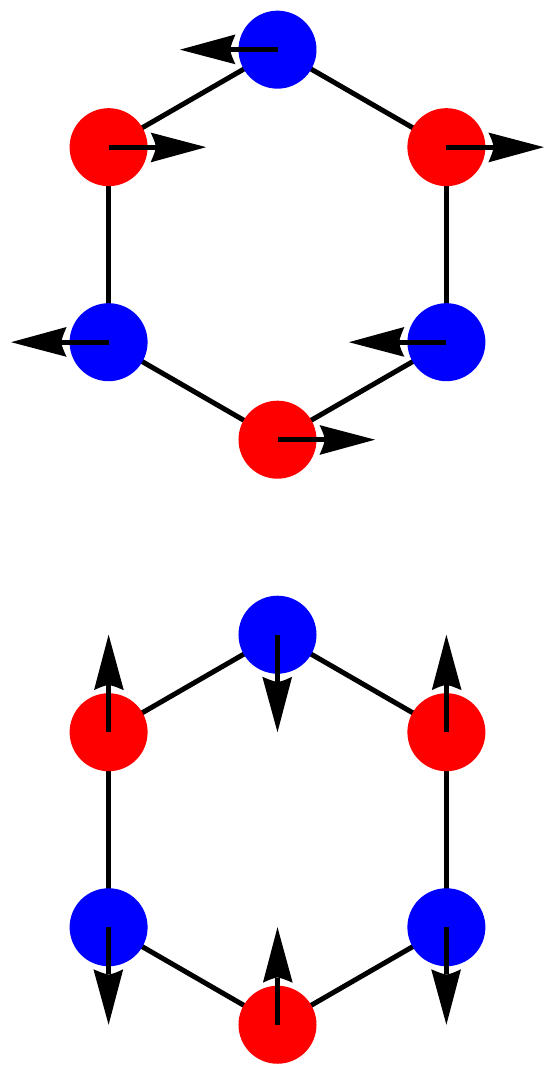}&\includegraphics[width=0.16\columnwidth]{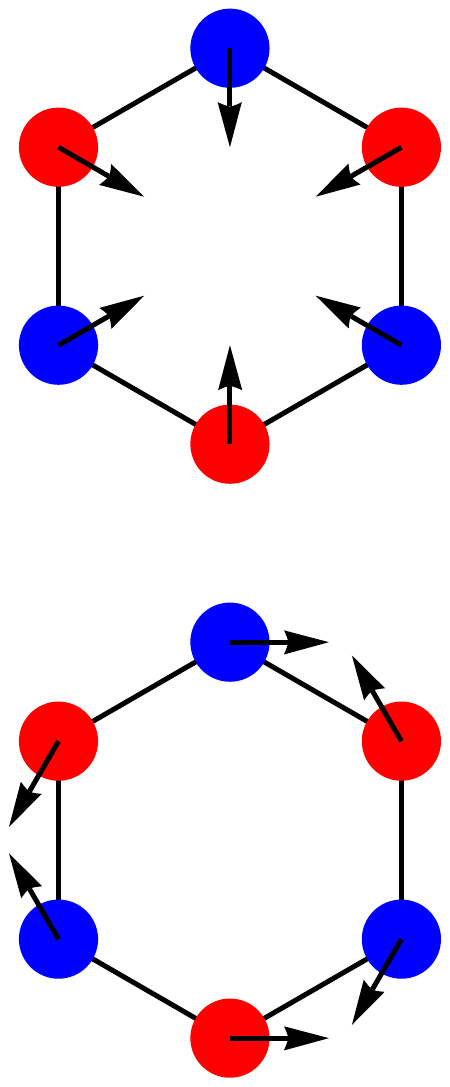}&
\includegraphics[width=0.19\columnwidth]{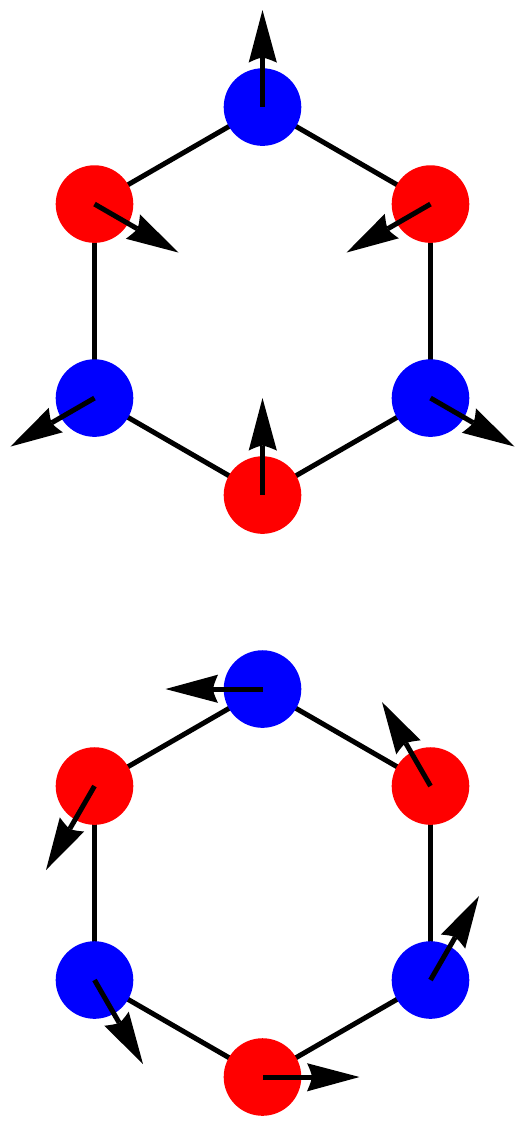}
&\includegraphics[width=0.15\columnwidth]{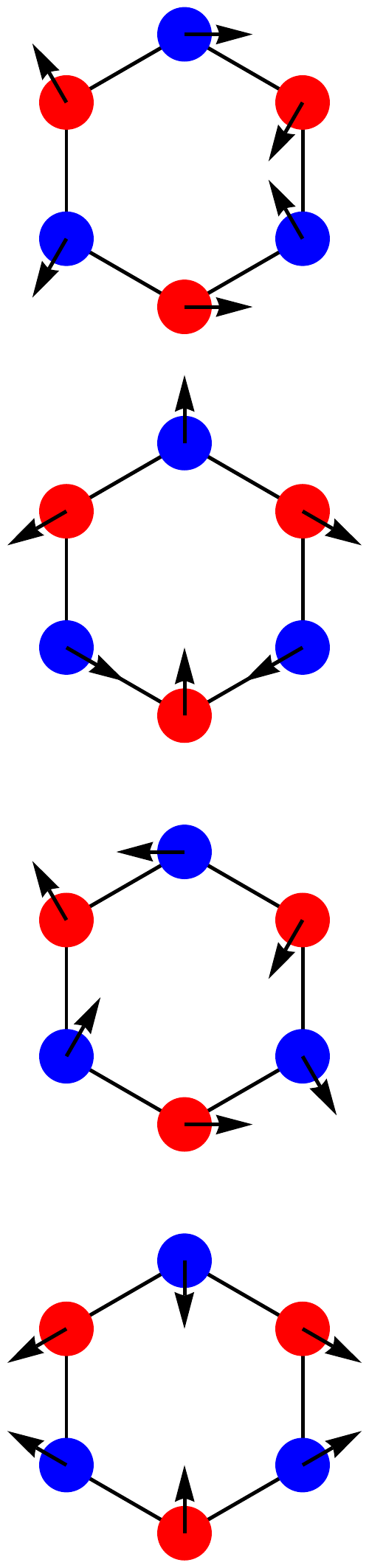}\\
\hline
\end{tabular}
\caption{Classification of in-plane phonon modes according to the irreducible representations of $C_{6v}''$.}
\label{Tab:phononsIN}
\end{centering}
\end{table}

\begin{table}
\begin{centering}
\begin{tabular}{|c|c|c|}
\hline
$A_1$&$B_2$&$G'$\\
\hline
&&\\
&&\includegraphics[width=0.19\columnwidth]{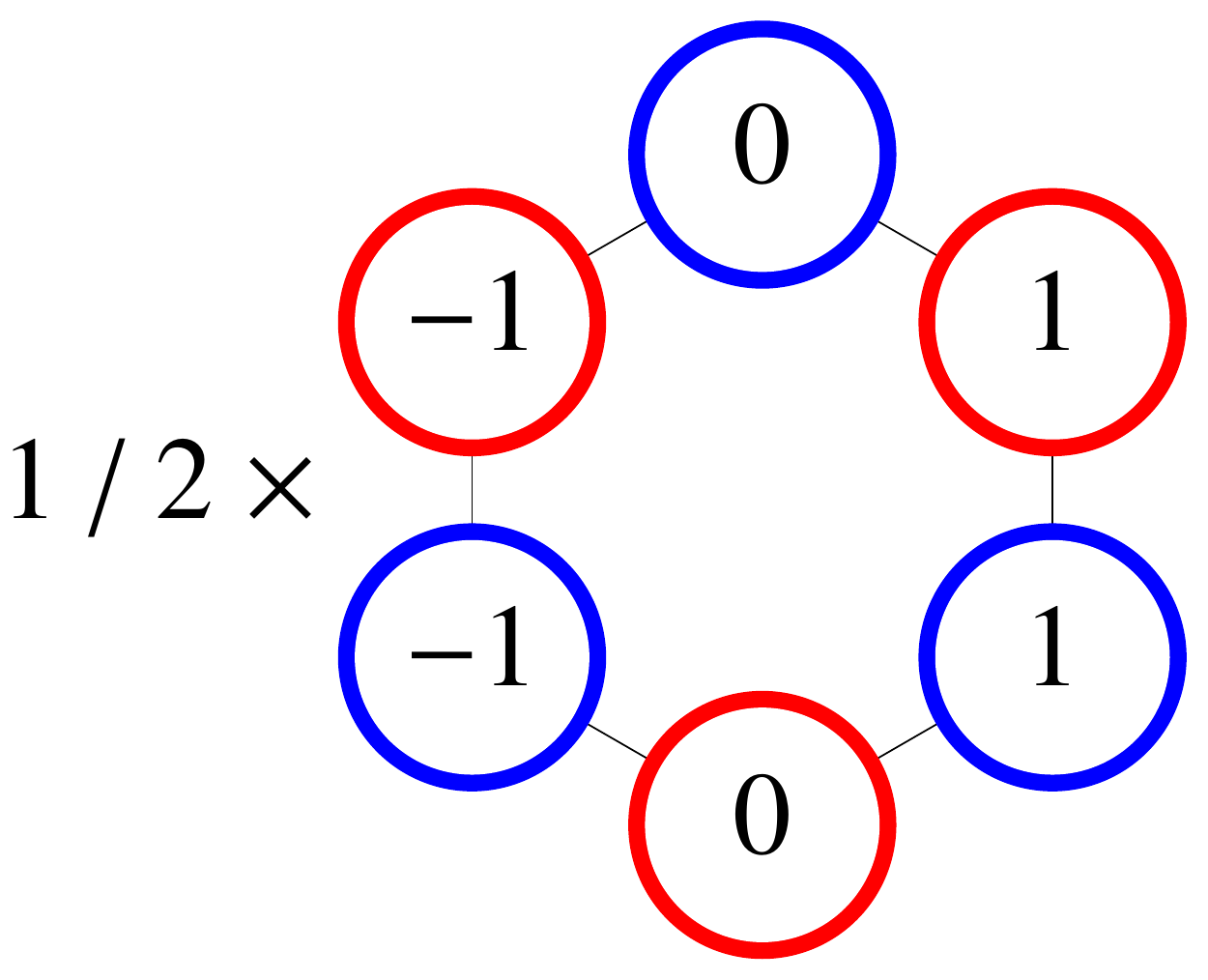}\includegraphics[width=0.21\columnwidth]{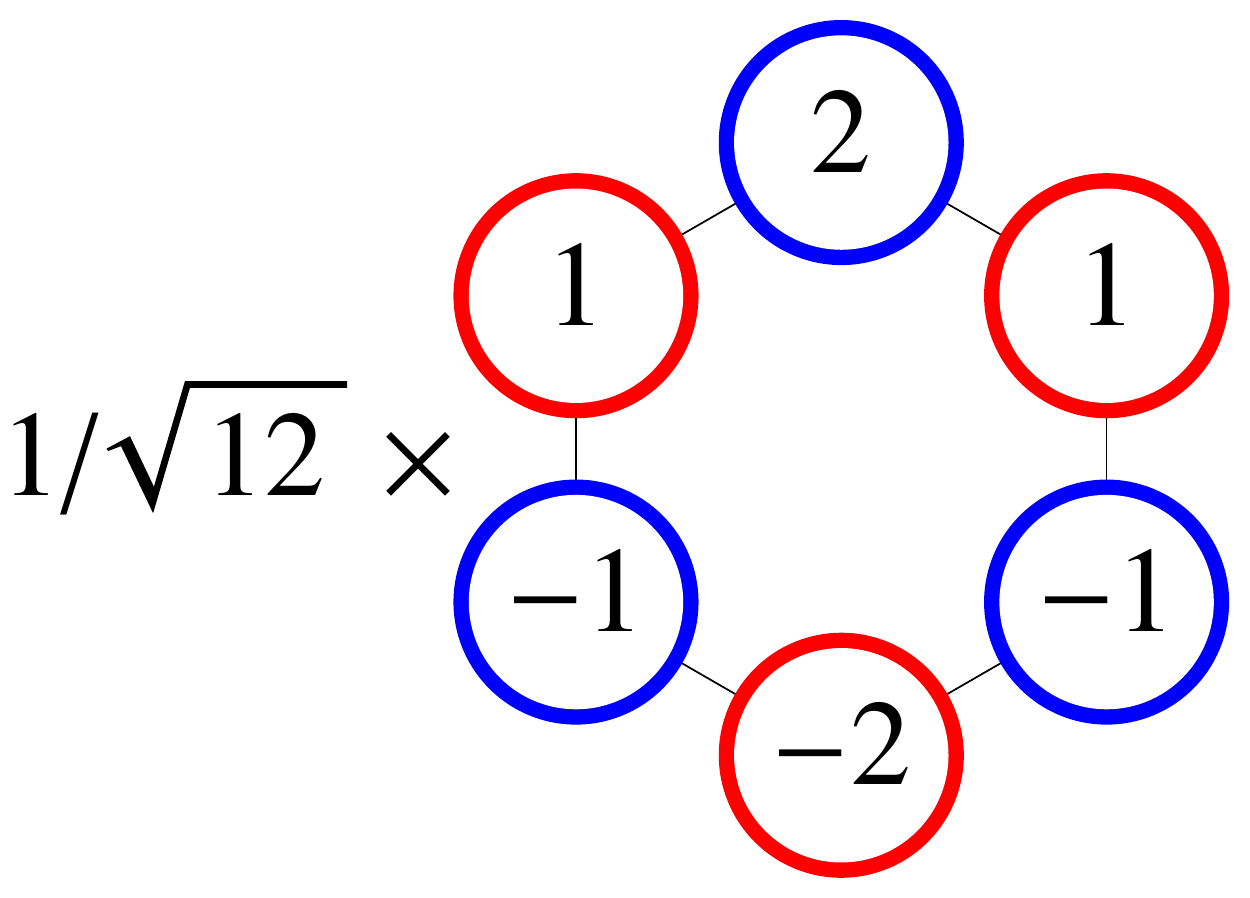}\\
\includegraphics[width=0.2\columnwidth]{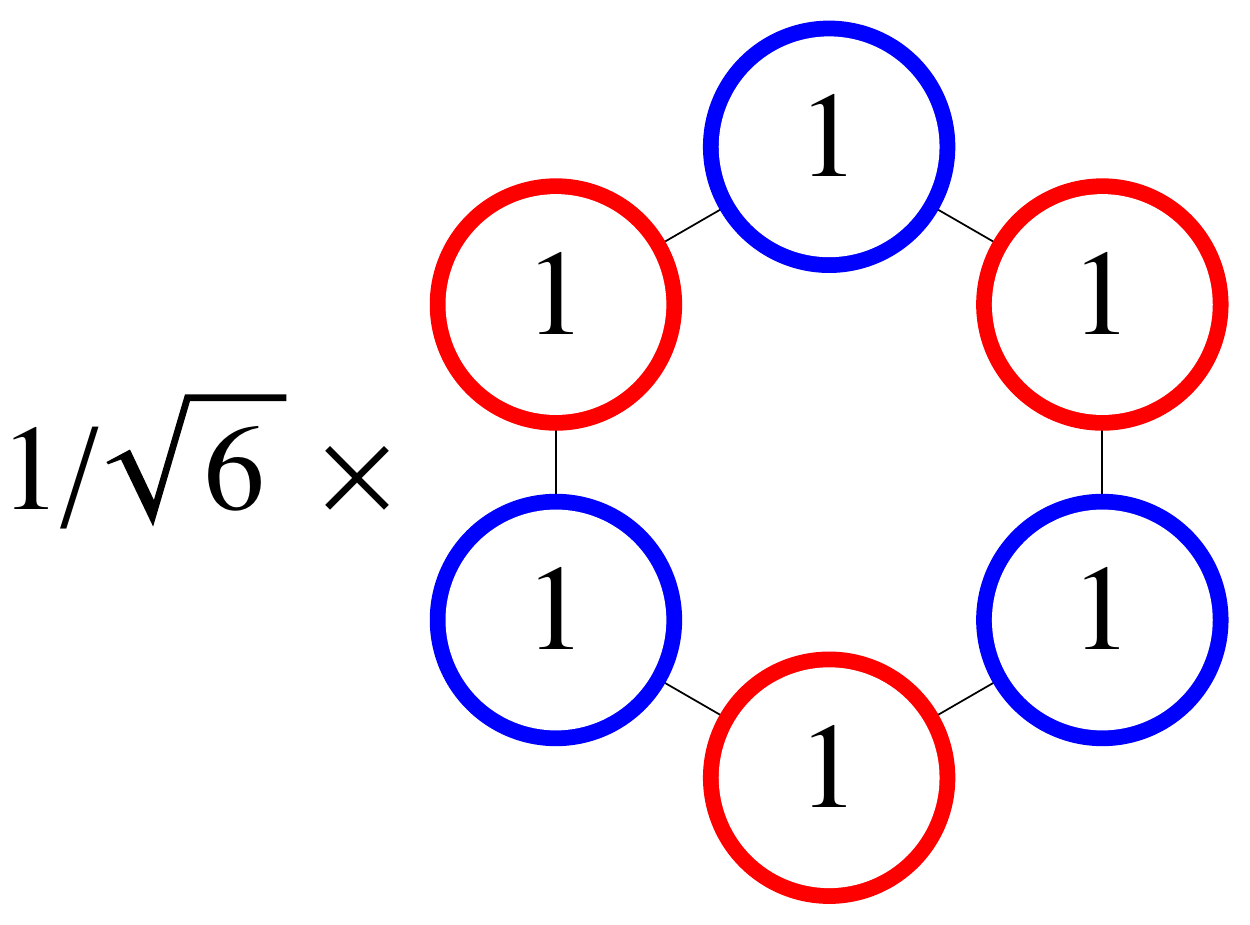}&
\includegraphics[width=0.2\columnwidth]{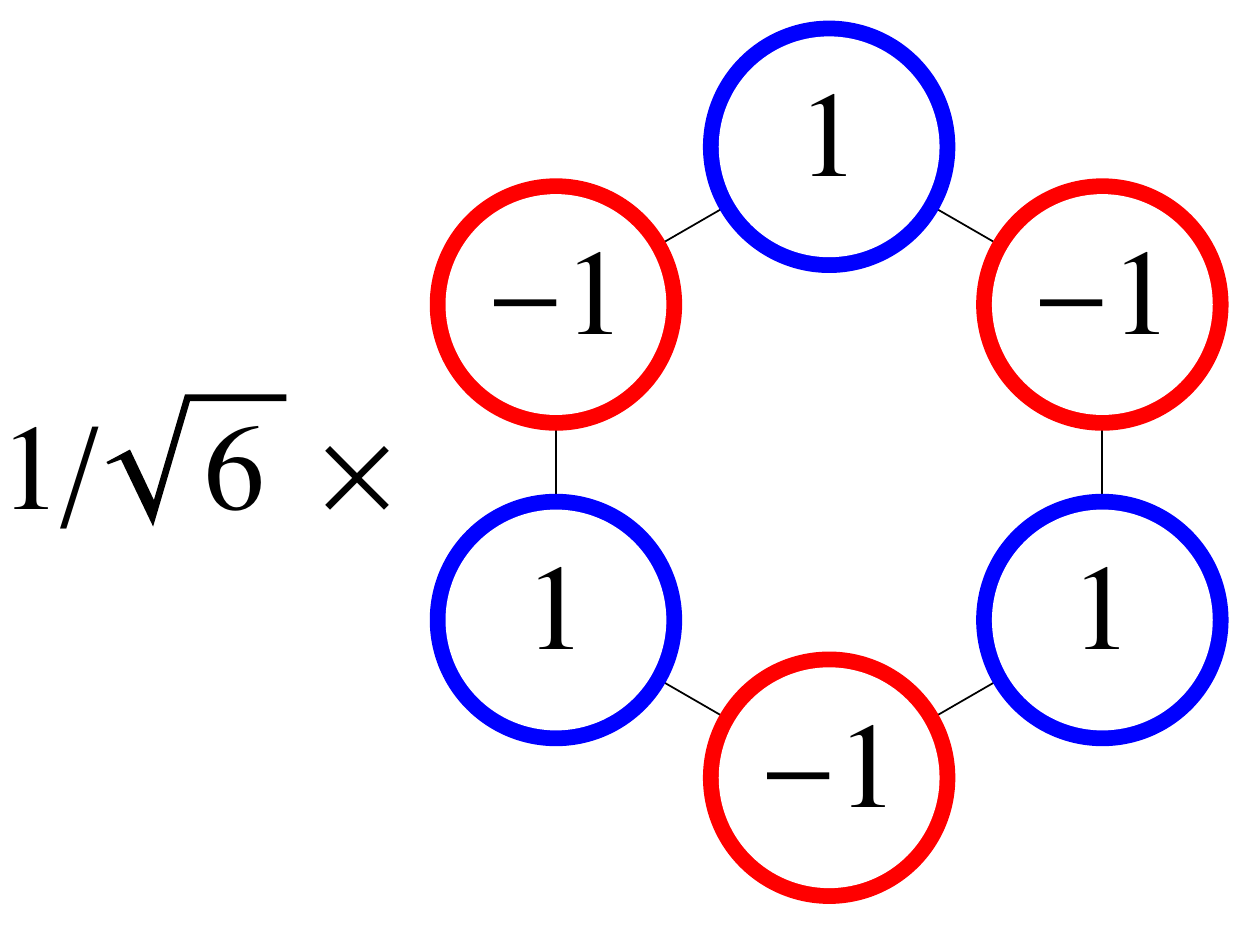}&
\includegraphics[width=0.19\columnwidth]{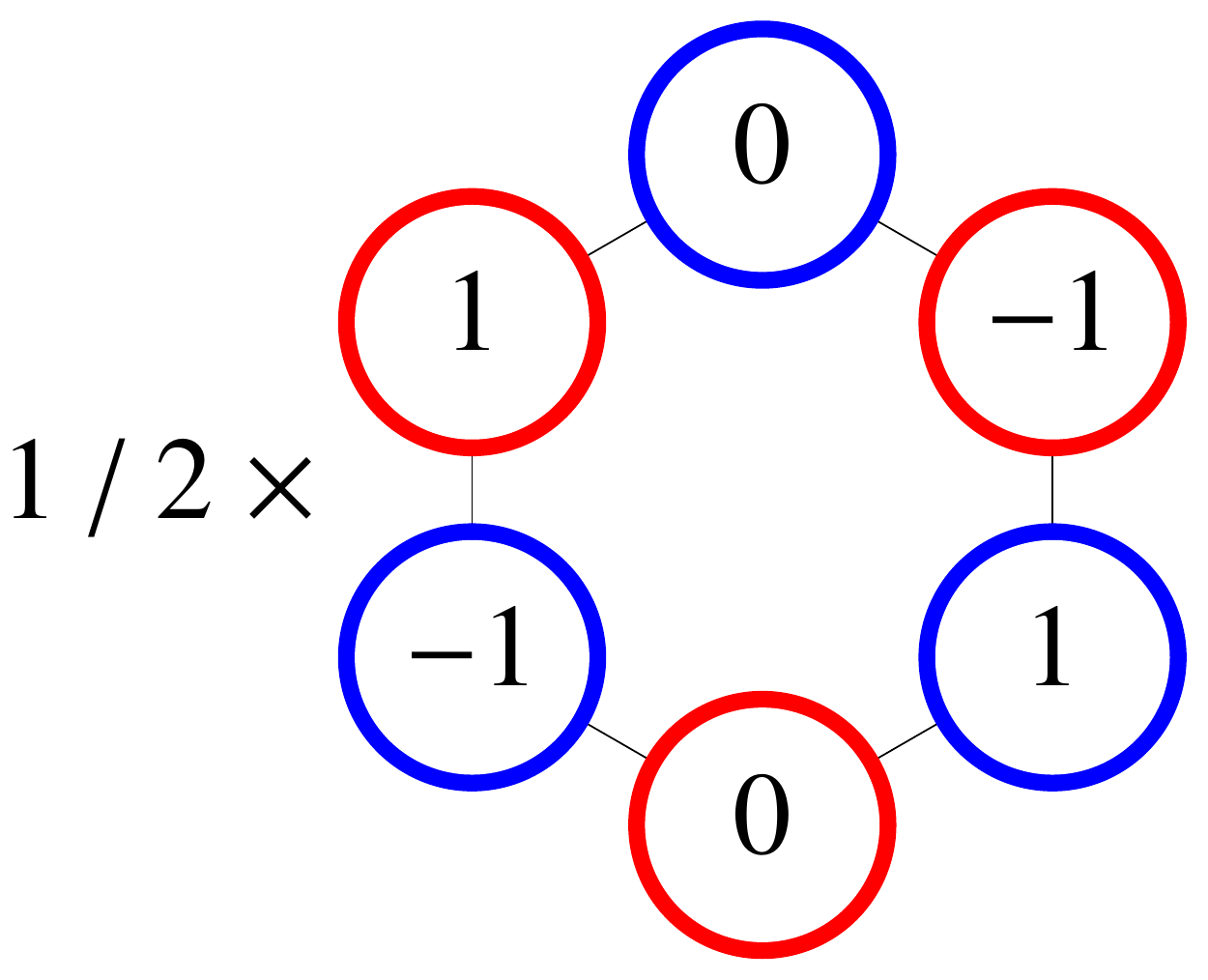}\\
&&\includegraphics[width=0.2\columnwidth]{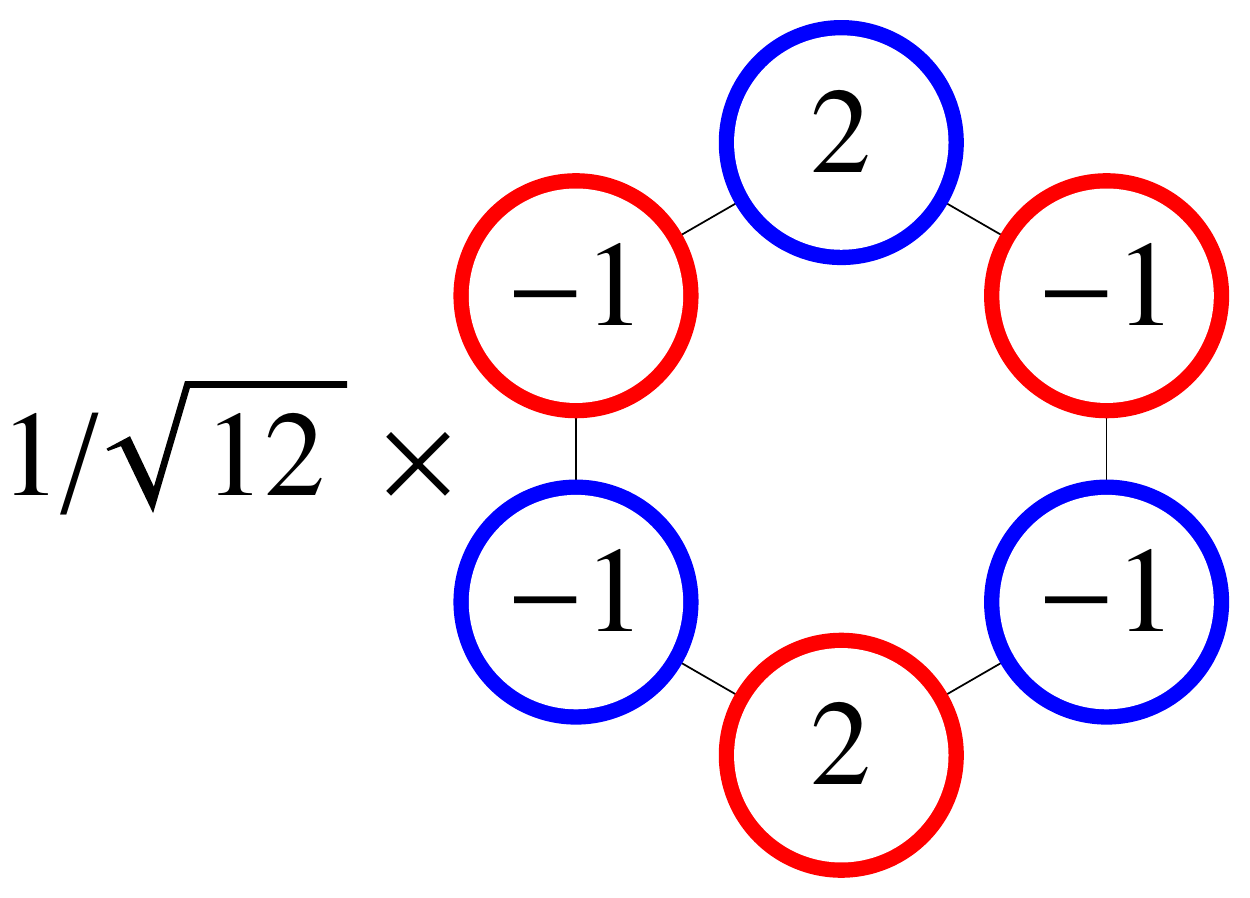}\\
\hline
\end{tabular}
\caption{Classification of flexural phonon modes according to the irreducible representations of $C_{6v}''$.}
\label{Tab:phononsOUT}
\end{centering}
\end{table}

\subsection{Theory of elasticity: graphene as a crystalline membrane}

\label{Sec:elasticity}

The dynamics of low energy vibration modes of graphene are governed by its elastic constants, and a 
long-wavelength description in terms of a theory in the continuum is valid since we focus on 
in-phase displacements only. The mechanics of solids, regarded as continuous media, is the subject of the 
theory of elasticity \cite{LL_59,CL_00}, which can be constructed by purely geometrical means with the only constraint imposed by the symmetry of the underlying crystalline structure. The unit 
cells of the graphene crystal can be seen as points on a surface embedded in $\mathbb{R}^3$, each of 
them labelled by a 3D vector $\mathbf{r}$ (throughout the text we will denote 
3D vectors with upright bold font and 2D vectors with an arrow). Therefore, the elastic energy 
associated with such an 
object can be split into two terms with a different geometrical and physical origin, which we write 
as
\begin{equation}
 F_{\text{elastic}}=F_{\text{st}} + F_{\text{b}},
\end{equation}
where $F_{\text{st}}$ is a stretching energy term that accounts for the energy cost due to 
relative changes in the distances and in-plane bond angles between atoms, and 
$F_{\text{b}}$ is a bending energy term which penalizes deviations from a flat configuration.  

The stretching energy depends on the first fundamental form, which is nothing but the pull-backed 
metric of the embedding space projected onto the surface,\begin{align}
g_{ij}=\partial_{i}\mathbf{r}\cdot\partial_{j}\mathbf{r},
\label{Eq:metric}
\end{align}
where $\partial_{i}$, $i=1,2$, is the derivative with respect to the internal coordinates 
which parametrize the surface. Note that this is a completely intrinsic quantity. In the Monge 
representation, carbon atoms at equilibrium, flat configurations are assumed to occupy positions 
labelled by 
$\mathbf{r}_0=\left(x_0,y_0,0\right)$, where $x_0$, $y_0$ correspond to the intrinsic coordinates 
that parametrize the surface. We consider deviations from these equilibrium positions given by a 
vector of displacements $\mathbf{r}=\mathbf{r}_0+\mathbf{u}$. The displacements are fields which 
depend on the position of the unit cell, $\mathbf{u}\left(x_0,y_0\right)$. The first fundamental 
form is given by
\begin{align}
g_{ij}=\delta_{ij}+2u_{ij},
\label{Eq:strain}
\end{align}
where $u_{ij}$ is the strain tensor. To the lowest order in the displacements, 
$\mathbf{u}=\left(u_x,u_y,h\right)$, this reads 
\begin{equation}
\label{usual} 
u_{ij}= \frac{1}{2}\left(\partial_i u_j +\partial_j u_i+ \partial_i h \partial_j 
h\right)\;\;\;,\;\;\;i,j=x,y. 
\end{equation}
Note that $u_{ij}$ is quadratic to the lowest order in the out-of-plane displacements, $h$, as a 
manifestation of $\sigma_h$ symmetry. We write the linear strain tensor only due to in-plane 
deformations as $u_{(i,j)}=\left(\partial_i u_j +\partial_j u_i\right)/2$ and the in-plane 
displacement as $\vec{u}=(u_x,u_y)$. The energy cost due to stretching can be written as a 
quadratic form in the strain tensor components \cite{CL_00,LL_59},\begin{align}
 F_{\text{st}}=\frac{1}{2}\int d^2\vec{r}\mathcal{C}^{ijkl}u_{ij}u_{kl},
\label{Eq:free}
\end{align}
where repeated indices are summed over and $\mathcal{C}^{ijkl}$ is the elastic 
constants tensor. Since $u_{ij}=u_{ji}$ by definition, in principle there are only 6 independent 
elastic constants in a 2D solid. Additionally, the elastic energy density must be invariant under 
the point group symmetries of the crystal. Note that Eq. (\ref{Eq:free}) is not a complete invariant 
under coordinate transformations since for that the integration measure should be completed with 
the factor $\sqrt{\mbox{det} g}\approx 1 + u_{ii}$. The description in the continuum of a 2D 
hexagonal 
crystal is essentially different from a fluid or any continuous media in the sense that the action 
must not be invariant under diffeomorphisms given that the atoms in the solid define a special 
coordinate system \cite{AL_88}.

For isotropic media, only two elastic constants are independent. Since in that case the response of 
the solid must be independent of the direction, the only tensor available to construct higher order 
tensors is the Kronecker delta $\delta^{ij}$. The only 4-rank tensors formed by $\delta^{ij}$ 
satisfying the symmetries of $\mathcal{C}^{ijkl}$ are $\delta^{ij}\delta^{kl}$ and 
$\delta^{ik}\delta^{jl}+\delta^{il}\delta^{jk}$, therefore
\begin{align}
\mathcal{C}^{ijkl}=\lambda \delta^{ij}\delta^{kl}+\mu \left(\delta^{ik}\delta^{jl}+\delta^{il}
\delta^{jk}\right),
\label{Eq:stress_tensor}
\end{align}
where $\lambda$, $\mu$ are the Lam\'e coefficients. 
%(the subscript indicates that these 
%correspond to the bare elastic constants, which are renormalized due to the anharmonic coupling 
%with out-of-plane displacements as dicussed in Sec. \ref{sec_anharmonic}). 
In 2D, this is precisely the case of hexagonal crystals. In the particular case of graphene, the 3 
independent components of $u_{ij}$ transform according to
\begin{gather}
u_{xx}+u_{yy}\sim A_1,\nonumber\\
\left(\begin{array}{c}
u_{xx}-u_{yy}\\
-2u_{xy}
\end{array}\right)\sim E_2.
\label{Eq:strain_sym}
\end{gather}
Then, among the six independent combinations of the form $u_{ij}u_{kl}$, only two of them form 
scalars belonging to $A_1$ irreducible representations. This is inferred from the algebraical 
reductions\begin{gather}
A_1\times A_1\sim A_1,\nonumber\\
A_1\times E_2\sim E_2,\nonumber\\
E_2\times E_2\sim A_1+A_2+E_2.
\label{Eq_red2}
\end{gather}
Therefore, for crystals with $C_{6v}$ there are only 2 independent elastic constants corresponding 
to the Lam\'e coefficients. Thus, the stretching energy for graphene has the explicit form
\begin{equation}
 F_{\text{st}}=\frac{1}{2} \int d^2 \vec{r} \left( \lambda u_{ii}^2 + 2 \mu u_{ij} u_{ij} 
\right). 
\label{eq:stretching_energy}
\end{equation}
Notice that inclusion of the quadratic term $\partial_i h\partial_j h$ in $u_{ij}$ makes the 
theory defined by the above equation an interacting theory. For free membranes this has important 
consequences as will be seen in Section~\ref{sec_anharmonic}. If we discard these anharmonic terms 
in Eq.~(\ref{eq:stretching_energy}), we obtain two in-plane acoustic phonons 
with linear dispersion relations. At low momentum these modes can be classified as longitudinal and 
transverse, and the respective dispersion relations are given by
\begin{align}
 \omega_{\vec{q}}^{L} & = \sqrt{\frac{\lambda + 2\mu}{\rho}} \left|\vec{q}\right| = v_L 
\left|\vec{q}\right|,\label{eq:LA_dispersion}\\
 \omega_{\vec{q}}^{T} & = \sqrt{\frac{\mu}{\rho}} \left|\vec{q}\right| = v_T 
\left|\vec{q}\right|,\label{eq:TA_dispersion}
\end{align}
where $\rho\approx 7.6 \times 10^{-7}$ kg/m$^2$ is the graphene's mass density and $v_{L/T}$ is the 
longitudinal/transverse sound velocity. Typical values for the elastic constants of graphene are 
$\mu \approx 3\lambda \approx 9$ eV \AA$^{-2}$ \cite{Kudin_etal,LWK08,ZKF09}, leading 
to $v_{L/T} \sim 10^4$ m/s.

Information about the bending of the membrane, or in other words, its embedding in 3 dimensional space, such as changes from 
point to point of the normal vector to the 
surface, 
$\hat{\mathbf{n}}=\partial_1\mathbf{r}\times\partial_2\mathbf{r}/|\partial_1\mathbf{r}
\times\partial_2\mathbf{r}|$, are described by the second fundamental form,
\begin{align}
\mathcal{F}_{ij}=\hat{\mathbf{n}}\cdot\partial_{i}\partial_{j}\mathbf{r}.
\label{Eq:curvature1}
\end{align} 
%The graphene membrane is also free to oscilate in the out-of-plane direction. The first 
%geometrical consequence is that the normal to the surface is not globally defined since 
To lowest order in the displacement fields, and assuming 
smooth out-of-plane displacements, the normal vector at each point of the membrane is 
given by
$\hat{\mathbf{n}}\approx\left(-\partial_x h, -\partial_y h, 1\right)$. In this approximation the 
second 
fundamental form is given by
\begin{align}
\mathcal{F}_{ij}\approx\partial_i\partial_j h.
\label{Eq:curvature2}
\end{align}
%To stabilize the solid with respect to fluctuations in the out-of-plane direction an extrinsic term 
%must be added to the elastic energy. Such term reflects the energy cost due to the bending of 
%graphene, and depends explicitly on the embedding, therefore on the components of the second 
%fundamental form. 
Two scalars may be constructed from the second fundamental form: the mean extrinsic curvature, 
$\mathcal{F}_{ii}$, and the intrinsic or Gaussian curvature $\mbox{det}\mathcal{F}$, but note that 
the latter is a purely topological term which does not affect the 
equations of motion. The bending energy is then a quadratic form of the former \cite{NPW04},
\begin{align}
F_{\text{b}}=\frac{\kappa}{2}\int d^2\vec{r} 
\left(\mathcal{F}_{ii}\right)^2\approx\frac{\kappa}{2}\int d^2\vec{r} \left(\nabla^2 
h\right)^2.
\label{Eq:bending}
\end{align}
Here $\kappa \approx 1$ eV \cite{ZKF09} is the (bare) bending rigidity. The 
dispersion relation of out-of-plane or flexural acoustic phonon modes of a free membrane is 
then quadratic,
\begin{align}
\omega_{\vec{q}}^{F}=\sqrt{\frac{\kappa}{\rho}}\left|\vec{q}\right|^2.
\label{Eq:flexural}
\end{align}
However, a rotational symmetry breaking tends to linearize the dispersion relation, for instance, 
due to the presence of strain. On the other hand, these modes are strongly suppressed in supported 
samples, where part of the spectral weight is transferred to a hybrid state with the surface 
Rayleigh mode of the substrate \cite{AG_13}.

Putting together the two contributions to the elastic energy of the membrane we obtain
\begin{equation}
 F_{\text{elastic}}=\frac{1}{2}\int d^2 {\vec r} \left(\kappa \left(\nabla^2 h\right)^2 + \lambda 
u_{ii}^2 + 2\mu u_{ij} u_{ij} \right).
\label{eq:elastic_energy_full}
\end{equation}
This equation has the same functional form as the usual von 
K{\'a}rm{\'a}n free energy for thin plates \cite{LL_59}. However, two differences must be pointed 
out. First, for a thin plate the bending rigidity, $\kappa$, is related to the 
elastic constants $\lambda$ and $\mu$ and the thickness of the plate. The same is not true 
for atomically thin membranes, for which it is not possible to define a thickness. Therefore, 
$\kappa$ is an independent parameter of the model for a membrane. Secondly,   
Eq.~(\ref{eq:elastic_energy_full}) corresponds to the potential energy of a Hamiltonian, not a 
free energy. The difference becomes important when thermal fluctuations, discussed in 
Section~\ref{subsec:membrane_classical}, are taken into account. For $\kappa$ sufficiently large, 
as in a plate, anharmonic interactions are suppressed and the free energy is given by the same 
expression as the potential energy. If $\kappa$ is small, the case of a membrane, then anharmonic 
interactions become dominant and the free energy can have a very different form than the potential 
energy.

\subsection{Electron-phonon coupling}

\label{Sec:electron-phonon_coupling}
The symmetry approach has been widely applied to the problem of
strained graphene ({\it e.g.} \cite{Manes07},\cite{Winkler10},\cite{Linnik12}).  With the help of
group analysis it is possible to construct a systematic derivative
expansion of the low energy effective Hamiltonian of graphene in the
presence of non-uniform elastic deformations. For the sake of
simplicity we will consider only the case of spinless graphene. The
spinful case and the role of strain in spin relaxation are discussed in
Refs.~\cite{H1,H2}. We restrict ourselves to linear order in the
electron momentum $k$ (first derivatives), so that the effective
Hamiltonian will be a function of the strain tensor $u_{ij}$ and its
derivatives, and the electron fields $\psi$ and $\psi^{\dag}$. Beyond
the Dirac approximation the procedure can be straightforwardly
extended to higher order in $k$.  In this way, the systematic
expansion is characterized by two integers, $(n_q,n_k)$, the order of the
derivatives of the strain tensor and the electron fields respectively.

Since we are interested in long wavelength deformations that are not able to couple different 
valleys, in what follows we will construct the possible interactions allowed by symmetry based on   
the little group
approach~\cite{brad}. 
%
%\textcolor{red}{The graphene Fermi surface at half filling consist of only two non equivalent Fermi points (Dirac points $K_2=-K_1$)~\cite{NGetal09}. We will consider all possible allowed terms for the little group of one of these two point, lets say $K_1$, and mapping these terms with the remaining operation of the point group we will obtain the corresponding allowed therms for the other Dirac point $K_2$. This mapping can be done alternatively by time reversal operation $\theta$. Putting all these terms together we can obtain the desired point group invariant Hamiltonian.  The little point group of $K_1$ is given by $C_{3v}$. $C_{3v}$ is a subgroup of $C_{6v}$ with only $6$ elements: rotations by multiples of $2\pi/3$ around the $OZ$ axis and reflections by three vertical planes. Since both $C_2$ (the extra operation belonging to $C_{6v}$) and time reversal $\theta$ maps one Dirac point into the other, besides $C_{3v}$, $K_1$ is also invariant under the combined operation $C_2 \theta$.}
%
As explained in Ref.~\cite{MJSV13} it is
sufficient to consider terms invariant under $C_{3v}$, the little
group of one of the inequivalent Dirac points, together with the combined operation $\mathcal{T} C_2$, where $C_2$ is a $\pi-$rotation around the $z$ axis. Once the degrees of freedom are established and the symmetries of the
system are determined, all we need to know is the way they
transform under the relevant group operations and how these can be decomposed as a sum of
irreducible representations. Using standard group character operation \cite{brad} it
is possible to determine how many and which are the terms allowed by symmetry
at a given order $(n_q,n_k)$.  As explained in Ref.~\cite{MJSV13},
up to first order in derivatives, $n_q + n_k \leq 1$, there are only
six terms involving the strain tensor. These six
terms corresponding to $K_+$ are listed in Tab.~\ref{Tab:tH}
together with their physical significance and the corresponding
relative sign at the inequivalent Dirac Point $K_-$.

\begin{table}[h]
\begin{tabular}{|c || c |c | c | c | }
\hline
$H_i$ & $(n_q, n_k)$ & Interaction term & Physical interpretation & $K_-$\\
\hline\hline
$H_1$ & (0,0) & $(u_{xx}+u_{yy})\mathbbmtt{1}$ &  Position-dependent electrostatic pseudopotential & $+$\\
 % \hline
  $H_2$ & (0,0) & $(u_{xx}-u_{yy})\s_x-2u_{xy} \s_y$& Dirac cone shift or U(1) pseudogauge field 
$(\mathcal{A}^{el}_x , \mathcal{A}^{el}_y)$ & $-$ \\
 \hline
$H_3$ & (0,1) & $\big[(u_{xx}-u_{yy})k_x-2u_{xy} k_y\big]\mathbbmtt{1}$ & Dirac cone tilt & $-$\\
% \hline
$H_4$ & (0,1) & $(u_{xx}+u_{yy})( \s_x k_x+\s_y k_y)$ & Isotropic position-dependent Fermi velocity& $+$\\
 %\hline
$H_5$ & (0,1) & $u_{ij} \s_i k_j\;\; ;\;\; i,j=x,y$& Anisotropic position-dependent Fermi velocity & $+$\\
 \hline
$H_6$ & (1,0) & $\;\;\;\;\big[\p_y(u_{xx}-u_{yy})+2\p_x u_{xy} \big]\mathbf\s_z\;\;\;\;$ & Gap opening by non-uniform strain & $-$\\
\hline
\end{tabular}
\caption{ Effective low energy terms in the Hamiltonian for the electron--strain
  interactions allowed by symmetry. The role of the gamma matrices as
 projectors into irreducible representations is clarified in \cite{CGetal13}.}
\label{Tab:tH}
\end{table}

We can now write the most general Hamiltonian in the presence of
non-uniform strain, up to first derivatives of the strain tensor
$u_{ij}$ (\ref{usual}), and up to first order in the derivatives of the electron
fields $\psi$ and $\psi^{\dag}$.  It is important to note that $u_{ij}$
is a function of two independent variables $u_i$ and $h$. In order to
keep the necessary number of independent coupling constants, we will
introduce for each $H$ term, function only of $u_{ij}$, an extra term $\tilde{H}$, function only of $h$;

%The undetermined coupling constants should be fixed setting up a microscopic model.

\beq
H=H_0+\sum_{i=1}^6 g_i H_i+\sum_{i=1}^6 \tilde g_i \tilde H_i.
\label{Eq:genham}
\eeq 
The first term is the standard Dirac contribution $H_0= v_F( \pm \s_x k_x+\s_y k_y)$, and the terms $H_i$ are given
in Tab.~\ref{Tab:tH}, and described below. The terms $\tilde H_i$ are obtained from those in
Tab.~\ref{Tab:tH} through the substitution $u_{ij}\to \p_i h\p_j h$.
Of course, the symmetry machinery can give us all the possible terms up
to a given order in derivatives, but  the
coupling constants remain undetermined and have to be fixed with a concrete microscopic model. 

The second-quantized Hamiltonian operator  is  given by 
$\mathcal{H} = \int d^2 \vec{r}\, \psi^\dagger H \psi$, where the symmetric convention for the 
derivatives acting on the electron fields is assumed, i.e.,   
 \hbox{$\psi^\dagger k_i\psi\!\to\! -i/2(\psi^\dagger\overleftrightarrow{\p_i}\psi)\!\equiv\!-i/2(\psi^\dagger\p_i\psi -\p_i\psi^\dagger\psi)$}.
In  more general terms,  $\partial_j$ acts only on the electron fields. The construction of second quantized hermitian operators becomes simpler using the symmetric derivative convention. Note that Tab.~\ref{Tab:tH}
gives the orders of the derivatives when terms are written with the
symmetric convention.

%Table \ref{tH}  displays  all the hermitian,  symmetry-allowed terms of given orders  $(n_q,n_k)$ in the derivatives of the electron fields ($n_k$) 
%and strain ($n_q$), as indicated in the second column. The remaining columns give their physical interpretation and the relative sign of the  
%couplings at the two non-equivalent Dirac points. In what follows we will comment briefly on the physical significance of the various terms which, 
%with the exception of $H_6$, have already been given~\footnote{Note that the term in Eq.~(45) of ref.~\cite{WZ10} is proportional to the combination 
%$2H_5\!-\!H_4$} in refs.~\cite{WZ10,Lin12}:

\begin{itemize}

\item[$\bullet$] $H_1=(u_{xx}+u_{yy})\mathbbmtt{1}$: This term has been already
  described in Ref. \cite{SA_02}. It is a scalar potential
  \begin{align}
  V_s\left(\vec{r}\right)=g_1\left(u_{xx}\left(\vec{r}\right)+u_{yy}\left(\vec{r}\right)\right).
  \label{Eq:deformation_potential}
  \end{align}
   The coupling $g_1$ is called deformation potential and its strength
   has been estimated to be of the order $4$ eV for single layer graphene
   \cite{CJS10}, although other works \cite{PBS14,SCP14} argue that
   screening leads to a strongly suppressed $g_1\approx 0$. Its
   physical consequences have been explored in
   Ref.~\cite{LGK11} and will be discussed in Section \ref{sec_electronic}.

\item[$\bullet$] $H_2=(u_{xx}-u_{yy})\s_x-2u_{xy} \s_y$: This term represents  a
  shift in momentum space of the Dirac cone. It corresponds to
  the well known $U(1)$ elastic pseudo-gauge field
  \begin{align}
  \vec{\mathcal{A}}^{\mathit{el}}\left(\vec{r}\right)=\pm 
g_2\left(u_{xx}\left(\vec{r}\right)-u_{yy}\left(\vec{r}\right),-2u_{xy}\left(\vec{r}\right)\right).
  \label{Eq:gauge_field}
  \end{align}
   In the tight binding formalism $g_2 \sim \beta=-\partial( \log t)/\partial
   (\log a)$, where $t$ is the hopping integral between nearest neighbours and $a$
   is the lattice constant. It is at the basis of most of the results in the 
   literature related to  strain engineering and there are experimental realizations 
     \cite{VKG10}. It has also been used to
   explain data in artificial graphene \cite{GMetal12}.

\item[$\bullet$] $H_3=\big[(u_{xx}-u_{yy})k_x-2u_{xy}
  k_y\big]\mathbbmtt{1}$: Dirac cone tilt. This term appears naturally in
  the description of the two dimensional organic superconductors
  \cite{KKetal07} which are described by anisotropic Dirac
  fermions. It also arises when applying uniaxial strain in the zigzag
  direction, a situation that has been discussed at length in the
  literature \cite{GFetal08,WGS08,MPetal09,CJS10}. A novel physical manifestation of 
  the Dirac cone tilt has been described recently in \cite{L14}.

\item[$\bullet$] $H_4=(u_{xx}+u_{yy})( \s_x k_x+\s_y k_y)$: Isotropic
  position-dependent Fermi velocity \cite{JSV12}.

\item[$\bullet$] $H_5=u_{ij} \s_i k_j\;\; ;\;\; i,j=x,y$: Anisotropic
  position-dependent Fermi velocity \cite{JSV12}. This term, together
  with $H_4$, has been predicted to arise within the geometric modelling of
  graphene based on techniques of quantum field theory in curved space
  \cite{JCV07}. It was later obtained in a tight binding model by
  expanding the low energy Hamiltonian to linear order in $\vec{k}$ and
  $u_{ij}$ \cite{JSV12,JMV13}. Since the Fermi velocity is arguably the most
  important parameter in graphene physics, this term affects all
  the experiments and will induce extra spatial anisotropies in
  strained samples near the Dirac point
  \cite{GKV08,BP08,ZBetal09,G__09c,PAP11,GTetal12}. This term will be 
  described in detail in Section \ref{Sec:QFT}.

\item[$\bullet$] $H_6=\big[\p_y(u_{xx}-u_{yy})+2\p_x u_{xy}
  \big]\mathbf\s_z$: This is a very interesting term that suggests a
  new gap-opening mechanism \cite{MJSV13}. It can be seen as the
  Zeeman coupling of pseudospin to the associated pseudomagnetic field
  $B_z= \p_x \mathcal{A}^{el}_y-\p_y \mathcal{A}^{el}_x$ \cite{ZLetal13}. The magnitude of this gap
  has been estimated to be $7$ meV \cite{MJSV13} .

%\item[$\bullet$] To first order in the derivative expansion we can also construct an invariant involving the antisymmetric derivativeof the 
%displacement vector $\om=\p_x\xi_y-\p_y\xi_x$:   \beq\label{om01} \om(k_x\s_y-k_y\s_x)=\omu_{ij}k_i\s_j, \eeq  but, as shown in Ref.~\cite{JMV13}, 
% it can be eliminated by a local rotation of the pseudospinor $\psi\to\exp(-\frac{i}{2}\om \s_z)\,\psi$. Thus the effective 
% hamiltonian~\eqref{genham} does not  depend on $\om$.

\end{itemize}

It is important to notice that because of the symmetric convention,
some non-spurious terms seem to be missing, as it is the case of the
vector field $\Gamma_i$ \cite{JSV12}. This is just an artefact of the
symmetric convention and all the relevant terms
will appear in the equation of motion. In the case of $\Gamma_i$, its
presence becomes evident by integrating by parts $H_4$ and $H_5$.

%Finally, for completeness, we show in Table \ref{Tab:tlab} an estimation
%of the value of the coupling constants using a generalized tight
%binding calculation \cite{MJSV13}. The microscopic model assume a
%hopping $t_2$ between next to nearest neighbors and a contribution $V$
%of a nearest neighbour potential to the on-site energy of an electron
%in a $p_z$-orbital. The primes denote derivatives respect to the position
%evaluated at the atomic equilibrium position.

%\pagebreak

\vskip0.5cm
%\begin{table}[!h]
%\begin{tabular}{||c | c || c | c || }
%\hline 
%$g_1$ & $\;\;\;\frac{3\sqrt{3}}{2}t_2'a+\frac{3}{2}V' a\;\;\;$ &  $\tilde g_1$ & $0$\\

%$g_2$ & \;\;$\frac{\beta}{2a}$& $\tilde g_2$ & $ 0$\\

%$g_3$ & \!\!\!$-\frac{9\sqrt{3}}{4}t'_2a^2$ & $\tilde g_3$ & $ 0$\\

%$g_4$ & $\frac{\beta}{4}v_F$ & $\tilde g_4$ & $0$\\

%$g_5$ & $(\frac{\beta}{2}+1)v_F$& $\tilde g_5$ & $\;\;-v_F\;\;$\\

%$g_6$ & $\frac{3}{8}V'a^2$ & $\tilde g_6$ & 0\\
%\hline
%\end{tabular}
%\caption{Couplings for the effective Hamiltonian (in the Lab frame\cite{JMV13}).}
%\label{Tab:tlab}
%\end{table}
%\end{document}

\subsubsection{Experimental confirmation of the elastic pseudomagnetic fields}
\label{Sec:expgauge}
The term $H_2$ representing fictitious magnetic fields coupling with opposite sign to the two 
valleys was the first to be identified in the early publications and its physical consequences were 
discussed at length in the review \cite{VKG10}. The theoretical suggestion   that Landau levels can 
form associated to strain in graphene was done in \cite{GKG10,GGetal10}. The simple observation that 
applying a trigonal strain gives rise to uniform pseudomagnetic fields in a region of the sample, 
gave birth to real strain engineering. The theoretical proposal was soon followed by the 
experimental confirmation. A Landau level structure was observed in a STM experiment on graphene 
samples with triangular nanobubbles in \cite{LBM10}. Fields of up to 300 Tesla were estimated in 
the central region of the bubbles. This pioneer experiment was followed by many reports of elastic 
Landau levels in graphene samples \cite{AAG11,Ketal12,MHetal13}, cold atoms \cite{TGetal12}, and 
artificial 
graphene \cite{PGLetal13}.
Elastic gauge fields have also been described in bilayer graphene \cite{MPO12}, other 2D 
semiconductors \cite{RCC15} and lately in the recently discovered three dimensional Weyl semimetals 
\cite{CFLV15}.
The effects of the elastic pseudomagnetic fields on various transport coefficients have been 
analysed in numerous publications \cite{Gradinar:PRB12,Cosma:PRB14,PSLetal10} and will be described 
in more detail in Section \ref{sec_electronic}.

\subsubsection{Experiments probing local variations in \texorpdfstring{$v_{F}$}{Lg}}
\label{Sec:expvfermi}
There are several ways to measure the Fermi velocity experimentally. In this section, we concentrate on measurements that are local and can thus 
probe the spatial variation of the Fermi velocity predicted by Eq.~\eqref{Eq:vfermi}. The optimal experimental technique to do this is scanning 
tunnelling spectroscopy (STS), which is sensitive to the local density of states (LDOS)\footnote{See 
section~\ref{Sec:superlattices} for a discussion of how superlattices may also induce local variations of the LDOS that is seen by STM.}.  Since the LDOS depends on $v_F$ as  $\rho(E) = 2 |E|/\pi 
v_F^2$, if the strain variation is smooth one may replace $v_F\rightarrow v_F(x)$ and fit the slope of the measured LDOS to this expression to obtain 
a local measurement of $v_F$. 

Two recent experiments have reported evidence of a spatially varying $v_F$ using this method. The first was performed on graphene grown on a Rh 
foil~\cite{YCY13}, which develops quasiperiodic ripples. Local variations of $v_F$ on different parts of the ripples were reported with a rather wide 
distribution of values, consistent with the strain induced by the ripples. In the second, graphene was grown on BN~\cite{JKS14}, where strained ridges 
were randomly formed. In a more quantitative comparison, the LDOS measured in these ridges reveals a spatially dependent $v_F$ which is shown to be 
correlated with the height of the ridges. This represents convincing evidence that $v_F$ fluctuations indeed originate from strain. 

An alternative way to measure $v_F$ locally is with Landau Level spectroscopy. In the presence of a magnetic field, the LDOS is rearranged into a 
series of well defined peaks at the Landau Level energies, given by $E_n = \mathrm{sgn}(n)v_F 
\sqrt{2 e \hbar B n}$. A measurement of $v_F$ can thus be obtained by fitting 
this expression as well. This type of experiment has been performed with randomly strained graphene as grown on SiO$_2$, and the measurement indeed 
reveals a local variation of $v_F$ of 5-10 $\%$ \cite{LLA11}.

\subsection{Quantum field theory in curved spaces}
\label{Sec:QFT}
One of the features that makes graphene unique is the way the electronic degrees of freedom couple 
to structural deformations of the lattice, and how this allows to modify its electronic properties 
in interesting ways. In the previous section, we have shown how the deformation of the lattice is 
described in terms of a strain tensor $u_{ij}$, and what are the possible strain couplings allowed 
by the discrete symmetry of the lattice in the low energy Dirac Hamiltonian. In this section, we 
discuss a complementary approach to the study of the effects of deformations which provides 
insightful connections to the geometry of the problem. This approach originates in the idea that, 
since local distances in the strained lattice can be described in terms of a metric (the first 
fundamental form introduced in \eqref{Eq:metric}), a natural model to describe the coupling of 
electrons to the deformation should be the Dirac equation in curved space. The formalism in full 
glory is taken from quantum field theory (QFT) in curved space \cite{BD82} and was explained in 
detail in the review article \cite{VKG10}. In this section we describe the modern uses of the metric 
(unrelated to strain deformations) to generate response functions in condensed matter theory.

While this approach is phenomenological rather than microscopic in nature, it offers an interesting 
geometrical interpretation of the different couplings, and it is a well understood formalism. The 
Dirac equation in curved space is completely determined by the metric  $g_{ij} = \delta_{ij} + 
2u_{ij}$ and takes the form
\begin{equation}
\mathcal{H} =- i \int d^2 \vec{x} \sqrt{g} \bar{\psi} \gamma^a e_a^{i} (\partial_{i} +
\Omega_{i}) \psi,
\label{curveH}
\end{equation}
where $a=1,2$ and $i=1,2$ run over the space dimensions, $\bar{\psi}= \psi^{\dagger} \gamma^0$, $\gamma^0\gamma^i = \sigma^i$ are the Pauli matrices, 
$e_a^i$ are the tetrads, $\Omega_i$ is the spin connection and $\sqrt{g}$ is the volume factor. These three geometric quantities can be obtained from 
$g_{ij}$, see Ref.~\cite{JSV12} for explicit expressions. Many authors have studied this model in connection with different problems, which include 
the spectrum of fullerenes \cite{GGV93}, the modelling of disclinations\cite{CV07,CV07a}, graphene wormholes  \cite{GH10} proposals of metrics to 
generate in optical lattices \cite{BCL11}, the holography conjecture \cite{K13} and the Unruh effect \cite{IL12,CG12,I14}. This model has also been 
applied to the study of the optical conductivity in graphene \cite{SSZ11,CTO14}, to the scattering by smooth deformations in topological insulator 
surfaces\cite{DHA10,PLV11}, and to study the Quantum Hall effect \cite{L09}. 

To be able to compare this approach with the standard tight-binding results, one proceeds by expanding  Eq.~\ref{curveH} to first order in $u_{ij}$ 
for small strains, which can be written as
\begin{align}
\label{Hcurved}
H =  -i [v_{ij}({ x}) \sigma_i \partial_j +  \sigma_i\Gamma_i],
\end{align}
where 
\begin{align}
v_{ij} =& v_F(\delta_{ij} + \delta_{ij} u_{kk} - u_{ij}) \\
\Gamma_i =& \frac{1}{2} \partial_j v_{ij}
\end{align}
where $v_F$ is the Dirac fermion velocity in the absence of distortion. The most interesting feature of this model is the presence of the matrix 
$v_{ij}$, which may be interpreted as a tensorial, spatially dependent Fermi velocity. This interpretation was first put forward in Ref.~\cite{JCV07}, 
where it was shown that it can be responsible for local modulations of the density of states (DOS) that are correlated with strain. The model also 
displays a spin connection, which can be responsible for pseudo-spin precession, for example. It can be seen that the two terms in the expression for 
the Fermi velocity are consistent with those allowed by the discrete symmetry of the lattice, terms $H_4$ and $H_5$ in Section 
\ref{Sec:electron-phonon_coupling}. 

The model just presented has many nice features and has proven useful, but it has the drawback that it has not been derived from a microscopic 
Hamiltonian. The simplest microscopic derivation of the coupling of electrons to strain can be done with a tight-binding model, assuming that the 
hopping parameters change locally with distance, $t_n = t (1- g_2 |\Delta u_n|)$, where $g_2$ is 
the dimensionless electron phonon coupling defined 
in the previous section. This model has been used extensively across the graphene literature to show that, to lowest order, strain couples as a gauge 
field of the form $H_2$ in the low energy Hamiltonian. The space-dependent Fermi velocity is also present in this model, when the expansion is taken 
to first order both in strain and in momentum \cite{JSV12}, which gives a Hamiltonian
\begin{equation}
H_{TB} =  -i [v_{ij}^{TB}({ x}) \sigma_i \partial_j +  v_F \sigma_i
\Gamma_i  + i v_F \sigma_i \mathcal{A}^{\mathit{el}}_i],
\label{HTBcomplete}
\end{equation}
where the field $\mathcal{A}^{\mathit{el}}_i$ is the one given in the Hamiltonian $H_2$ of Section 
\ref{Sec:electron-phonon_coupling}, and $v_{ij}^{TB}$ is the Fermi velocity tensor obtained from 
tight-binding, which is 
\begin{equation}
v_{ij}^{TB} = v_F[\delta_{ij} - {\frac{g_2}{4}}(2 u_{ij} + \delta_{ij} u_{kk})],
\label{Eq:vfermi}
\end{equation}
with $v_F = 3/2ta$. Again, the two terms found from the microscopic approach are consistent with those found in the symmetry analysis, $H_4$ and 
$H_5$. One may observe, however, that the coefficients obtained for these terms in tight-binding are not the same as those in the curved space model. 
The physical predictions of these two models are however qualitatively similar. The fact that the tight binding model gives the functional form of the 
Fermi velocity tensor that is different from the one postulated by curved space model was recently discussed in Ref.~\cite{Y14}. 

Analytical expressions for the dependence of the Fermi velocity on strain from which the coefficients of the terms $H_4$ and $H_5$ can be extracted 
were also reported in Refs.~\cite{PRP10,PAP11}. The anisotropy in the Fermi velocity was demonstrated numerically in tight-binding 
studies~\cite{PCP09,PAP10} and in \emph{ab initio} calculations~\cite{CJS10}. The corrections to the Fermi velocity tensor to second order in strain 
were computed in Ref.~\cite{RMP13}. The approach of Refs.~\cite{ON13,VZ14} based on discrete differential geometry can also be used to compute higher 
order corrections. The tight binding model with space-dependent velocity has also been used to study bound states in strain superlattices \cite{PAP12} 
and to study the modulation of persistent currents in graphene rings\cite{FLU13}.  The possibility of tuning the strain to get a given velocity 
profile has been suggested in \cite{JP14}.

It is worth emphasizing that the Fermi velocity tensor, as well as the other terms obtained from the symmetry analysis, was computed by expanding the 
Hamiltonian around the high symmetry point $K$. Only in this way one can classify the different terms under the symmetry of the little group at $K$. 
It has recently been argued that, in the case of constant strain, since the Dirac point is shifted due to the constant vector potential, the Fermi 
velocity tensor has a different form when the expansion is performed around the shifted Dirac point \cite{ON13,VZ14}.
 
There are several works that have addressed the modelling of a spatially dependent Fermi velocity with a different phenomenological model, which 
amounts to the direct replacement of $v_F$ by a scalar function $v(x)$ in the Hamiltonian \cite{P09,RPA10,CT10,YCZ11,LM15}. This model is equivalent 
to considering just the isotropic coupling to the strain $H_4$ with a scalar function $v(x) = 
v_F u_{kk}$. 

The motivation to study the case when the only term affecting the Fermi velocity is the isotropic 
one is to analyse how to control wave propagation 
by velocity changes in analogy with the modulation of the refraction index in optics. These works studied the propagation of Dirac electrons in 
different scalar velocity profiles, concluding that the transmission through a velocity barrier has a strong dependence on the incidence angle, but is 
always equal to one at normal incidence. When $v_F$ in the barrier is larger than outside it, there is a critical angle above which no transmission 
is possible, in analogy with the Brewster angle.

\subsubsection{Uses of the metric as an auxiliary field to generate response coefficients}
\label{Sec:metricuses}
We here review some uses of the metric
and the spin connection as 
auxiliary tools to derive exotic transport properties of materials. 

In an effective action framework transport coefficients can be computed from the
partition function of a system by coupling appropriate source terms to the operators whose response we want
to analyse. The best example is that of the electromagnetic currents obtained by coupling a
background electromagnetic potential (even in the absence of actual fields). Then Kubo formulae
can be used to compute the coefficients of the current-current correlation functions.
%The basis is the path integral formalism used in statistical mechanics (SM), quantum field theory (QFT) condensed matter, or any branch of physics:
%\beq
%{\cal Z}[\Phi]\sim e^{\alpha S[\Phi]},
%\eeq
%$\alpha$ can be $1/\beta$ in SM or $-i\hbar$ in QFT, etc. The response functions associated to a given operator $O_i$ constructed as a polynomial in the fields $\Phi$ are obtained by functional derivatives of the generating functional:
%\beq
%S[\Phi]=\int d^n x dt{\cal L}[\Phi]
%\eeq

Thermal transport is one of the most difficult aspects to address in materials. In a seminal 
work J. M. Luttinger \cite{Lut64} proposed the use of the metric as an auxiliary tool to analyse 
thermoelectric  transport coefficients in solids. In the remarkable footnote 7 he writes: {\it In fact, if the gravitational field didn't exist, one could invent one for the purpose of this paper.} The  aim was to derive an expression for the thermal conductivity 
similar to the Kubo formula of the electrical conductivity. The main difficulty with thermal transport 
coefficients is that there is no Hamiltonian which describes a thermal gradient since the temperature is a
statistical property of the system.  
Inspired by the relation between the metric and the energy-momentum or stress tensor ($T^{\mu\nu}$, 
with Greek indices running over space-time coordinates) that holds in
general relativity:
\beq
T^{\mu\nu}=\frac{2}{\sqrt{-g}}\frac{\delta {\cal S}}{\delta g_{\mu\nu}}.
\label{Eq:EMtensor}
\eeq
($\mathcal{S}$ is the action of the matter)
he proposed to couple the electron system to an auxiliary field  (an inhomogeneous gravitational 
field) which causes energy or heat currents to flow. The interaction term was $S_{int}=\int  h(r) 
\Phi(r,t)$ where $h(r)$ is the Hamiltonian density of the matter system and $\Phi(r,t)$ is a 
``gravitational potential". A varying $\Phi$ will give rise to a varying energy density which in 
turn will correspond to a varying temperature. No matter how ambiguous the original work can look 
today, it paved the way to very important recent developments in condensed matter.
An attempt to put this work on firmer grounds by deriving expressions similar to \eqref{Eq:EMtensor} for a non-relativistic system has appeared recently in \cite{GA14b}. The proposed formalism involves the Riemann Cartan geometry, which, in addition to curvature, includes torsion in the  background geometry (for a "primer" on torsion in condensed matter applications see \cite{JCV10}). The viscosity,  another important transport coefficient of the electronic fluid, also involves the metric in its definition as will be discussed in the next subsection.

\subsection{Topological aspects and time--dependent deformations}
\label{Sec:topo}

Topological phases of matter are a new paradigm in condensed matter
physics. These phases add to the standard classification
in terms of local order parameters and broken symmetries new quantum numbers associated
to topological invariants \cite{HK10,QZ11}. Since the early developments of the field and due to its peculiar properties, the honeycomb lattice 
has been used as the ideal model to demonstrate topological properties. The pioneer works proposing this lattice  as a ``condensed matter simulation 
of a three-dimensional anomaly"  \cite{S84,H88}, together with the analysis of the spin orbit in graphene done in  Refs. \cite{KM05,KM05b} opened the 
modern field of topological insulators. Most of the
new topological materials (graphene, topological insulators and superconductors, 
Weyl semimetals \cite{XTVS11,BB11}) share the property that their low energy electronic properties are described by Dirac fermions \cite{VV14,GM14} in one, two or three spacial dimensions  enlarging and widening the bridge between high energy and condensed matter in this century.

At a zeroth order approach to topological aspects, Dirac points in (2+1) dimensions are usually 
protected by a winding number: the Berry phase acquired by
the spinor wave function when circling around a Dirac point in momentum space. Time reversal symmetry
implies that Dirac points in crystals arise in pairs with opposite winding numbers
\cite{NN81}. 
Lattice deformations move the Dirac points that eventually can merge giving rise to interesting 
topological phases. The existence and merging of Dirac points in crystals was studied  in the early work
\cite{MGV07} and a very complete analysis of the issue has appeared in the
recent review \cite{GM14}.

In this section we review the late developments on the influence of lattice deformations on
the topological properties of graphene and related materials. The topological phases induced by a combination of strain and interaction will be discussed in Section \ref{Sec:topointeractions}.

\subsubsection{Hall viscosity}
\label{sec:Hallv}
In a similar way as done by Luttinger with the heat transport, the metric is being used more recently to analyse properties such as the 
Hall viscosity \cite{ASZ95} in topological insulators  and in quantum Hall fluids \cite{HLF11,HS12,BGR12,HLP13}. The analogue of friction in fluid dynamics is provided by viscosity, which causes dissipation of the energy (momentum) flow. In elasticity theory we can define (see Section \ref{Sec:elasticity})
\beq
T^{ij}=\mathcal{C}^{ijkl}u_{kl}+\eta^{ijkl}\dot u_{kl},
\eeq
where $u_{ij}$ is the strain tensor, and $\mathcal{C}^{ijkl}$ and $\eta^{ijkl}$ are the elastic and 
viscosity coefficients. 
Viscosity  can then be defined as the response of the system 
to a time-varying strain. In general, this viscosity tensor possesses both symmetric and 
antisymmetric components under the permutation of pairs of indices. While the symmetric part is 
generally associated to dissipation and vanishes at zero temperature, the antisymmetric part arises 
when time reversal symmetry is broken and was first described in the quantum Hall fluid 
\cite{ASZ95}. A stress-stress form for the response function 
yields a  Kubo formula for the viscosity. 

The Hall viscosity can be derived  by coupling the system to an external metric $g_{ij}$. Under a small time-dependent shape
deformation which preserves the volume: $g_{ij}=\delta_{ij}+\delta g_{ij}(t)$, the stress
tensor is
\beq
T^{ij}=P\delta^{ij}-\frac{1}{2}\eta^{ijkl}\partial_t \delta g_{ij},
\label{eq:hallv2}
\eeq
where $P$ is the pressure and the tensor $ \eta_{ijkl}=-\eta_{klij}$ is the Hall viscosity.
This transport coefficient is particularly interesting in the case
of fractional quantum Hall fluids (FQHF). Unlike the Hall conductivity which 
is quantized and dimensionless, the Hall
viscosity has units of $1/L^2$.
In a quantum Hall system the Hall viscosity, defined as the response of the electronic fluid to geometric shear distortions, is proportional to the
density of intrinsic angular momentum: $\eta_A = 1/2 s n$
($n$ is the particle number density and $s$ is the average spin
per particle, so $s n$ is the area density of intrinsic angular
momentum). 
Computation of the Hall viscosity in topological matter and 
its relation  to the total angular momentum is one of the most active areas of research nowadays  involving 
all the geometrical tools concerning the study of QFT in curved background fields including torsion \cite{HLF11,HLP13,CYF14}. In a very recent publication \cite{HKO14}, Ward identities have been used to 
relate Hall viscosities, Hall conductivities and the angular momentum.

In the early publication \cite{WZ92} it was recognized that Hall fluids
couple also to the curvature of the background surface, a fact that manifests itself by the presence of a shift
in the relation between the number of electrons $N_e$, and the number
of magnetic flux quanta $N_\phi$ going through the manifold. In a flat system they are related by
$\nu N_\phi=N_e$, where $\nu$ is
the filling factor.  In a compact curved manifold such as 
a sphere there is a shift ${\cal S}$ such that:
\beq
N_\phi=\nu^{-1}N_e-{\cal S}.
\eeq
It was observed that the shift on a sphere and in a torus were different (it was zero in a torus) what led  
the authors to conjecture that the Hall fluid did couple to the spin connection vector form defined as
$\Omega_\mu=\omega_\mu^{ab}\varepsilon_{ab}$. Coupling this vector as a source in an effective action 
formalism allows to obtain the shift as the coefficient of a mixed Chern-Simons electromagnetic--gravitational term of the form 
$S_{WZ}\sim \varepsilon^{ijk}\omega_i\nabla_j a_k$, known as the Wen--Zee term.
The effective Chern Simons action describing the geometrical factors in Hall fluids has been put in firmer grounds recently in \cite{CYF14}.

The Wen--Zee action can also be seen as an example of anomaly related transport, a field that is also investigated very actively these days in both condensed matter and high energies in the context of the quark-gluon plasma \cite{Satz11}. The most commonly cited example of the new non-dissipative transport phenomena occurring in the quark--gluon plasma
is the chiral magnetic effect \cite{FKW08} that refers to the generation of an electric
current parallel to a magnetic  field whenever an imbalance between the number of right 
and left-handed fermions is present. Another interesting example
is the axial magnetic effect (AME) associated with the generation of an energy current parallel to
an axial magnetic field, i. e. a magnetic field coupling with opposite signs to right and left handed
fermions. These new transport phenomena are specially relevant to the (3+1) dimensional Dirac materials, particularly in Weyl semimetals. The axial magnetic effect in Weyl semimetals as a fingerprint of the gravitational anomaly has been proposed in \cite{CCetal14}.
Of particular interest is the part associated to the anomalous transport coefficients generated by the gravitational anomaly \cite{Stone12,LMP11,GA14}.

\subsubsection{Transverse conductivity from time-dependent deformations}
\label{Sec:time}

The valley degree of freedom in graphene has been proposed as a novel ingredient to encode
information, its control and manipulation led to the concept of valleytronics \cite{RTB07}.
As mentioned in Section \ref{Sec:electron-phonon_coupling}, lattice deformations and defects couple to the
electronic current in graphene in the form of vector fields that couple with opposite
signs in the two valleys. This fact has given rise to several interesting proposals 
related to valley physics in strained graphene \cite{JLetal13}.  
It was shown in \cite{GKV08} that the combination of real and
fictitious magnetic field induces a valley asymmetry in the Landau level structure. 

Following Eq.~\eqref{Eq:gauge_field}, a time-dependent  deformation gives rise to a 
``synthetic electric field" $E^{\mathit{el}}_i=\partial_0 \mathcal{A}^{\mathit{el}}_i$ 
that also couples with opposite signs to the two Fermi points. 
The observable consequences of the ``synthetic electric field" in  graphene
were described  in the early publication \cite{OGM09}. It was recognized 
that, due to the opposite signs of the 
coupling at the two valleys, this synthetic electric field 
can drive charge--neutral valley currents 
in carbon nanotubes and graphene that are not subjected to 
screening and hence can have a stronger influence on the electronics
than the deformation potential. 
\begin{figure}[t]
\begin{center}
\includegraphics[angle=0,width=0.4\linewidth]{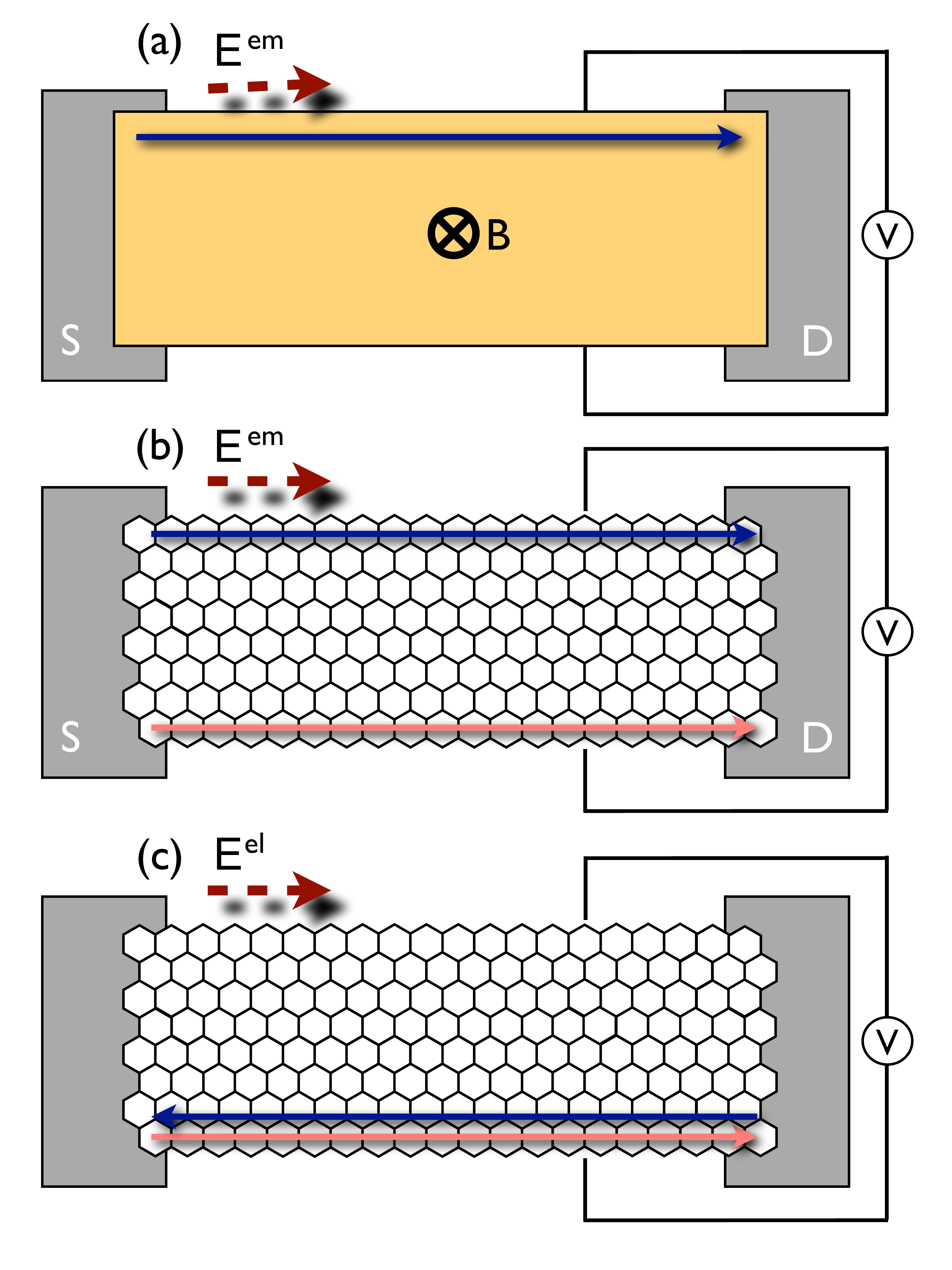}
\end{center}
\caption{(colour online) Comparison between the Hall effect (upper), the valley Hall effect 
(middle), 
and the effect proposed in the text (lower) following the scheme of the original implementation of 
the experiment first proposed by Hall. The arrows indicate the flow of valley polarized electrons 
under the action of an external electric field or voltage (upper and middle) and of a time 
dependent strain (lower part). In the later case the total current is
$<J^i_K+J^i_{K'}>\sim E^{el}_j$.(Fig. adapted from ref. \cite{VAA13}).}
\label{fig1}
\end{figure}

An interesting topological response due to the interplay of electromagnetic and a time-dependent  
elastic vector field was studied in \cite{VAA13}. The total Chern number - and hence the Hall 
conductivity - of a  time-reversal invariant system is zero. The topological signals in these 
systems, such as the spin Hall effect, have associated dissipationless transverse currents involving 
degrees of freedom -- spin, valley, layer \cite{CGV10} other than the electric charge. It was shown 
in \cite{VAA13} that the interplay of electromagnetic and elastic vector fields can give rise to a 
transverse  electric current. The argument goes as follows:
In the presence of an external electromagnetic and elastic field the gapped graphene Hamiltonian reads:
\begin{eqnarray}
\mathcal{H}=\sum_{\tau}\psi^{\dagger}_{\tau,\vec{k}}\left(\tau\sigma_{x}k_{x}+\sigma_{y}k_{y}
\right)\psi_ { \tau
,
\vec{k}}
&-&\sum_{\tau}[\psi^{\dagger}_{\tau,\vec{k}}\tau\sigma_{x}\left(eA^{em}_{x}+\tau g_2
\mathcal{A}^{el}_{x}\right)\psi_{\tau,\vec{k}}]-\nonumber\\
&-&\sum_{\tau}[\psi^{\dagger}_{\tau,\vec{k}}\sigma_{y}\left(e A^{em}_{y}+\tau g_2 
\mathcal{A}^{el}_{y}\right)\psi_{\tau,\vec{k}}+
m\psi^{\dagger}_{\tau,\vec{k}}\sigma_{3}\psi_{\tau,\vec{k}}],
\label{Eq:intham}
\end{eqnarray}
where $\tau=\pm1$ denotes the valley degree of freedom, $e$ is the electric charge, $\vec{A}^{em}$ 
and $\vec{\mathcal{A}}^{el}$ stand for the electromagnetic and the elastic vector fields 
respectively, and 
the parameter $g_2>0$ 
encodes the strength of the elastic coupling. The coefficient $\tau$ in front of the coupling 
of the electronic current to the elastic vector field marks the difference with the electromagnetic gauge field that couples with the same sign to the two valleys. In the presence of a time-reversal preserving mass $m$ (having the same sign in both valleys) as the one considered in  \eqref{Eq:intham} there is a non zero Chern number which  takes opposite values at the two Fermi points
$
C_{\vec{K}}=-C_{\vec{K}^{\prime}}=\sgn(m)/2.
$ 
In the absence of elastic deformations the  response to an external electric field ${\bf E}^{em}$ 
is a transverse charge current with opposite signs at each Fermi point: 
$
\langle J^{i}_{\tau}\rangle=e^{2}C_{\tau}\varepsilon^{ij}E^{em}_{j},
$ 
so the total charge current $\langle J^{i}_{\vec{K}}+J^{i}_{\vec{K}^{\prime}}\rangle$ vanishes.
The synthetic elastic field will induce a charge response in the system at each Fermi point: 
$
\langle J^{i}_{\tau}\rangle\sim\tau C_{\tau}\varepsilon^{ij}E^{el}_{j}.
$  
Now, because $C_{\vec{K}}=-C_{\vec{K}^{\prime}}$ the total net charge current is  non zero and its 
value 
is twice larger than the value at each Fermi point: 
\beq 
\langle J^{i}_{\vec{K}}+J^{i}_{\vec{K}^{\prime}}\rangle\sim2C_{\vec{K}}\varepsilon^{ij}E^{el}_{j}.
\label{Eq:totcurr}
\eeq

This result can also be understood  from the mixed Chern-Simons action arising from the combined
vector field $V_i=A^{em}_i+\tau \mathcal{A}^{el}_i$. 
Within a functional integral approach one can integrate out the fermionic degrees of freedom 
$\psi_{\tau,\vec{k}}, \psi^{\dagger}_{\tau,\vec{k}}$ in the action derived from Eq. 
\eqref{Eq:intham} and write the odd part of the effective Lagrangian:
\begin{eqnarray}
\mathcal{L}_\mathit{eff}&=&2 e g_2 
C_{\vec{K}}\varepsilon^{\mu\rho\nu}A^\mathit{em}_{\mu}\partial_{\rho}\mathcal{A}^\mathit{el}_{\nu}
+2e g_2 
C_{\vec{K}}\varepsilon^{\mu\rho\nu}\mathcal{A}^\mathit{el}_{\mu}\partial_{\rho}A^\mathit{em}_{\nu} 
+\nonumber\\
&+&J^{\mu}A^\mathit{em}_{\mu}+J^{\mu}_{el}\mathcal{A}^\mathit{el}_{\mu},
\label{Eq:effaction}
\end{eqnarray} 
where we have added to the Chern--Simons action the external sources:
$J^{\mu}$ is the total charge density current that naturally couples to the electromagnetic field, 
and $J^{\mu}_{el}$ is a classically conserved current associated to the elastic field 
$\mathcal{A}^{el}_{\mu}$. 
Notice that the standard Chern--Simons term bilinear in $A^{em}$ or $\mathcal{A}^{el}$ vanishes for 
the total action due to the opposite value of $C_K$ at the two valleys. Only the mixed term 
survives. The total charge current density ($\mathcal{S}_\mathit{eff}=\int 
d^{3}\mathcal{L}_\mathit{eff}$) is:
\begin{equation}
\langle J^{i}\rangle =\frac{\delta\mathcal{S}_{eff}}{\delta A^{em}_i}= 2 e\hat{\beta} C_{K} 
\varepsilon^{ij}\dot{\mathcal{A}}^{el}_{j}\equiv 2 e g_2 \frac{m}{|m|} \varepsilon^{ij}E^{el}_{j},
\label{Eq:chargeresponse2}
\end{equation} 
where  for simplicity it has been assumed that $\mathcal{A}^{el}_0=0$ and have replaced 
$\varepsilon^{ij0}=\varepsilon^{ij}$. Notice that the generation of this transverse electric current 
in a time reversal invariant system
does not contradict the general result of "no Hall current in T-invariant systems" because the Hall current is the response to an electric field while the present current is the response of the system to a time-dependent elastic deformation. A schematic description of the various possible valley currents is presented in Fig.~\ref{fig1}. The action discussed here is similar to the Wen--Zee action described in 
Section \ref{sec:Hallv} but there the additional vector field that couples to the electromagnetic field  
in the mixed Chern Simons term was induced by the geometric spin connection (higher order in derivatives, see
discussion in Section \ref{Sec:electron-phonon_coupling}) while here it is the elastic
gauge field. The result in both cases is similar: an electromagnetic response induced by 
a time--dependent deformation. This can also be seen as a quantum analogue of standard piezoelectricity.

Elastic electric fields have also been analysed recently in \cite{TUT13,FF14,YPZ14}.
As it is known, what prevents graphene from having observable topological properties is the the small value of the spin orbit coupling and the difficulty to open a gap in the spectrum. It has recently observed that depositing graphene on hexagonal BN opens a sizeable gap and a valley Chern invariant arises in commensurate
stacking  \cite{GSetal14}. These type of heterostructures are discussed at length in Section \ref{Sec:superlattices}.

\section{Lattice anharmonic effects in crystalline membranes}
\label{sec_anharmonic}

\subsection{Theory of anharmonic crystalline membranes \label{sec:membrane_theory}} 
 
\subsubsection{Classical theory for ideal free, flat membrane \label{subsec:membrane_classical}} 

The usual starting point for the study of lattice properties of a solid is the harmonic theory for lattice vibrations.
The harmonic theory is valid provided that the fluctuations of the atoms around the equilibrium positions are much smaller than the average
interatomic distance. In bulk three-dimensional crystals, the harmonic theory, or minimal extensions of it as the quasi-harmonic approximation, 
is usually a sufficient approximation for most applications. However, the situation is quite different in two-dimensional crystalline membrane 
(membranes with a finite shear modulus, also referred to as solid membranes), such as graphene and other two dimensional crystals, where, due to the 
low 
dimensionality, atomic fluctuations are not small.
As a matter of fact, it was argued by Peierls \cite{P__34} and Landau
\cite{L__37} that two dimensional crystals should not exist at any finite temperature. Mermin
\cite{M__68} formalized this argument and proved that, provided the pair interaction
between two atoms decays faster than $r^{-2}$ ($r$ is the distance
between atoms), there is no positional long-range order. However, there is a residual quasi-long range order, characterized by 
correlations with a logarithmic behaviour with distance, which is typical of two dimensional systems. Nevertheless, Mermin also pointed out that 
orientational long-range order can still exist.  

For a free two dimensional crystalline membrane embedded in a three dimensional space, the situation is potentially worse since now the membrane  has 
the possibility of fluctuating in the third dimension. Therefore the study of properties of the flat phase in crystalline membranes, its 
thermodynamic stability and the study of in-plane and out-of-plane positional and  orientational orders
became a topic of intensive research in the context of statistical mechanics (for an in depth discussion on the physics of membranes, the interested 
reader can refer to Refs.~\cite{NPW04,GK_97,BT_01}). It turns out that, although a free crystalline membrane does not have positional 
long range order, the anharmonic interaction between in-plane and out-of-plane modes is essential in reducing fluctuations and protecting the 
long-range in-plane and out-of-plane orientational orders, such that a flat configuration of the membrane is thermodynamically stable. 
\cite{NP_87,AL_88,AGL89}. 

Much of the physics of flat crystalline membranes can be understood using the simple classical model of continuous local elasticity with a potential 
energy described by Eq.~(\ref{eq:elastic_energy_full}). \cite{NP_87,AL_88} As already discussed in Sec.~\ref{Sec:elasticity}, the inclusion of the 
term $\partial_i h \partial_j h$ in the strain tensor $u_{ij}$ makes the theory defined by Eq.~(\ref{eq:elastic_energy_full}) interacting. Neglecting 
anharmonic effects, in-plane and out-of-plane modes are decoupled and at classical level the theory is described by the correlation functions
\begin{align}
 \langle h(\vec{q}) h(-\vec{q}) \rangle_{0}         & = \frac{k_B T}{\rho (\omega_{\vec{q}}^{F})^2} = \frac{k_B T}{\kappa \lvert \vec{q} 
\rvert^{4}},\\
 \langle u_{i}(\vec{q}) u_{j}(-\vec{q}) \rangle_{0} & = \frac{k_B T}{\rho (\omega_{\vec{q}}^{L})^2} P_{ij}^{L}(\vec{q}) + \frac{k_B T}{\rho 
(\omega_{\vec{q}}^{T})^2} P_{ij}^{T}(\vec{q}) \nonumber  \\ 
                                                &  = \frac{k_B T}{(\lambda + 2\mu)\lvert \vec{q} \rvert^{2}} P_{ij}^{L}(\vec{q}) + \frac{k_B T}{\mu 
\lvert \vec{q} \rvert^{2}} P_{ij}^{T}(\vec{q}),
\end{align}
where $P_{ij}^{L/T}(\vec{q})$ is the longitudinal/transverse projector along the momentum $\vec{q}$, $\omega_{\vec{q}}^{F}$, $\omega_{\vec{q}}^{L}$ 
and $\omega_{\vec{q}}^{T}$ are, respectively, the dispersion relations for the flexural, in-plane 
longitudinal and in-plane transverse phonons (see Eqs.~(\ref{eq:LA_dispersion}), 
(\ref{eq:TA_dispersion}) and (\ref{Eq:flexural}) of Sec.~\ref{Sec:elasticity}), and 
$ \langle \rangle_{0}$ is a thermal average 
with respect to the harmonic theory. Using the harmonic correlation functions to compute the on-site 
displacement-displacement correlation functions one obtains
\begin{align}
\left\langle \vec{u}^{2}\right\rangle _{0} & =   \int\frac{d^{2}\vec{q}}{(2\pi)^2} \langle 
u_{i}(\vec{q}) u_{i}(-\vec{q}) \rangle_{0} \sim\log\left(L/a\right),\label{eq:divergence_u}\\
\left\langle h^{2}\right\rangle _{0} & =  \int\frac{d^{2}\vec{q}}{(2\pi)^2} \langle h(\vec{q}) 
h(-\vec{q}) \rangle_{0} \sim 
L^{2},\label{eq:divergence_h}
\end{align}
where $L$ is the linear size of the membrane and $a$ is an interatomic
distance. The logarithmic dependence of $\left\langle \vec{u}^{2}\right\rangle _{0}$ with the 
system size is the expected behaviour of quasi-long range 
order in two dimensional systems. However, the linear growth of the out-of-plane fluctuations with system size, $\sqrt{\left\langle h^{2}\right\rangle 
_{0}} \sim L$, indicates that at harmonic level a flat membrane is not possible, and that the membrane is instead crumpled. There is however a long 
range in-plane orientational order, which can be seen by looking at \cite{M__68}
\begin{equation}
\left\langle \partial_{i}u_{j}\partial_{k}u_{l}\right\rangle _{0} = 
\int\frac{d^{2}\vec{q}}{(2\pi)^2} q_{i}q_{k} \langle 
u_{j}(\vec{q}) u_{l}(-\vec{q})\rangle_{0},
\end{equation}
which remains finite. There is also a quasi-long range order of the local normals of the membrane, 
$\hat{\mathbf{n}}(x)$,
\begin{equation}
\left\langle \delta\hat{\mathbf{n}}\cdot\delta\hat{\mathbf{n}}\right\rangle_{0} \simeq 
\int\frac{d^{2}\vec{q}}{(2\pi)^2} \lvert\vec{q}\rvert^{2} \langle h(\vec{q}) 
h(-\vec{q}) \rangle_{0} \sim\log\left(L/a\right),
\end{equation}
where, to lowest order in the displacements, 
$\delta\hat{\mathbf{n}}(x)=\hat{\mathbf{n}}(x)-\hat{\mathbf{z}} \simeq -\left(\nabla h(x),0 
\right)$,
with $\hat{\mathbf{z}}$ the unit vector pointing in the reference out-of-plane direction. Nelson 
and Peliti \cite{NP_87} were the first to argue that the 
interactions between
the in-plane and out-of-plane fluctuations present in Eq.~(\ref{eq:elastic_energy_full})
would give origin to a long-range interaction between local Gaussian curvatures (which to lowest order in the displacements reads 
$-\nabla^{2}P_{ij}^{T} \partial_i h \partial_j h /2$), that would stabilize the flat phase. Since Eq.~(\ref{eq:elastic_energy_full}) is quadratic in 
the in-plane displacements, $u_{i}$, these can be integrated out, and the effective action describing the dynamics of the out-of-plane mode is given 
by \cite{NP_87}
\begin{equation}
\label{eq:effective_action}
 S_{\text{eff}}[h]= \frac{\kappa}{2}\int d^2 \vec{r} \left( \partial^{2} h \right)^{2} + 
\frac{Y}{8}\int d^2 \vec{r} \left(P_{ij}^{T} \partial_i 
h\partial_j h \right)^2,
\end{equation}
where $Y = 4\mu (\lambda + \mu )/ (\lambda + 2\mu)$ is the two dimensional Young modulus of the membrane. Computing to lowest order in perturbation 
theory the self-energy correction to the out-of-plane mode correlation function, $\langle h(\vec{q}) h(-\vec{q})\rangle = k_B T \left(\kappa 
\lvert\vec{q}\rvert^4 + \Sigma(\vec{q})\right)^{-1}$, due to the anharmonic terms in Eq.~(\ref{eq:effective_action}) one obtains 
\cite{NP_87,NPW04,ZRFK10,ARC14,ARC15}
\begin{equation}
\Sigma(\vec{q})= \frac{3 Y k_B T}{16 \pi \kappa}\lvert \vec{q} \rvert^2.
\label{eq:Ginzburg_momentum}
\end{equation}
As we go to low momentum we have that $\Sigma(\vec{q})\gg \kappa \lvert\vec{q}\rvert^4$, which shows that perturbation theory breaks down and that a 
crystalline membrane is a strongly interacting system. Using a Ginzburg like  criterion, from  the condition $\Sigma(\vec{q}) \simeq \kappa 
\lvert\vec{q}\rvert^{4} $, the momentum scale below which anharmonic effects become dominant is given by\cite{NP_87,NPW04,MvO08,ZRFK10,ARC14,ARC15}
\begin{equation}
\label{eq:q_Ginzburg}
q^{*}=\sqrt{\frac{3k_{B}T Y}{16\pi\kappa^{2}}},
\end{equation}
which for graphene has the value $q^* \simeq 0.2$ \AA$^{-1}$. This momentum scale can be translated into a length scale $L^{*}=2\pi / q^{*} \simeq 3$ 
nm, that is typical size of a membrane above which anharmonic effects become dominant. For large membranes and below the anharmonic scale $q^{*}$, 
Nelson and Peliti \cite{NP_87} showed that a non-perturbative treatment calculation of the self-energy leads to a correlation function for the 
out-of-plane displacement characterized by an anomalous momentum dependence. A renormalization group calculation \cite{AL_88} showed that the in-plane 
correlation functions also acquire an anomalous momentum dependence. In general, at small momenta the behaviour of the displacement correlation 
functions is characterized by two exponents $\eta_h$ and $\eta_u$, such that
\begin{align}
\langle h(\vec{q})h(-\vec{q}) \rangle          & \sim q^{-4+\eta_h},\\
\langle u_{i}(\vec{q}) u_{j}(-\vec{q}) \rangle & \sim q^{-2-\eta_u}.
\end{align}
This behaviour can also be seen as the effective elastic constants becoming momentum dependent at low momenta
\begin{align}
\kappa_\text{eff}(q) & \sim q^{-\eta_h},\\
\lambda_\text{eff}(q), \mu_\text{eff}(q) & \sim q^{\eta_u}. \label{eq:Lame_eff}
\end{align}
Using the renormalization group method, it was shown that the exponents $\eta_h$ and $\eta_u$ are not independent, but are related by 
\cite{AL_88,AGL89}
\begin{equation}
\eta_{u}+2\eta_{h}=4-D,\label{eq:scaling}
\end{equation}
where $D$ is the dimension of the membrane, with $D=2$ for physical
membranes. Provided $\eta_{h}>0$ and $\eta_{u}<2$ (for $D=2$, these are equivalent conditions, provided Eq.~(\ref{eq:scaling}) holds), there is 
in-plane and out-of-plane long-range orientational order, since taking into account the momentum dependence of the effective elastic constants yields
\begin{align}
\langle \partial_{i}u_{j}\partial_{k}u_{l} \rangle  & \sim  L^{\eta_{u}-2},\\
\langle \delta\hat{\mathbf{n}}\cdot\delta\hat{\mathbf{n}} \rangle  & \sim   L^{-\eta_{h}},
\end{align}
and therefore a flat phase for the membrane is indeed stable. It was argued in 
Refs.~\cite{DG_88,AGL89} that the lower critical dimension above which a flat 
phase for a crystalline membrane is possible should be smaller than $2$. 
Anharmonic effects also reduce the crumpling of the membrane, since the height fluctuations scale with 
the system size as
\begin{equation}
\langle h^2 \rangle \sim L^{2 \zeta},
\label{eq:roughness_exp}
\end{equation}
where $\zeta = 1-\eta_{h}/2$ is the roughness exponent of the membrane. At the same time, for $\eta_{u}>0$, the anharmonic effects generated 
by the out-of-plane fluctuations lead to a degradation of the positional quasi-long range order of the membrane, since
\begin{equation}
\langle \vec{u}^2 \rangle \sim L^{\eta_u}.
\end{equation}

The anomalous dependence on momentum of the effective elastic constants was confirmed using the self-consistent screening
approximation for the continuous model described by Eq.~(\ref{eq:elastic_energy_full}) \cite{DR_92,G__09,ZRFK10}, Monte Carlo and molecular dynamics 
simulations \cite{ZDK93,BCF96,BCF97,LFY09,T__13,T__15} and functional renormalization group methods \cite{KM_09,BH_10,HB_11,EKM14}. The relation 
between anomalous exponents Eq.~(\ref{eq:scaling}) was also derived using the self-consistent screening approximation \cite{AN_90b,DR_92} and was
further tested using Monte Carlo simulations \cite{BCF96,BCF97}. 

Although it is believed that the anomalous exponents should be universal quantities, there is some dispersion on the reported values. An analytical 
result obtained using the self-consistent screening approximation (which becomes exact in the limit of large codimension of the membrane, $d_{C}$, 
which for a physical membrane if given by $d_{C}=3-2=1$) gives a value for $\eta_h$ of $0.821$ \cite{DR_92}, an improved self-consistent 
calculation to next leading order in $1/d_{C}$ gives a value of $0.789$ \cite{G__09}, functional 
renormalization group calculations give $0.849$ \cite{KM_09,EKM14} and $0.85$ \cite{BH_10,HB_11},  molecular dynamics simulations give 
$0.81$ \cite{ZDK93}, Monte Carlo simulations give $0.72$ \cite{BCF96,BCF97}, $0.85$ \cite{LFY09} and $0.795$ \cite{T__13}, while renormalization 
group Monte Carlo gives $0.79$ \cite{T__15}.

\subsubsection{Deviations from the ideal case: defects, corrugations and residual strains \label{subsec:membrane_deffects}}

The previous section concerned ideal membranes without defects and unconstrained by any external stress or fixed boundary condition. In an 
experimental setup  graphene samples are either supported on a substrate or suspended over a trench. In supported samples coupling to the substrate 
leads to the opening of a gap in the out-of-plane phonon dispersion relation and, therefore, anharmonic effects should be severely quenched 
\cite{AG_13}. A suspended sample is much closer to the ideal case. Nevertheless, even a suspended sample is still subject to constrained boundary 
conditions, residual stresses, impurities and other sources of disorder. Therefore, it is important to understand how the previous discussion is 
affected once we consider these effects.

The possibility of buckling due to constrained boundary conditions was studied in Refs.~\cite{GDL88,GDL89}. The effect of residual stresses was 
studied in Ref.~\cite{RFZK11}, where it was found that the anomalous momentum dependence of the correlation functions survived for small 
externally applied stresses. It was estimated that, in graphene, an applied stress corresponding to a strain of $0.25\%$ can completely quench 
anharmonic effects. 

The effect of  disorder in the stability of flat membranes was studied in Refs.~\cite{NR_91,RN_91,ML_92,LR_93}, where 
disorder was simulated by random strain and curvature fields. It was found that disorder can indeed destabilize the flat phase and that this 
instability can still be present at finite temperature, depending on the type of disorder. This possibility was verified with  
atomistic simulations of graphene \cite{TMM09}, where it was found that a 20\% adsorption of OH molecules leads
to crumpling of the membrane. Membranes with a frozen non-flat background metric characterized by out-of-plane variances with a power law 
dependence on momentum, called warped membranes, were studied in Refs.~\cite{KN_13,KN_14}. In these 
works it was found that for strong enough 
fluctuations of the background metric the anomalous exponents can differ from the ones due to 
thermal fluctuations in a flat membrane. The effect of short range quenched curvature 
disorder was also recently studied in Ref.~\cite{GKM15} using a renormalization group analysis. 
There it was found that the experimental results of Refs.~\cite{MGK07} and \cite{KDT11}, 
for the average size of ripples and the mean deviation of the normals in suspended graphene 
samples, compare well with values obtained with the renormalization group analysis for weak 
curvature disorder.

\subsubsection{Quantum theory \label{subsec:membrane_quantum}}

The previous discussion was based on a classical description of membranes. The original theoretical works on crystalline membranes were motivated 
by the study of biological amphiphilic membranes \cite{NPW04}. Since such objects are formed by very large molecules, one can expect that their 
mechanical behaviour will be governed by classical statistical mechanics at room temperature. The same assumption is not easily justifiable in 
atomically thin membranes such as graphene. A simple estimation of graphene's Debye temperature, gives us $T_{D} \sim 1000$ K, considerably larger 
than room temperature. In this scenario one expects that quantum effects will still be important at room temperature.

The quantum treatment of flexural phonons was first performed in Ref.~\cite{SGG11}, in the context of the study of the combined effects of 
phonon-phonon and electron-phonon interactions in graphene. Performing a self-consistent calculation, it was found that at zero temperature and in the 
absence of electron-phonon interactions, anharmonic effects lead to logarithmic corrections to the bending rigidity of graphene. The effect of 
quantum fluctuations at zero temperature on the Young modulus was studied in Ref.~\cite{GLW14} using an expansion in the number of out-of-plane 
directions of the membrane (which for a physical membrane is one). It was found out that quantum fluctuations contribute to a reduction of the Young 
modulus at small momentum scales, just as thermal fluctuations, although being a much smaller effect. It was estimated that thermal fluctuations are 
dominant with respect to quantum fluctuations for momentum scales smaller than  the thermal wavelength for flexural phonons
\begin{equation}
q_{Q}(T)= \left( \sqrt{ \frac{ \rho}{\kappa} } \frac{k_B T}{\hbar }\right)^{1/2}.
\end{equation}
Anharmonic crystalline membranes at zero temperature were also studied in Ref.~\cite{KL14} by performing a renormalization group study. It was 
found that zero temperature fluctuations lead to an increase of the effective coupling constant between in-plane and out-of-plane phonons which leads 
to a destabilization of the flat phase of the membrane. This behaviour is to be contrasted with the effect of thermal fluctuations, which lead to an 
increase of the bending rigidity at large length scales, and therefore stabilize the flat phase. Refs.~\cite{SGG11,GLW14,KL14} did not take 
into account retardation effects for the in-plane phonons, effectively treating them as classical, on the basis that typical frequencies of
in-plane phonons are much larger than the frequencies of out-of-plane phonons at low momenta. The effect of retardation of the in-plane phonons was 
studied in Ref.~\cite{ARC14,ARC15} by means of a perturbative calculation of the out-of-plane phonon self-energy. It was found that 
such retardation effects, change the corrections to the bending rigidity from being logarithmic to a power law in momentum, contributing to 
an increase of the bending rigidity at long wavelengths, a scenario that is very similar to the one due to thermal fluctuations. Furthermore, it was 
estimated that the effect of quantum fluctuations is dominant for temperatures smaller than
\begin{equation}
 T^{*} \simeq \frac{2 \hbar}{3 k_B \rho^{1/2} \kappa^{1/2}} \frac{(\lambda+2\mu)^2}{Y}f,
\end{equation}
where $f$ is a dimensionless cutoff dependent quantity. For graphene it is estimated that $T^{*} 
\simeq 70 - 90$ K. The momentum scale below which 
anharmonic effects dominate, in the low temperature regime, was found to be smaller but of the same order of magnitude as in the high temperature 
limit Eq.~(\ref{eq:q_Ginzburg}). Quantum corrections to thermodynamic and elastic properties of graphene were also studied in 
Ref.~\cite{SCT14}, by means of a self-consistent calculation of the density matrix, where it was found that quantum and anharmonic effects 
become comparable for temperatures smaller than $100$ K, while quantum fluctuations still play a significant role for temperatures as high as $1000$ 
K.

The contradictory nature of the results published so far does not allow one to access what is the exact role and importance of quantum effects in the 
physics of crystalline membranes, such as graphene.

\subsubsection{Anharmonic effects and electron-phonon interaction \label{subsec:membrane_electron_phonon}}

As we have seen in the previous sections, the flat phase of a crystalline membrane is very fragile and anharmonic effects play a very important role. 
In membranes that have low energy electronic excitations, such as graphene, one might ask if these electronic degrees of freedom can have a 
significant impact in the physics of crystalline membranes. 

This question was first addressed in Ref.~\cite{G__09b}. There it was argued that the deformation 
potential interaction between graphene electrons and phonons, Eq.~(\ref{Eq:deformation_potential}), gives origin to 
an effective strain of the form $\sigma = g_{1}  n $, where $n$ is the electronic carrier 
density. This coupling is asymmetric with respect to electron or hole doping. While for electron doping,  $ n >0$, the 
membrane would be subject to a tension, in the case of hole doping, $ n <0$, the membrane would be subject to a compressive 
strain, which would lead to the buckling of the membrane. Such a  structural deformation induced by electron-phonon interaction bears some 
resemblance with the physics of the Peierls transition and formation of charge density waves. The 
role of electron-phonon interactions in undoped 
graphene was studied in Ref.~\cite{GP_09}, where it was shown that the decay of flexural phonons into electron-hole pairs, leads to the 
disappearance of the flexural mode branch below a critical momentum. It was speculated that such disappearance could be associated with the formation 
of ripples. A perturbative evaluation of the normal-normal correlation function in the classical limit performed in Ref.~\cite{G__09c}, also 
displayed a peak at finite momentum induced by the electron-phonon interaction. Such peak was interpreted as signalling the onset of ripples. The same 
problem was studied at zero temperature in Ref.~\cite{SGG11}, via a self-consistent evaluation of flexural phonon correlation function and it 
was found that for large values of the deformation potential, $g_{1} \simeq 23.1$ eV, the effective bending rigidity vanishes at a finite momentum 
scale of the order of $q \sim 0.1$ \AA. Similar conclusions were found in Ref.~\cite{GLW14}, where, by treating electrons and phonons on 
equal footing, it was found that electron-phonon interaction can drive the system towards an instability in the charge carrier density - Gaussian 
curvature channel at finite momentum, that was speculated to be associated with the spontaneous and simultaneous formation of structural ripples and 
charge puddles. This mechanism is suppressed  by the presence of an electronic bandgap, and should therefore be more relevant in graphene than in 
semiconductor transition metal dichalcogenides or boron nitride. The nature of this electron induced instability was further studied in 
Ref.~\cite{G__14}, by means of a self consistent calculation and a renormalization group calculation. It was found that the vanishing of the 
effective bending rigidity and the condensation of the mean curvature (which is given by $\partial^2 h$) are simultaneous and manifestations of the 
same transition at finite momentum. This transition should occur even in the absence of an external tension and does not involve an in-plane 
distortion. Nevertheless, the nature of the relevant operators at this critical point is still not clear, and a condensation of the Gaussian curvature 
could not be ruled out. Therefore, further analysis is needed to clearly identify the electron-phonon interaction as a candidate to the formation of 
ripples in graphene.

\subsection{Mechanical properties \label{sec:membrane_mechanical}}

\subsubsection{Structure of suspended two dimensional crystals \label{subsec:membrane_structural} }

One of the most important quantities in the study of the structure of matter is the static structure factor 
$S(\mathbf{q})=\sum_{n,m} \left\langle e^{i\mathbf{q}\cdot \left(\mathbf{R}_n - \mathbf{R}_m 
\right) }\right\rangle 
$, where $\mathbf{R}_n$ are the atomic positions (in 3D space), 
$\mathbf{q}=(q_x,q_y,q_z)$ is the transferred momentum and the average is to be understood as a 
time 
or ensemble average. As discussed in 
Sec.~\ref{subsec:membrane_classical}, in a crystalline membrane displacement-displacement correlation functions have an anomalous momentum 
dependence, which will affect the structure form factor. In a scattering experiment, Bragg peaks will no longer be sharp, but will instead be 
broadened with an algebraic decay.  It is in this weaker sense, that a two dimensional crystal is defined \cite{K__12,KA_13}.  The structure factor 
of rough, self-affine membranes and surfaces \footnote{A surface is said to be self-affine if its roughness is described by a random variable with 
probability distribution that is invariant under anisotropic scaling transformations of the form $x \rightarrow b x$ and $z \rightarrow b^{\zeta} 
z$}, 
characterized by power law displacement variances: $\left\langle \left| h(\vec{x})-h(\vec{x}^\prime) \right|^{2} \right\rangle = A 
\left|\vec{x}-\vec{x}^\prime \right|^{2 \zeta}$ and $\left\langle \left| \vec{u}(\vec{x})-\vec{u}(\vec{x}^\prime) \right|^{2} \right\rangle = B 
\left|\vec{x}-\vec{x}^\prime \right|^{\eta_u}$, has been theoretically studied in 
several works \cite{SSG88,AN_90,AN_90b,AG_92,GLM92,YLW93}. A semi-analytical expression for the structure factor of flat membranes subject to thermal 
fluctuations
was obtained in Refs.~\cite{AN_90,AN_90b,AG_92}. For $q=0$, (with $q = \sqrt{q_x^2+q_y^2}$) it 
was found that the structure factor has a 
power law behaviour $S(q=0,q_z)\sim q_z^{-2/\zeta}$, while $S(q,q_z=0)$ did not show a 
well defined power law behaviour. In 
Ref.~\cite{GLM92} the orientationally averaged structure factor, $S(\rvert \mathbf{q} \lvert)$, was 
studied in detail. It was found that for $1/L \ll q \ll 1/\ell$, 
where $L$ is the membrane size and $\ell$ is the characteristic length scale of out-of-plane or 
in-plane fluctuations, one obtains $S(\rvert \mathbf{q} \lvert)\sim \rvert \mathbf{q} \lvert^{-2}$ 
as in a completely flat, non-fluctuating membrane \cite{SSG88}. For $1/\ell \ll q \ll 1/a$, where $a$ is the lattice spacing, it was obtained 
$S(\rvert \mathbf{q} \lvert)\sim \rvert \mathbf{q} \lvert^{\zeta -3}$, for the case where in-plane 
fluctuations are small compared to 
out-of-plane, and $S(\rvert \mathbf{q} \lvert)\sim \rvert \mathbf{q} \lvert^{\eta_u 
/\zeta-2/\zeta-2}$, in the opposite case.

The roughness exponent, $\zeta$, of red blood cell membrane skeletons was obtained from the structure factor 
measured with light and x-ray scattering \cite{SSL93}. The roughness exponent was extracted by 
fitting the orientationally averaged structure factor 
to the power law behaviour $S(\rvert \mathbf{q} \lvert)\sim 
\rvert \mathbf{q} \lvert^{\zeta -3}$, having been obtained a value of $\zeta = 0.65 \pm 0.10$, 
which is 
in good agreement with the theoretical prediction for thermal 
fluctuations in a flat crystalline membrane (see Eq.~(\ref{eq:roughness_exp}) and text below). The exponent $\eta_{h}$ was also measured in 
amphiphilic films \cite{GDB97}, by fitting the structure factor measured with x-ray scattering, with a continuous model that neglects in-plane 
fluctuations. The best fit was obtained for $\eta_{h}=0.7\pm0.2$, once again in good agreement with the theory of thermal fluctuations in a flat 
membrane. 

It was experimentally found that suspended samples of both graphene and bilayer graphene display roughness \cite{MGK07,MGK07b}. Transmission electron 
microscopy revealed that samples with lateral sizes of $\sim 500$ nm display broadened Bragg peaks with Gaussian shape. These roughness showed no 
preferred 
orientation, and is associated with static ripples. It was estimated that these ripples have a characteristic lateral size of 2 - 20 nm and a mean 
deviation of the normals of the membrane of $5\%$ in single-layer and $2\%$ in bilayer graphene. The typical height of the ripples was 
estimated to be of 1 nm. Similar corrugations have also been observed in suspended samples of MoS$_{2}$ \cite{BAK11}. The roughness of suspended 
graphene was further studied using low-energy electron microscopy and diffraction in Ref. \cite{LKC10}. It was found that at small length 
scales, suspended graphene appears to be a self-affine surface characterized by a correlation length of $\xi \simeq 24$ nm and a roughness exponent of 
$\zeta \simeq 0.5$. At larger length scales, graphene seemed to present a preferred ripple 
wavelength of $\lambda_\text{ripple} \simeq 60$ nm. Existence of the two 
length scales $\xi$ and $\lambda_\text{ripple}$ indicate that at large length scales graphene 
deviates from a purely self-affine surface and seems to be a mounded 
surface.\cite{LKC10} It was also found that the roughness exponent becomes smaller with increasing temperature and with increasing exposition time to 
an electron or photon beam. Furthermore, it was found that suspended bilayer and trilayer graphene present a roughness exponent of $\zeta \simeq 
0.8$. Scanning tunnelling microscopy studies \cite{ZMB12} further confirmed the static nature of 
ripples in suspended graphene. The ripples found in 
Ref.~\cite{ZMB12} had a characteristic lateral size of 5 nm and showed to be static for imaging times of about 5 minutes. The used samples were 
subject to annealing and reported to be free of contamination over areas of tens of nm$^2$. The amplitude of the corrugation of graphene was further 
studied in Ref.~\cite{KDT11} with transmission electron microscopy, where, by analysing the width of diffraction peaks, it was obtained a height and 
normal variances of $\sqrt{\left\langle h^2 \right\rangle }=1.7$ \AA~and  $\sqrt{\left| \nabla h 
\right|^2 }=0.011$, with a characteristic length 
scale of 10 nm, for suspended samples with a size of 0.5-5 $\mu$m. It was also found that, at least for length scales smaller than 100 nm,  
the height 
fluctuation decreases with increasing temperature.

The exact origin of the static corrugation presented in graphene samples is still unclear. While the temperature dependence of the corrugation 
measured in Refs. \cite{LKC10,KDT11} indicates that thermal activation of phonons should play a role, the theory of flat anharmonic crystalline 
membranes alone is unable to explain the static nature of the observed ripples in \cite{MGK07,MGK07b} and the non-universal roughness exponents 
measured in Ref. \cite{LKC10}. A possible origin of the ripples is that in a suspended graphene sample, its edges are constrained and therefore the 
membrane is unable to freely expand or contract, leading to buckling and the formation of ripples \cite{GDL88,GDL89}. Therefore, 
residual thermal or external stresses could lead to the formation of ripples. Thermal and external stresses have indeed been used to form ripples in 
graphene suspended over trenches as reported in Ref.~\cite{BMF09}. This has been explained with the theory from Ref. \cite{CM_03} based 
on classical elasticity, but seems unrelated to the short wavelength corrugations reported in Refs.~\cite{MGK07,MGK07b,LKC10,KDT11}. Another 
possibility is that in the 
fabrication process of a suspended graphene sample, relaxation of a constrained graphene membrane after removal of the substrate could lead to 
rippling, or the ripples might themselves be inherited from the substrate \cite{GHD08}. Electron-phonon interaction in graphene might also lead to a 
crumpling instability of the membrane \cite{G__09b,SGG11,GLW14}. Disorder has also been proposed as a source of ripples. Indeed, the numerical 
simulations from Ref.~\cite{TMM09} show that a 20\% adsorption of OH molecules leads to the crumpling of a graphene membrane compatible with amplitudes and typical sizes of corrugation observed in experiments. Scanning transmission 
electron microscopy also shows that the presence of point defects as vacancies and adatoms, such as hydrogen or carbon atoms, favours rippling 
\cite{BGB09,BGB09b}. These disorder induced ripples might still display dynamics due to migration of the defects. 

Scanning tunnelling microscopy studies were also capable of probing time-dependent height fluctuations of 
suspended graphene \cite{XNB14}. At small bias voltages ($\sim 0.01$ V) and currents ($0.2$ nA), 
random on-site fluctuations were attributed to thermal fluctuations. These displayed a variance of 
$\sqrt{\left\langle h^2 \right\rangle } = 1.47$ nm, with a characteristic decay time of $\tau = 8 s$ 
($\left\langle h(t)h(0) \right\rangle \propto  e^{-t/\tau}$). It was also found that for larger 
currents and gate voltages, effects of thermal expansion due to the local heating of graphene and 
the electrostatic forces induced by the tip can lead to periodic oscillations of the membrane and 
to mirror buckling \cite{XNB14,NXS14,SXM15}.

\subsubsection{Anomalous elasticity \label{subsec:membrane_elasticity}}

The Young modulus of graphene was first measured using a blister test
in Ref. \cite{LWK08}, where an atomic force microscope is used to perform
a nanoindentation in a graphene membrane suspended over a circular
hole, with sizes of $1$ and $1.5$ $\mu$m in diameter. The two-dimensional
Young modulus, $Y_{\text{eff}}$,
was determined by fitting the force--indentation curve by \cite{KBS05,GWD05,C__08,JW_08}
\begin{equation}
F=\sigma_{0}\pi \delta h + Y_{\text{eff}}\vartheta^{3} 
\frac{(\delta h)^3}{d^2},\label{eq:Force-indentation}
\end{equation}
where $F$ is the applied force, $\delta h$ is the displacement of
the membrane at its centre, $\sigma_{0}$ is a pretension,
$d$ is the membrane diameter and $\vartheta=\left(1.05-0.15\nu-0.16\nu^{2}\right)^{-1}$, where $\nu$ 
is the Poisson ratio of the membrane 
\cite{KBS05,C__08}. By taking $\nu=0.165$, the obtained value for the Young modulus of graphene was $342$ N/m  \cite{LWK08}, a value 
that makes graphene the stiffest material to have ever been measured.  

Equation (\ref{eq:Force-indentation}) is derived from the F{\"o}ppl--von K{\'a}rm{\'a}n plate theory, which is described by a free energy of the 
form of Eq.~(\ref{eq:elastic_energy_full}) \cite{LL_59}. However, while the high bending rigidity of a plate strongly suppresses thermal 
fluctuations, which can therefore be neglected, on a membrane long wavelength vibrations are thermally excited and interact due to strong 
anharmonic effects. As discussed in 
Sec.~\ref{subsec:membrane_classical} this leads to an anomalous dependence on momentum of the displacement correlation functions, which translate 
into a scale dependence of the elastic constants. As a matter of fact, while the starting potential energy that describes the membrane is a local 
functional of the displacements, the free energy will become non-local  \cite{BH_10,HB_11}. Assuming that the strain induced by the indentation is 
nearly uniform over most of the sample, we can assume that Eq.~(\ref{eq:Force-indentation}) is still valid, but with a scale dependent Young modulus 
$Y_{\text{eff}}(\ell) \propto \ell^{-\eta_u}$ (Eq.~(\ref{eq:Lame_eff})), where $\ell$ is the characteristic length scale of the problem (which can be 
the size of the membrane, a length scale imposed by the induced strain, or a length scale due to disorder or an initial corrugation of the 
membrane). This justifies the use of Eq.~(\ref{eq:Force-indentation}).

\begin{figure}[h!]
\begin{center}
\includegraphics[height=6.5cm]{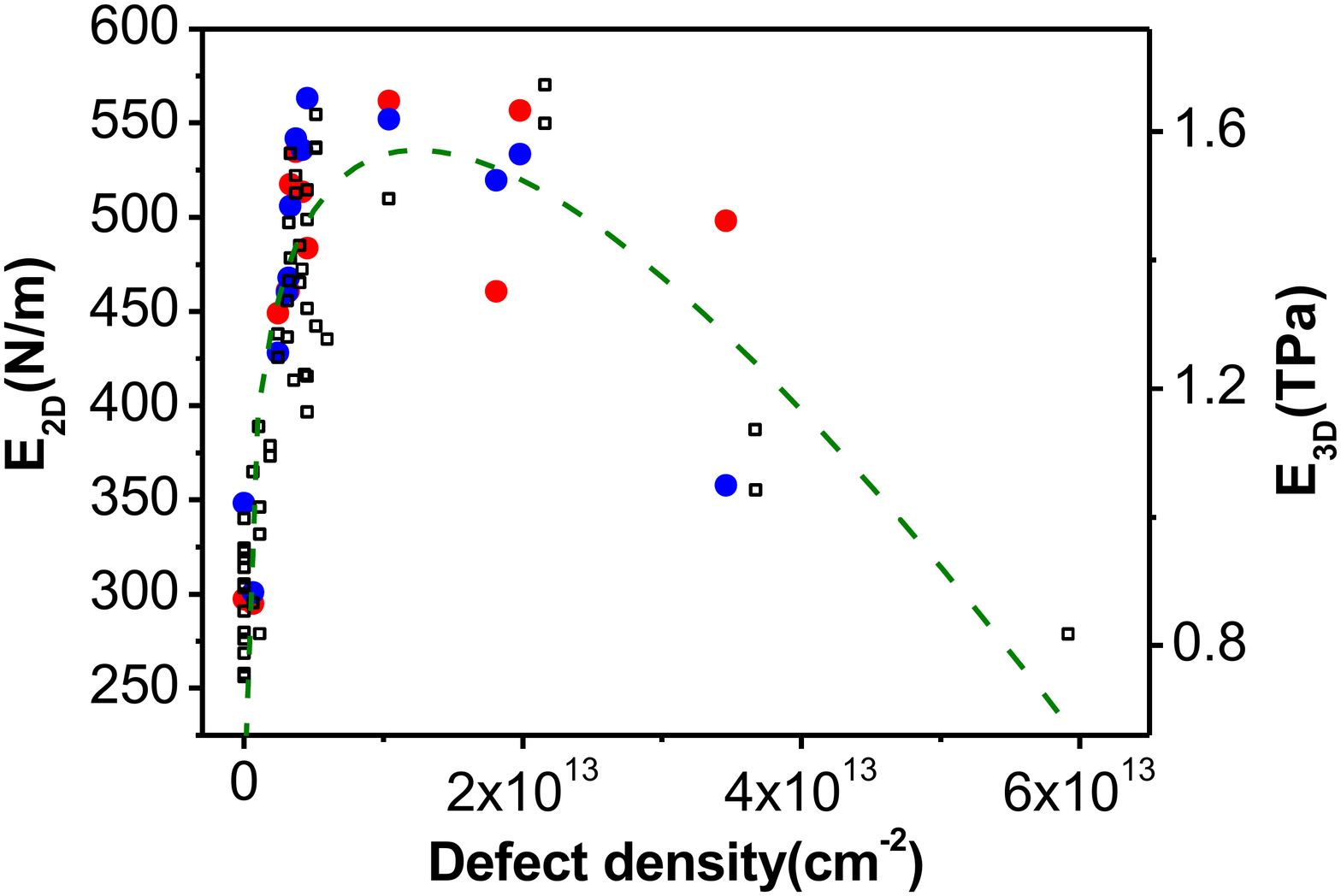}
\end{center}
\caption{Young Modulus as a function of the defects density\cite{LGP14} (see main text). The broken green line represents the best fitting to the 
data via Eq.\eqref{eq:fit}. }
\label{fig:young_defects}
\end{figure}

It has been experimentally found that the Young modulus of graphene increases with the density of defects created by bombardment of Ar$^{+}$ ions 
\cite{LGP14}, provided the density is not too high. The measured Young modulus, changed from 250 - 360 N/m for pristine graphene, to 550 N/m for 
graphene with an average defect distance of 5 nm (0.2 \% coverage of defects). The increase of the Young modulus
has been explained in terms of the classical theory of flat anharmonic crystalline membranes described in Sec.~\ref{subsec:membrane_classical}. 
According to it, the Young modulus becomes smaller at larger length scales. The creation of defects would introduce a new length scale, the average 
distance between defects, which acts as an effective membrane size from the point of view of the graphene phonons. Being smaller than the membrane 
size, this new scale would restore the value of graphene's Young modulus to a higher value than in the absence of defects. The Young modulus in 
Ref.~\cite{LGP14} was measured with a nanoindentation technique and by fitting the obtained force-displacement curve to 
Eq.~(\ref{eq:Force-indentation}) for different graphene drums. In Fig.~\ref{fig:young_defects} the experimental value of the Young modulus obtained 
in this way is plotted versus the density of defects for the different drums. The Young modulus grows up to twice its pristine value for all the 
drums with the same law; the experimental curves then show a crossover to a decreasing  regime as the density of defects is increased, as expected 
from 
first principle calculations\cite{FPF12, JNX12, NP10}. These competing behaviours can be encoded in 
a qualitative fitting to the experimental data
\begin{equation}\label{eq:fit}
Y_{\text{eff}}=K \left(b+\frac{1}{l_0^2}+n_i\right)^{\eta_{u}} \left(1-c\left(\frac{1}{l_0^2}+n_i\right)\right),
\end{equation}
where $n_i$ is the density of defects, $K$ and $c$ are constants to be evaluated from the fit, $b$ is a geometrical parameter of the order of the 
inverse of the area and $l_0$ is the localization length for the flexural phonons in pristine graphene.
Best fitting with the experimental data, shown in Fig.~\ref{fig:young_defects} with the broken green line, is given by the value of the parameters 
$K=1.5\times10^9$N m$^{\eta_{u}-2}$, $l_0=70$ nm, $\eta_{u}=0.36$ and $c=1.2\times10^{-18}$ m$^2$. The localization length $l_0$ takes into account 
other sources of disorder that can introduce an effective length in the problem. The radii used in Ref.~\cite{LGP14} vary from 500 nm to 3 
$\mu$m and thus the estimated value is such that $1/l_0^2\gg b$. This justifies the independence of 
the growth of the Young modulus observed. 
The above interpretation is supported by recent experimental measurements of the evolution of 
the Young modulus with external strain \cite{LJG15}, where it was found that the in-plane elastic 
constant increases by a factor of 2 when the membranes are under tension. This effect is expected 
from the elastic theory of membranes, which predict that external strain suppresses thermal 
out-of-plane fluctuations, with the corresponding enhancement of the Young modulus \cite{RFZK11}. 
Recent atomistic Monte Carlo simulations have shown that external tension can indeed cause an 
increase of the Young modulus in a factor of 2 \cite{LFK15}, in agreement with the experiments 
\cite{LJG15}. However, the discrepancy between the Young modulus measured by indentation experiments 
on tensioned membranes, and that obtained by {\it ab initio} calculations is an open question.  
Another experimental evidence of the importance of anharmonicities on the elastic properties of 
atomically thin membranes has been reported by Blees {\it et al.} \cite{BBR15}, who have shown that 
the bending rigidity of micro-ribbons of graphene is three order of magnitude larger than its 
microscopic value. These results have been studied theoretically in the framework of a 
renormalization group analysis of flexural phonons on elongated ribbons, which are enough to explain 
the qualitative behaviour \cite{KN15}. 
Nevertheless, further studies, both experimental and theoretical, are required in order to fully 
understand the elastic behaviour of graphene.
%, in particular effects of temperature,
%sample size, defects and possible presence of ripples.

\subsection{Thermodynamic properties \label{sec:membrane_thermo}}

\subsubsection{Thermal expansion \label{subsec:membrane_expansion}}

Thermal expansion is an intrinsically anharmonic effect. In bulk three-dimensional crystals, thermal expansion is well described with the 
quasi-harmonic approximation. This is essentially a minimal mean field extension of the 
harmonic theory, where the phonon frequencies depend on the lattice expansion. The dependence of the phonon frequencies on the expansion is encoded in
the so called Gr{\"u}neisen parameters. For crystalline membranes anharmonic effects will be too strong to treat in this matter. 
At not too high temperatures, thermal expansion will be controlled by the low energy out-of-plane phonons and can be studied using the elasticity 
model (\ref{eq:elastic_energy_full}). In a quasi-harmonic approach, the Gr{\"u}neisen parameter for out-of-plane
phonons is given by \cite{AGK12}
\begin{equation}
\gamma(q)=-\frac{\lambda+\mu}{2\kappa q^{2}},
\end{equation}
which diverges for small momenta. The areal thermal expansion is defined as being the change with temperature of the membrane area projected onto the 
reference plane. Within elasticity theory the relative change of projected area is given by $\left\langle 
\partial_i u_i \right\rangle$. In terms of the out-of-plane mode Gr{\"u}neisen, the thermal expansion in the quasi-harmonic approximation is given 
by \cite{AGK12,K__12}
\begin{equation}
 \alpha_{A}= \frac{k_{B}}{\lambda + \mu}\int \frac{d^2\vec{q}}{(2\pi)^2} \left(\frac{\hbar 
\omega_{\vec{q}}^{F}}{2 k_{B}T}\right)^2 \frac{\gamma(q)}{\sinh 
^2 \left( \frac{\hbar \omega_{\vec{q}}^{F} }{2 k_B T}\right)}.
\end{equation}
From the previous expression, it is clear that in the quasi-harmonic approximation the thermal expansion for an infinite membrane will be divergent 
at any finite temperature \cite{AGK12,ARC14}. Inclusion of anharmonic effects beyond the quasi-harmonic approximation corrects this unphysical 
result. In the context of the
model described by Eq.~(\ref{eq:elastic_energy_full}), the areal thermal
expansion can be exactly written as a normal-normal correlation function  \cite{AGK12,AG_13,ARC14}
\begin{equation}
\alpha_{A}=-\frac{1}{2}\frac{\partial}{\partial T}\left\langle \partial_i h(x) \partial_i h(x)\right\rangle . \label{eq:thermal_expansion}
\end{equation}
This is the formal expression of the membrane effect \cite{L__52}: the intuitive picture that the thermal excitation of out-of-plane vibrations
should lead to a reduction of the area of the membrane. Equation~(\ref{eq:thermal_expansion}) always predicts a negative thermal expansion, being a 
limitation of the model (\ref{eq:elastic_energy_full}). Higher order anharmonic effects, such as cubic terms in $\partial_i u_j$, would give origin 
to competing terms, that at high enough temperature, could change the sign of the thermal expansion. Nevertheless, at low temperature, Eq. 
(\ref{eq:thermal_expansion}) is the main contribution.  If the average
in Eq. (\ref{eq:thermal_expansion}) is taken with respect to the harmonic theory,
one recovers the result from the quasi-harmonic approximation. Going one step
beyond the quasi-harmonic approximation, one can estimate the contribution due to strong anharmonic effects at low momentum by taking the Ginzburg
momentum scale (\ref{eq:q_Ginzburg}) as a low momentum cutoff,
obtaining \cite{AGK12} 
\begin{equation}
\alpha_{A}  \sim-\frac{k_{B}}{4\pi\kappa}\log\left(\Lambda/q^{*}\right),
\end{equation}
where $\Lambda$ is a large momentum cutoff, which one takes to be
the thermal wavelength for flexural phonons at intermediate temperatures
and the Debye momentum at high temperatures. Using typical values
for graphene one obtains a thermal expansion of $\alpha_{A} \sim-10^{-5}\text{ K}^{-1}$. This estimation is in agreement with more involved 
calculations using a quasi-harmonic approximation based on \textit{ab initio} calculations
\cite{MM_05}, classical Monte Carlo simulations \cite{ZKF09}, non-equilibrium
Green's functions method \cite{JWL09}, \textit{ab initio} molecular
dynamics \cite{PAL11,GH_14} and self-consistent-field methods \cite{SCT14}. The thermal expansion at very low temperature is dominated by flexural 
phonons, and it was found that in the quantum limit, inclusion of anharmonic effects renders the thermal expansion finite in the thermodynamic limit, 
behaving with temperature as $\alpha_A \propto - T^{2 \eta_{h}/(4-\eta_{h})}$, for flexural phonons with dispersion relation of the form 
$\omega_{\vec{q}}^{F} \propto q^{2-\eta_{h}/2}$ \cite{ARC14,ARC15}.

There is a debate regarding the behaviour of the thermal expansion
coefficient of graphene with temperature. From the theoretical point of view, while Refs.~\cite{MM_05}
and \cite{PAL11} predict an always negative thermal expansion (Ref.~\cite{PAL11}
predicts this only for large enough samples, while for small samples
there is a change of sign), Refs.~\cite{ZKF09,JWL09,GH_14} predict a change
of sign in the thermal expansion for temperatures in the range $600-900$
K. This possible change of sign occurs due to a competition of thermal
contraction due to the anharmonic effects involving the out-of-plane
mode and thermal expansion due to the in-plane modes. It is very important
to emphasize that the thermal expansion of graphene greatly depends
on whether it is free or if it is supported on a substrate \cite{PAL11,AG_13,GH_14}.
It was also found that the calculated thermal expansion is dependent
on the size of the simulated system \cite{PAL11,GH_14}, which can
be understood since the anharmonic coupling between out-of-plane and
in-plane vibrations lead to long range interactions \cite{NP_87}
which might be difficult to simulate numerically. The discrepancy of results can be attributed to the different approximations involved in the 
calculation. An accurate description of the physics would have to take into account quantum effects, anharmonic effects in a 
non-perturbative way and should be extensible to the thermodynamic limit (a non-trivial task, from the numerical point of view, since anharmonic 
effects are stronger at larger length scales).

Experimentally, the thermal expansion coefficient of graphene has
been measured in suspended samples \cite{BMF09}, and was found out
to be negative up to $350$ K with a value of $\alpha_{A} \simeq-7\times10^{-6}\text{ K}^{-1}$ at $300$ K. The thermal expansion coefficient was also 
measured in
supported samples by Raman spectroscopy\cite{YYG11} and was found to be negative in the range $200-400$ K, with an obtained value
of $\alpha_{A}\simeq-8\times10^{-6}\text{ K}^{-1}$ at room temperature. The data from Ref.~\cite{BMF09} was successfully described by the theory of 
Ref.~\cite{SCT14}, which is formulated in the thermodynamic limit, taking into account anharmonic effects in a self-consistent way and 
including lowest order quantum corrections to the classical result.

The areal thermal expansion discussed so far, being a change of the project area of the membrane, is related to the change of the lattice parameter. 
Another interesting quantity is the change in nearest neighbours carbon-carbon distance. There is also no consensus regarding the behaviour 
of this quantity with temperature. Monte Carlo simulations indicate that for small temperatures there is also a reduction in the carbon-carbon 
distance, reaching a minimum at a temperature of $T \simeq 700$ K \cite{ZKF09}. On the other hand, \textit{ab initio}  molecular dynamics simulations 
show that the carbon-carbon distance always increases with temperature up to $T=1000$K, even when the lattice parameter contracts \cite{PAL11}. 
Experimentally it is verified that the carbon-carbon distance increases in graphene supported by an iridium substrate over the temperature range of 
$300 - 1200$ K \cite{PAL11}. However, one must point out, that theoretically it is also expected that the carbon-carbon distance should increase for 
supported 
graphene \cite{PAL11} and therefore, it is not clear if the experimental result on supported graphene can be extrapolated to the 
suspended case.  

%Once again, the continuum model of Eq.~(\ref{eq:anharmonic_hamiltonian}) can provide some intuition. While the relative change of projected area of 
%the membrane is described by $\partial_i u_j$, the actual change of distance between two points of the membrane in the three spatial dimensions is 
%encoded in the strain tensor $\epsilon_ij$. Therefore, we can estimate the relative change of distance between carbon atoms as $\left\langle 
%\epsilon_ii\right\rangle = \left\langle \partial_i u_i\right\rangle+ \left\langle \partial_i h \partial_i h\right\rangle/2$. Since for the theory 
%described by Eq.~(\ref{eq:anharmonic_hamiltonian}) one can prove that $\left\langle \partial_i u_i\right\rangle = -\left\langle \partial_i h 
%\partial_i h\right\rangle/2$, one expects that the change in the carbon-carbon distance should be zero or at least very small.  

\subsubsection{Specific heat \label{subsec:membrane_specific_heat}}

Another thermodynamic quantity that is affected by anharmonic effects is the specific heat. Anharmonic effects are significant both in the high 
and low temperature limits. At high temperature, anharmonic effects lead to a violation of the Dulong--Petit law: instead of a saturation of the 
specific heat to a value of $k_B$ per degree of freedom, one obtains a linear correction on temperature for high enough temperature. This was 
investigated via Monte Carlo simulations in Ref.~\cite{ZKF09}, where it was found that the specific heat at constant volume in graphene per 
carbon atom  for temperatures $T>800$ K is approximately given by
\begin{equation}
c_A = 3 k_B \left(1 + \frac{k_B T}{E_0} \right),
\end{equation}
with $E_0=1.3$ eV. Inclusion of quantum effects to lowest order \cite{SCT14} leads to a specific heat of the form
\begin{equation}
c_A = 3 k_B \left(1 - \frac{k_B T}{E_1} - \left(\frac{E_2}{k_B T}\right)^2 \right),
\end{equation}
with $E_1 \simeq 0.053$ eV and $E_2 \simeq 0.022$ eV. The second term of the previous expression corresponds to the first quantum correction to the 
classical anharmonic result. 

At low temperatures, anharmonic effects also lead to a correction to the power law dependence of the specific heat with temperature, since the low 
energy acoustic phonons acquire a renormalized dispersion relation. This was studied in Ref.~\cite{ARC14,ARC15}, where it was found that at 
very low temperatures, where the main contribution to  the specific heat is given by the flexural phonons, the harmonic result for the specific heat, 
$c_p \propto T$, is corrected to $c_p \propto T^{4/(4-\eta_{h})}$, for flexural phonons characterized by a dispersion relation of the form 
$\omega_{\vec{q}}^{F} \propto q^{2-\eta_{h}/2}$.

\subsection{Topological defects}
\label{sec_defects}

A closely related problem to strain in graphene is the presence of topological defects in the otherwise perfect crystalline system. The literature of 
topological defects in graphene is so vast that one can consider that this topic might deserve a review by itself. However, there are several reasons 
that motivate the inclusion of a section about these defects in the present review. If we think on a graphene layer as a crystalline membrane, 
structural defects play an important role in the way a membrane melts. Just as elastic deformations affect the electronic properties of 
graphene at low energies, so will topological defects, but in a different way.

The role of topological defects in the melting and crumpling thermal phase transitions in two 
dimensional membranes was envisaged and developed long time before the appearance of graphene\cite{HN78,NH_79}. In the melting scenario of 
Refs.~\cite{HN78,NH_79}  for strictly two dimensional crystals,  the positional quasi-long range order is destroyed at high enough temperature by the 
sequential release, first of dislocations and, at higher temperature, of disclinations. The possibility of buckling in the out-of-plane direction 
leads to a reduction of the formation energy of defects such as dislocations and disclinations \cite{NP_87,SN_88,N__95}.

In graphene, due to the covalent nature of the $\sigma$ bonds, responsible of the structural properties, the melting transition shows some 
differences both at qualitative and quantitative levels. Although in graphene the presence of topological defects made of a single heptagon-pentagon 
pair (dislocations) is allowed, it is the presence of Stone-Wales defects  (formed by two pentagons and two heptagons)  that plays a crucial 
role according to Monte Carlo simulations, since they are the structural defects with smallest formation energy (in practical terms, the 
Stone-Wales defect consists in just rotating one carbon-carbon link by 90 degrees). The melting transition temperature has been estimated to lie in 
the range $T_{c}\sim 4500- 5800 K$\cite{ZFL11,LZK15}. At the transition temperature, the system starts to form an amorphous network of one dimensional 
carbon chains. The same effect has been studied in carbon systems with small number of carbon atoms\cite{SNP13}. We should have in mind that the 
mentioned Monte Carlo simulations are performed keeping constant the number of carbon atoms during the simulation. It is quite likely that at lower 
temperatures than the ones mentioned above something more drastic, such as burning, will take place.

At low temperatures (well below any melting transition) topological defects still have an effect in the elastic properties of graphene. Early 
signatures of the role played by defects on the elastic and plastic properties of carbon-based structures appeared in the literature 
of nanotubes\cite{IBC96,BYB98}. It was explained that the way a carbon nanotube releases tension is to form Stone-Wales defects when the strain 
exceeds the critical value of five per cent. Quite remarkably, the Stone-Wales defects act as nucleation centres for the formation of dislocations, in 
the sense that when more tension is applied, the Stone-Wales defect splits and the two pentagon-heptagon pairs start to separate, and even to form 
more complex defects.

Back to graphene, it has recently been observed through high-resolution transmission electron microscopy that dislocations carry 
with them sizeable elastic deformations\cite{WMM12}. Interestingly, the dislocations and the associated elastic distortion are dynamical, that is, 
they can move with time in order to release stress. The time scales of these movements are of the order of seconds, too slow to have any impact in the 
electronic dynamics, but this defect movement might have an impact in the elastic dynamics of graphene as a membrane. Also, understanding the way 
these defects evolve constitutes an interesting problem on its own. In Ref.~\cite{WMM12} it has been observed that, as happens in nanotubes, 
under stress the plastic deformations start with the formation of Stone-Wales defects\cite{GMS10}. Increasing the applied 
stress, the two pentagon-heptagon pairs split apart\cite{CB08}. In the experiments shown in Ref.~\cite{WMM12}, the two heptagon-pentagon 
defects are separated by several lattice spacings. It has also been found experimentally that the dislocation movement is accompanied by the removal 
of carbon atoms (activation energy of the order of $5$ eV) to accommodate the energy irradiated by the transmission electron microscope. It has been 
also observed\cite{LKK13} that buckling has a profound impact in the dynamics of a dislocation from its origin, since for a membrane, 
distortions along the third dimension are allowed and graphene samples might also release elastic energy via buckling. This buckling 
process induces a long-ranged interaction between the heptagon-pentagon pairs, and the relative direction of the associated Burgers vectors 
determines the buckling profile, indicating an unexpected large feedback between the dislocation configuration, its associated buckling and the way 
the pentagon-heptagon pairs mutually interact through this buckling.

From the theoretical side, the dislocation dynamics in graphene is still an open problem. Significant advances have been achieved  
with discrete\cite{CB08} and continuous\cite{CC11,L13} modelling. Nevertheless, the fact that the defect dynamics seems not to be well described 
by the standard theory of dislocation dynamics indicates that the particular lattice topology of the honeycomb lattice might play an important role.

In the previous discussion, we only gave a small glimpse of the role of small topological defects on the elastic properties of a graphene  
membrane. However, it is by now apparent that in realistic samples extended defects are also present and they possibly have a more unexpected impact 
on the elastic properties of graphene. One-dimensional extended topological defects usually appear in chemically vapour deposited samples grown over 
metallic substrates, due to the fact that small graphene grains with different crystalline orientation serve as nucleation points and form large-area 
samples\cite{XWA09,YJQ11,HRC11}. 

Usually, the structure of such grain boundaries consists on heptagon-pentagon pairs along the grain boundary \cite{GSV10}. In Ref.~\cite{GSV10} it 
was stated that if the borders of adjacent graphene flakes are of zig-zag type, the Burgers vectors associated to each heptagon-pentagon pair are 
parallel, while if the borders are of armchair type, the associated Burgers vectors lose this constraint. Also, depending on the relative mismatch 
between the crystalline orders, the grain boundaries will be made of well separated heptagon-pentagon pairs (low tilt angles) or a dense array of 
them (high tilt angles). By using molecular dynamics plus \emph{ab initio} calculations, it has theoretically observed that the strength of the 
graphene sample  counterintuitively increases with the number of defects in the grain boundary, with the critical strain needed to 
break the sample growing with the number of pentagon-heptagon pairs present in the grain boundaries)\cite{GSV10}. Interestingly, calculations based 
on continuum mechanics fail to explain this behaviour, indicating that it is the microscopic lattice 
structure of these grain boundaries the ultimate 
responsible of the obtained behaviour. The reason behind this is that  what controls the failure 
under strain (which carbon bonds start to break 
first) is the way the strain is not uniformly accommodated along all the carbon bonds, and it tends to concentrate in the carbon bonds on the 
heptagon rings, being then the first to break. This indicates that the system releases strain by breaking carbon bonds around the heptagonal defects, 
leaving the rest of hexagonal rings much more free of strain.

This scenario has been partially questioned due to the fact that graphene can also release strain by buckling (or warping) as it has been mentioned 
previously. Other groups employing similar numerical techniques find that actually the stress-strain curve is not monotonic with the increase of 
defects, meaning that not always the strength of the sample increases with the number of defects at the grain boundaries\cite{HRC11}. In previous 
paragraphs we mentioned that the buckling around two close pentagon-heptagon pairs strongly depended on the relative orientation of their Burgers 
vectors, and the same applies here. The relevant point here is then to know if the relative orientation between the Burgers vectors plays the same role when the distance between heptagon-pentagon pairs is smaller than in the previous cases of a dilute density of dislocations. While for zig-zag borders this constraint seems to be quite general, we have much more freedom to choose the way the Burgers 
vectors are not parallel to each other. In this way, for armchair grain boundaries, many possible defect configurations with different Burgers vectors 
can be found, and accepting that the buckling profile depends on the Burgers vectors orientation, it is reasonable, at least qualitatively, to expect 
that the way the graphene sample will release strain by buckling (and the behaviour of the strength 
of the graphene sample) is not as uniform 
and universal as one might initially think.

\section{Electronic properties}
\label{sec_electronic}
% Sat Jan  3 17:18:11 CET 2015
%
%
%
%
%
%{\bf {\large Random strain fluctuations as a source of carrier scattering in graphene.}}
%
%
%%%%%%%%%%%%%%%%%%%%%%%%%%%%%%%%%%%%%%%%%%%%%%%%%%%%%%%%%%%%%%%%%%%%%%%%%%%%%%%%
\subsection{Strains and the mobility of carriers in graphene}
\label{Sec:general}
%%%%%%%%%%%%%%%%%%%%%%%%%%%%%%%%%%%%%%%%%%%%%%%%%%%%%%%%%%%%%%%%%%%%%%%%%%%%%%%%
Graphene is an excellent conductor, with carrier mobilities similar, or higher,
than those found in good metals, like copper or silver.  The lack of dependence of the mobility 
on carrier concentration in graphene strongly limits scattering mechanisms. Defects which induce a 
short range potential lead to a mobility proportional to the {\it inverse} of the concentration, 
which is incompatible with observations. Static strains are a possible source of disorder, and limit 
the mobility of the carriers. Other proposed mechanisms which lead to the correct dependence of 
mobility on carrier concentration are charged impurities in the substrate, and resonant 
scatterers\cite{Peres2010}. It is worth noting that thermally excited phonons lead to random strain 
distributions which can scatter electrons. The in-plane acoustic phonons of graphene are only 
relevant at high temperatures \cite{HS_08}, but out-of-plane flexural modes in suspended 
samples change significantly the mobility, being the primary electron scattering mechanism at room 
temperature \cite{MvO08}, giving origin to a $T^{2}$ dependence of resistivity \cite{Cetal10}. In 
supported samples, out-of-plane phonons become quenched and their contribution to resistivity is 
severely reduced \cite{AG_13}.

%
%
%%%%%%%%%%%%%%%%%%%%%%%%%%%%%%%%%%%%%%%%%%%%%%%%%%%%%%%%%%%%%%%%%%%%%%%%%%%%%%%%
\subsubsection{Strains and mobility.}
%%%%%%%%%%%%%%%%%%%%%%%%%%%%%%%%%%%%%%%%%%%%%%%%%%%%%%%%%%%%%%%%%%%%%%%%%%%%%%%%
In order for strains to give the correct (in)dependence of the mobility on carrier concentration, 
they should be correlated over long distances, as initially proposed in \cite{katsnelson_corrug}. 
In the following, we analyse the influence of strains in electronic transport, following recent 
experimental and theoretical work reported in \cite{Cetal2014}. As discussed there, it is natural 
to expect long range correlations in the strain distribution, and strains are the only mechanism 
compatible with weak localization experiments and the observed correlation between the mobility and 
the strength of puddles near the neutrality point.

The carrier mobility
$\mu_{c}$ is defined as the conductivity per carrier,
\begin{align}
\mu_{c} &= \frac{\sigma}{e n}
\label{eqn:paco_mu_def}
,
\end{align}
where $\sigma$ denotes the conductivity, $e$ is the electric charge, and $n$
the carrier density.  The highest mobilities are found in exfoliated flakes,
which are single crystals. They lie in the range $10^3$-$10^5$~cm$^{2}$/(V\,s),
and are independent of the carrier concentration
\cite{DYM10,XSB11,DWB11}.
Different substrates lead to different mobilities, and the highest values are
found in samples on boron nitride substrates and in suspended samples, where
mean free paths can be of the order of few microns.

If we describe the scattering of carriers by disorder in terms of an effective
scattering time $\tau$, we can write
\begin{align}
\mu_c &= \frac{e}{h}\frac{2 \pi v_F \tau}{k_F}
,
\label{eqn:paco_mu_tau}
\end{align}
where $h$ is Planck's constant, $v_F$ is the Fermi velocity, and $k_F$ is the
Fermi wavevector.
Below, we review some properties of the
mobility \cite{CGetal09,Peres2010_review,DasSarma2011_review}, and
(following the work in \cite{Cetal2014}) present arguments which suggest that
the dominant mechanism which determines the mobility is the presence of random
strains in the system.
%
%
%%%%%%%%%%%%%%%%%%%%%%%%%%%%%%%%%%%%%%%%%%%%%%%%%%%%%%%%%%%%%%%%%%%%%%%%%%%%%%%%
\subsubsection{Form of the disorder potential}
%%%%%%%%%%%%%%%%%%%%%%%%%%%%%%%%%%%%%%%%%%%%%%%%%%%%%%%%%%%%%%%%%%%%%%%%%%%%%%%%
%\subsubsection{Long range and short range scattering}
\label{subsubsec:paco_long_short}
The internal structure of the wavefunctions of carriers in graphene implies
that, within states in a given valley, an internal degree of freedom, the
pseudospin, can be defined. The value of the pseudospin is determined by the
carrier's momentum, and a variation of the pseudospin affects the phase of the
wavefunction in a similar way as that of a real spin in a material with strong
spin-orbit coupling. The effect of the pseudospin can be measured in weak
localization experiments, as they probe interference effects by applying a
small magnetic field which breaks the effective time-reversal symmetry of
electronic states around each $K$-point.  These experiments allow for the
definition of three different scattering times \cite{McCann2006,Morpurgo2006}:
(i)~$\tau_0$, a scattering time which changes momentum but does not influence
the phase of the wavefunction,
(ii)~$\tau_*$, which describes the suppression of interference associated to
random fluctuations of the pseudospin, and
(iii)~$\tau_i$, which gives the strength of the inter-valley scattering
processes (we neglect here the effect of the real spin-orbit coupling in
graphene, which is very small). The total scattering time, $\tau$, which enters
in the mobility is given by
$\tau^{-1} = \tau_0^{-1} + \tau_*^{-1} + \tau_i^{-1}$.

Weak localization experiments \cite{Metal06,Tikhonenko2008,Tikhonenko2009} show
that $\tau_*^{-1} \approx \tau^{-1} \gg \tau_i^{-1}$.
Thus, the random potential $V(\vec q)$ responsible for the scattering processes
in graphene must be described by a long range potential, which does not allow
for inter-valley scattering. In addition, this potential should couple to the
pseudospin of the wavefunction.

Further, the experimental observation of a mobility which is independent of
carrier concentration implies that the effective scattering time $\tau \propto
k_F$ [see Eq.~(\ref{eqn:paco_mu_tau})].  The scattering time is due to
processes which transfer a carrier from one position on the Fermi surface to
another.  We assume that the Fermi surface is a circle, and that the scattering
mechanism is isotropic.  Then, a scattering process by an angle $\theta$
involves a momentum transfer $\vec{q}$ such that $| \vec{q}| = 2 k_F \sin (
\theta )$, and the inverse of the scattering time (according to Fermi's golden
rule)  can be written as
\begin{align}
\frac{1}{\tau} &= \frac{2 \pi}{\hbar^2} \frac{N ( E_F )}{4 \pi^2} \times \nonumber \\ &\times 
\int_0^{\pi} d \theta \left[ 1 - \cos ( \theta ) \right] \left. \left\langle V ( \vec{q} ) V  ( - 
\vec{q} ) \right\rangle \right|_{| \vec{q} | = 2 k_F \sin ( \theta / 2 )}
,
\end{align}
where $N ( E_F ) \propto k_F / v_F$ is the density of states at the
Fermi energy $E_F$ and $\langle~\rangle$ denotes average over disorder. The $\tau \propto 
k_F$ dependence mentioned
above implies that
\begin{align}
\left. \left\langle V ( \vec{q} ) V  ( - \vec{q} ) \right\rangle \right|_{| \vec{q} | = 2 k_F \sin ( \theta / 2 )} &\propto \frac{1}{k_F^2}
.
\label{scaling}
\end{align}
Hence, the potential which scatters the carriers has to diverge like
$| \vec{q} |^{-1}$ as $k_F \rightarrow 0$.
%
%
%%%%%%%%%%%%%%%%%%%%%%%%%%%%%%%%%%%%%%%%%%%%%%%%%%%%%%%%%%%%%%%%%%%%%%%%%%%%%%%
\subsubsection{Correlation between mobilities and puddles at the Dirac point}
\label{subsubsec:paco_mobilities_puddles}
%%%%%%%%%%%%%%%%%%%%%%%%%%%%%%%%%%%%%%%%%%%%%%%%%%%%%%%%%%%%%%%%%%%%%%%%%%%%%%%
Experiments show a linear correlation between the mobility at high carrier
concentration and the minimum carrier density which can be measured near the
neutrality point \cite{Cetal2014}. Defects can shift locally the position of
the Dirac energy, so that, for nominal zero concentration, the carrier density
fluctuates, with a non zero absolute average. This effect (typically referred
to as the existence of puddles \cite{Martin2008}) leads to a lower bound in the
resolution of the carrier density, $n^*$, near neutrality.

The observed dependence is $1/\mu_{c} = C n^*$, where $C$ is a constant which does
not depend on the type of substrate or amount of disorder.
%
%
%%%%%%%%%%%%%%%%%%%%%%%%%%%%%%%%%%%%%%%%%%%%%%%%%%%%%%%%%%%%%%%%%%%%%%%%%%%%%%%
%\subsection{Models for scattering.}
\subsubsection{Carrier scattering by defects and impurities}
\label{subsubsec:Paco_models_for_scattering}
%%%%%%%%%%%%%%%%%%%%%%%%%%%%%%%%%%%%%%%%%%%%%%%%%%%%%%%%%%%%%%%%%%%%%%%%%%%%%%%
Apart from the aforementioned random strains, there are  various theoretical
models describing the charge scattering in graphene. The different microscopic mechanisms
relate to different scattering times $\tau$ and carrier mobilities $\mu$:
\begin{enumerate}[(i)] %% This needs `\usepackage{enumerate}' in the preamble
\item \emph{Weak defects:}
The simplest scattering mechanism is given by local scatterers which perturb
weakly the band structure of graphene. Each of these scatterers can be
described by a momentum independent potential $\bar{V}$, resulting in
\begin{align}
\tau^{-1} &\propto N ( E_F ) \bar{V}^2 \propto k_F
.
\label{eqn:Paco_tau_defect}
\end{align}
\item \emph{Charged impurities:}
The electrostatic potential induced by a point charge,
\begin{align}
V ( \vec{q} ) = V_C ( \vec{q} ) / \varepsilon ( \vec{q} )
,
\label{eqn:Paco_Velectrostat}
\end{align}
where $V_C ( \vec{q} ) = e^2 /(2 \varepsilon_0 | \vec{q} |)$ is the 2D Coulomb
potential (in SI units), $\varepsilon_0$ is vacuum permittivity and  $\varepsilon ( {\vec
q} ) = 1 + V_C ( \vec{q} ) N ( E_F ) = 1 + (e^2 k_F) / (2 \pi v_F \varepsilon_0 | {\vec
q} | )$
is the static dielectric function in graphene, satisfies the scaling law
%for the disorder potential in
Eq.~(\ref{scaling}).
The resulting scattering time $\tau$ and mobility are
\begin{align}
\tau^{-1} &\propto \frac{c_{ch} v_F}{k_F}\nonumber , \\ \mu_c &\propto \frac{e}{h} \frac{1}{c_{ch}}
,
\end{align}
where $c_{ch}$ is the concentration of charged impurities,
independent of the carrier density\cite{Nomura2007,Hwang2007,Adam2007_Pnas}.
This mechanism can be consistent with the correlation between $\mu_c$ and $n^*$
described in Sec.~\ref{subsubsec:paco_mobilities_puddles}.
\item \emph{Resonant scattering:}
A strong local potential, like the one induced by a vacancy or an adatom covalently bonded to a 
neighbouring carbon atom, induces a strong resonance near the Dirac 
energy\cite{Stauber2007,Ni2010b}. 
A potential of this type also satisfies the scaling behaviour in Eq.~(\ref{scaling}), and leads to 
a 
mobility
\begin{align}
\mu_c &\propto \frac{e}{h} \frac{1}{c_{res}}
,
\end{align}
where $c_{res}$ is the concentration of resonant scatterers.
\end{enumerate}
The scattering time Eq.~(\ref{eqn:Paco_tau_defect}) for weak defects is inconsistent
with the observed independence of the mobility on carrier density.
On the other hand, the long range potential
induced by charged impurities, Eq.~(\ref{eqn:Paco_Velectrostat}), does
not modify the pseudospin,
%and cannot explain the weak localization experiments.
while resonant scatterers have dimensions comparable to the lattice spacing, and can
thus be expected to induce inter-valley scattering.
This leaves mechanisms {(ii)} and {(iii)} at odds
with the results of weak localization experiments described in Sec.~\ref{subsubsec:paco_long_short}.
%
%
%%%%%%%%%%%%%%%%%%%%%%%%%%%%%%%%%%%%%%%%%%%%%%%%%%%%%%%%%%%%%%%%%%%%%%%%%%%%%%%
\subsection{Carrier scattering by random strain fluctuations}
%%%%%%%%%%%%%%%%%%%%%%%%%%%%%%%%%%%%%%%%%%%%%%%%%%%%%%%%%%%%%%%%%%%%%%%%%%%%%%%
%\subsubsection{General properties}
The effect of strains on electrons and holes in graphene can be divided
into a scalar potential,
$V_s ( \vec{r} ) = g_1 \left( u_{xx} (\vec{r} ) + u_{yy} (\vec{r}) \right) $,
and a gauge potential,
$\vec{\mathcal{A}}^{\mathit{el}} ( \vec{r} ) =
g_2 \left( u_{xx} ( \vec{r} ) - u_{yy} ( \vec{r} ) ,- 2 u_{xy} ( {\vec r} ) \right)$, see Eqs. 
\eqref{Eq:deformation_potential},\eqref{Eq:gauge_field}, where $u_{ij}$ is the strain tensor, and 
$g_1$ and $g_2$ are
constants \cite{VKG10} discussed in Sec. \ref{Sec:electron-phonon_coupling}.
%(ALSO --THROUGHOUT THIS SECTION-- UPDATE NOTATION FOR $g_{1,2}$ WITH THE FINAL FORM OF EQ \ref{Eq:VdA} TO BE DEFINED IN THE INTRODUCTION)
Random strains will give rise
to a potential which satisfies the scaling in Eq.~(\ref{scaling}) if
\begin{align}
\left. \left\langle u_{ij} ( \vec{q} ) u_{kl} ( - \vec{q} ) \right\rangle \right|_{| \vec{q} | = k_F} &\propto \frac{1}{k_F^2}
\label{scaling_strains}
\end{align}

While the gauge field $\vec{\mathcal{A}}^{el}$ does not break the electron-hole symmetry of the 
graphene
bands, so that it cannot induce density fluctuations near the neutrality point,
the scalar potential $V_s$ does break this symmetry, and can therefore lead to the
formation of puddles \cite{GTetal12}. On the other hand, the scalar potential
does not couple to the pseudospin of the wavefunction, while the gauge potential
does, and  hence can explain the weak localization experiments.
The two contributions to the potential lead to the scattering times
\begin{align}
\frac{1}{\tau_s} &=\frac{2 \pi}{\hbar^2} \frac{N(E_F)}{4\pi^2}
\nonumber
\\
&\times \int_0^\pi d\theta \left.
\frac{1-\cos^2(\theta)}{2}  \frac{\langle
V_s({\vec{q}})V_s({-\vec{q}})\rangle}{\varepsilon^2(\vec{q})} \right|_{|\vec{q}|=2k_F\sin(\theta/2)}
\\
\frac{1}{\tau_g} &=\frac{2 \pi}{\hbar^2} \frac{N(E_F)}{4\pi^2}
\nonumber
\\
&\times \int_0^\pi d\theta \left. [1-\cos(\theta)]  \langle
\vec{\mathcal{A}}^{el}_{\perp}({\vec{q}})\vec{\mathcal{A}}^{el}_{\perp}({-\vec{q}})\rangle 
\right|_{|\vec{q}|=2k_F\sin(\theta/2)}
\label{taug},
\end{align}
where $\vec{\mathcal{A}}^{el}_{\perp} ( \vec{q} )$ is the component of $\vec{\mathcal{A}}^{el}$ 
perpendicular
to $\vec{q}$, as $\vec{\mathcal{A}}^{el}_{\parallel} ( \vec{q} )$ can be gauged away. The
average puddle density due to the scalar potential is
\begin{equation}
n^*= \frac{1}{\pi} \frac{\langle V_s(\vec{r})^2\rangle }{(\hbar
v_F)^2}=\frac{1}{4\pi^3 \hbar^2 v_F^2} \int d^2\vec{q}\frac{\langle
V_s({\vec{q}})V_s({-\vec{q}})\rangle}{\varepsilon^2(\vec{q})}.
\end{equation}
The scalar potential is screened by the static dielectric function, $\varepsilon ( \vec{q} )$,
while the gauge potential is not. That implies that $\tau_g^{-1} \ll
\tau_s^{-1}$. Hence, when the scattering is induced by strains, $\tau \sim
\tau^* \sim \tau_g$, in agreement with the weak localization experiments
described above.

In the following, we discuss strain due to out-of-plane corrugations and to
in-plane displacements.
Their effect on spectral properties will be discussed in Sec. \ref{Sec:superlattices}.
Both types of strain lead to potentials consistent with the
scaling behaviour in Eq.~(\ref{scaling}), and with the observed correlation
between mobility and puddle density.
%
%
%%%%%%%%%%%%%%%%%%%%%%%%%%%%%%%%%%%%%%%%%%%%%%%%%%%%%%%%%%%%%%%%%%%%%%%%%%%%%%%%
\subsubsection{Random out-of-plane corrugations}
\label{subsubsec:Paco_out_of_plane}
%%%%%%%%%%%%%%%%%%%%%%%%%%%%%%%%%%%%%%%%%%%%%%%%%%%%%%%%%%%%%%%%%%%%%%%%%%%%%%%%
Random out-of-plane corrugations have been observed in graphene on different
substrates (where it is assumed that they follow the corrugations of the
substrate) \cite{XSB11} as well as in suspended samples \cite{MGK07}.
The fluctuations in the height of the graphene layer, $h ( \vec{r} )$, give
rise to strains $u_{ij} \propto \partial_i h \partial_j h$.
For the relation (\ref{scaling_strains}) to hold, we have
\begin{align}
\left\langle \left. \partial_i h \partial_j h \right|_{\vec{q}} \left. \partial_k h \partial_l h \right|_{-\vec{q}} \right\rangle &\propto \frac{1}{| \vec{q} |^2}
.
\end{align}
This momentum dependence can be obtained if \cite{katsnelson_corrug}
\begin{align}
\left\langle h ( \vec{q} ) h ( - \vec{q} ) \right\rangle &\propto \frac{1}{| \vec{q} |^4}
.
\label{height}
\end{align}
The correlation function (\ref{height}) describes the thermal distribution of
classical flexural modes in the harmonic approximation (see Section 
\ref{sec:membrane_theory}). Hence, if we assume that the corrugations in a
graphene flake are given by the quenched shape of that flake at some high
temperature, we obtain a mobility which is independent of the carrier
concentration.

Assuming a height scaling as in Eq.~(\ref{height}), the relation between
mobility and average puddle density satisfies
\begin{equation}
\frac{1}{\mu_{c}} =
n^*\frac{h}{4e}\left[\frac{\hbar^2v_F^2}{8e^4}+\frac{g_2^2(\lambda+\mu)^2}{g_1^2\mu^2}\right]
\frac{1}{\log[1/(k_F(n^*)a)]} ,
\label{munstar1}
\end{equation}
which, neglecting the last logarithmic factor, only depends on parameters
defined in the absence of disorder. For reasonable values of $g_1$ and $g_2$
this relation is in close agreement with the experimentally observed one
\cite{Cetal2014}.

\begin{figure}[h]
\begin{center}
\begin{tabular}{lr}
\includegraphics[width=4cm]{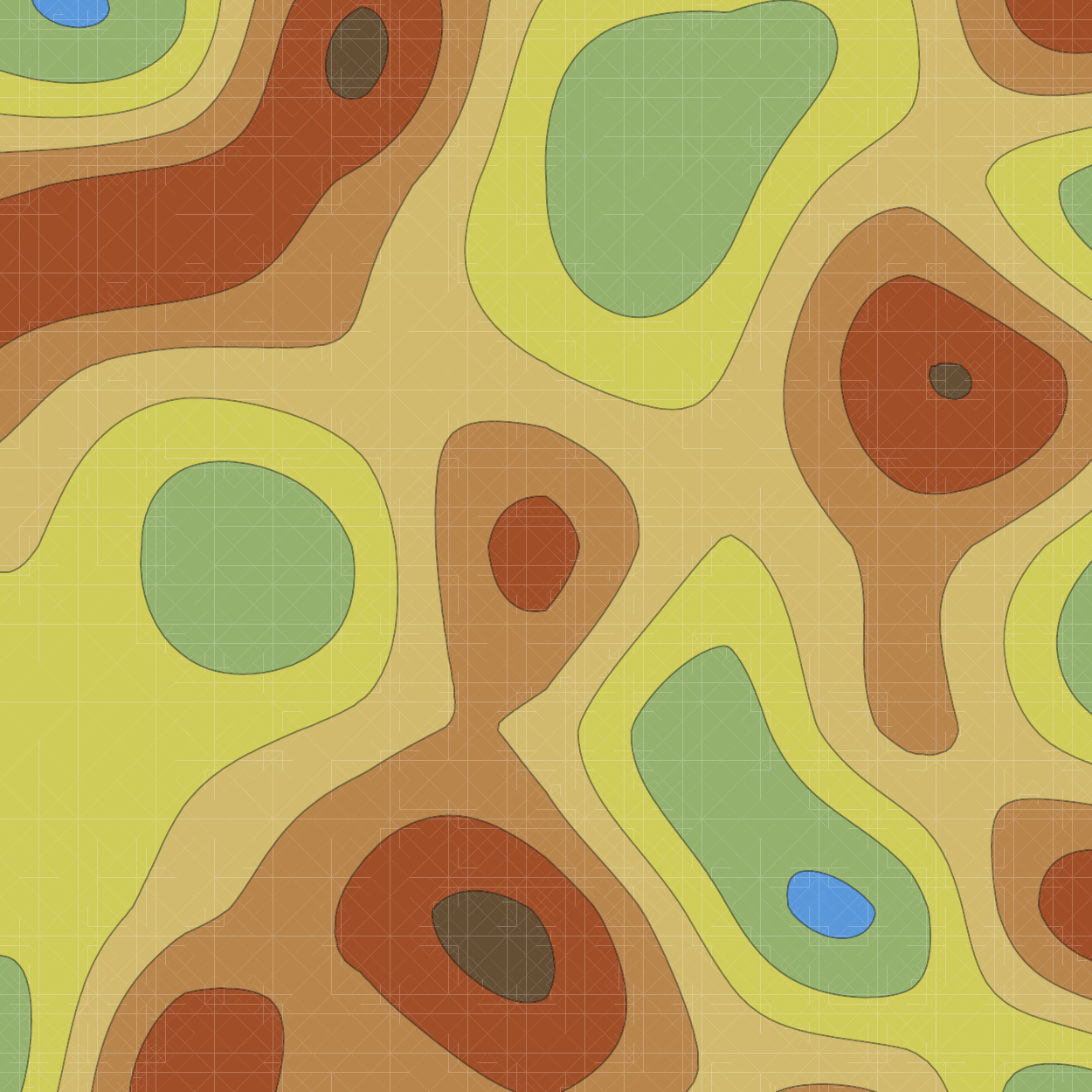}%
&
\includegraphics[width=4cm]{Figure4a}%
\end{tabular}
\end{center}
\caption[fig]{\label{fig:strains}(Colour online).
Sketch of the potential (left) and forces (right) induced by the substrate on a graphene layer.
To compare with a floating phase configuration, cf. Fig. \ref{Fig:deformations}.
}
\end{figure}

%
%
%%%%%%%%%%%%%%%%%%%%%%%%%%%%%%%%%%%%%%%%%%%%%%%%%%%%%%%%%%%%%%%%%%%%%%%%%%%%%%%%
\subsubsection{In-plane random displacements}
\label{subsubsec:Paco_in_plane}
%%%%%%%%%%%%%%%%%%%%%%%%%%%%%%%%%%%%%%%%%%%%%%%%%%%%%%%%%%%%%%%%%%%%%%%%%%%%%%%%
The substrate induces in-plane forces on the carbon atoms that constitute the graphene
layer, such that the atoms will be displaced towards positions which lower the potential
energy. This effect is responsible for the formation of Moir\'e patterns in
graphene on BN substrates \cite{WBE14,JRQM14,SGSG14}.
A sketch of a possible
random potential induced by the substrate, and the resulting displacements, is
given in Fig.~\ref{fig:strains}.
A detailed analysis of this issue concerning periodic
deformations is carried out in Sec. \ref{Sec:superlattices}, where a floating
phase is assumed. (See in particular the discussion in Section
\ref{Sec:spontaneous_deformations}.)

Forces $\vec{F}$ with a  modulation $\vec{K}$ similar to the
reciprocal vectors of the graphene lattice, $\vec{G}$, lead to smooth
displacement distributions, modulated by the wavevector $\vec{k} = \vec{K} -\vec{G}$:
\begin{align}
\vec{u} ( \vec{k} ) &\propto \frac{\vec{F}_\parallel ( \vec{k} )}{( \lambda + 2 \mu ) | \vec{k} |^2} + \frac{\vec{F}_\perp ( \vec{k} )}{\mu | \vec{k} |^2}
,
\end{align}
where $\vec{F}_{\parallel} ( \vec{k} )$ and $\vec{F}_{\perp} ( \vec{k} )$ denote the
components of $\vec{F} ( \vec{k} )$ parallel and normal to $\vec{k}$, respectively, and
$\lambda$ and $\mu$ are elastic Lam\'e coefficients, as discussed in Sec.~\ref{Sec:elasticity}. The 
linear strain tensor is given
by $u_{(i,j)} ( \vec{k} ) = i \left( k_i u_j ( \vec{k} ) + k_j u_i ( \vec{k }) \right) / 2$.

The forces can be written as a function of the potential between the substrate and the graphene layer
\begin{align}
\vec{F} ( \vec{K} )&= i \vec{K} V_{\vec{K}}
.
\end{align}
Hence, for $\vec{K} \approx \vec{G}$, we find
\begin{align}
\left\langle u_{(i,j)} ( \vec{k} ) u_{(l,m)} ( - \vec{k} ) \right \rangle &\propto \frac{ | \vec{G} 
|^2 \left\langle V_{\vec{G} - \vec{k}} V_{-\vec{G} + \vec{k}} \right\rangle}{| \vec{k} |^2}
.
\end{align}
We assume that the potential $V_{\vec{G} - \vec{k}}$ has random correlations
for $| \vec{k} | \ll | \vec{G} |$. Then $\lim_{| \vec{k} | \rightarrow 0}
\left\langle V_{\vec{G} - \vec{k}} V_{-\vec{G} + \vec{k}} \right\rangle = $
constant, and the correlation between strains follows the scaling in
Eq.~(\ref{scaling_strains}). This mechanism gives a mobility which is
independent of the carrier concentration. The relation between the mobility and
the puddles' density is, similar to that in Eq.~(\ref{munstar1}),
\begin{align}
 \frac{1}{\mu_{c}} &=  n^* \frac{h}{4e}
 \left[\frac{\hbar^2v_F^2}{16e^4} + \frac{ g_2^2}{g_1^2}
 \left( 1 + \frac{ ( \lambda + 2 \mu )^2}{\mu^2} \right) \right] \nonumber \\ &\times
 \frac{1}{\log[1/(k_F(n^*)a)]}.
 \label{munstar2}
 \end{align}
As in Eq.~(\ref{munstar1}), for reasonable values of the parameters of $g_1$ and $g_2$,
the relation between $\mu_{c}$ and $n^*$ agrees with experiments \cite{Cetal2014}.

%\subsubsection{Conclusions}
Thus, scattering of charge carriers by random strains, as detailed in Sections~\ref{subsubsec:Paco_out_of_plane}
and~\ref{subsubsec:Paco_in_plane}
above,  fulfils
the experimental constraints which a microscopic charge scattering mechanisms in graphene
must satisfy.
Moreover, these constraints, namely
(i)~the existence of a mobility independent of carrier
concentration, (ii)~the dominance of long range intra-valley scattering
processes coupled to the wavefunction's pseudospin,
%, as evidenced in weak localization experiments,
and (iii)~the correlation between mobility and
puddles' density near the neutrality point,
are not consistent with the other models for scattering discussed in Section~\ref{subsubsec:Paco_models_for_scattering} \cite{Cetal2014}.
%only consistent with scattering by random strains\cite{Cetal2014}.
Furthermore,
spatially resolved Raman measurements show a good correlation between
carrier mobilities and amplitudes of strain distributions \cite{Cetal2014}.
%
%
%%%%%%%%%%%%%%%%%%%%%%%%%%%%%%%%%%%%%%%%%%%%%%%%%%%%%%%%%%%%%%%%%%%%%%%%%%%%%%%%%%%%%%%%%%%%%%%%%%%%
\subsection{Electron pumping through mechanical resonance in graphene}
\label{subsec:mechanical_resonators}
%%%%%%%%%%%%%%%%%%%%%%%%%%%%%%%%%%%%%%%%%%%%%%%%%%%%%%%%%%%%%%%%%%%%%%%%%%%%%%%%%%%%%%%%%%%%%%%%%%%%
While electron scattering from random microscopic strain configurations or from thermally excited phonons contributes to
the resistivity of graphene sheets, macroscopic
deformations (with amplitudes of a few nanometres in samples of micrometres in
size) produced in graphene based  nano-electromechanical systems
(GNEMS)
%produced electrically or optically
can  generate electrical current by moving electrons coherently\cite{LJK12}.
A typical setting for a GNEMS is made up of a graphene flake of length $L$
suspended over a pre-patterned trench. The suspended flake serves as a mechanical resonator,
which can be actuated via a
radiofrequency gate-voltage
%$\delta V_g$
applied between graphene and the
substrate\cite{BSZ07,CLD13}.
%in addition to a constant gate voltage $V_g$ \cite{BSZ07,CLD13}.
In alternative, optical actuation can be achieved by modulating
the intensity of a diode laser shone over the flake\cite{BSZ07}.
Since their first experimental realization
\cite{BSZ07}, GNEMS have been fabricated in various ways: graphene flakes
produced by mechanical exfoliation\cite{BSZ07}, CVD\cite{ZBA10}, or by epitaxial
growth on SiC\cite{SBY09}, have been suspended on differently shaped substrates.

In the miniaturization of NEMS
to the nanoscale, which  calls for low oscillator masses and
high resonance frequencies \cite{C00,B04,ER05},
%Reducing devices dimensions to
%This justifies
%the interest on graphene as principal component for efficient NEMS
%(NanoElectroMechanical Systems).
%In the technological race toward miniaturization nanomechanical systems (NEMS),
%being in the nanoscale, costitute the next frontier. These devices have the
%purpose to take advantage of mechanical and electrical response of material in
%the submicrometer scale for the most different practical
%applications\cite{C00}\cite{B04}\cite{ER05}.  Reducing devices dimensions to
%such a small scale brings challenges in read-out and device performances.
%In this scenario, graphene has been proved to be once again an extraordinary
%material, reversing the usual fabrication scheme "room-bottom" to a "bottom-up"
%one.
graphene
with its low mass density, high
stiffness, high electronic mobility, and its chemical inertness is a promising material \cite{LJK12,CH13}.
Furthermore, graphene allows tunability of the resonance frequency $\omega_0$ on a wide range by means
of strain:
%In one-dimensional approximation, i.e.
Modelling the graphene membrane as a beam,
%and neglecting the bending contribution the resonance frequency
$\epsilon$ scales with the square root of the applied strain $\epsilon$,
\begin{equation}\label{eq:fundfrequency}
%\omega_0=\frac{8h_0}{L^2}\sqrt{\frac{\lambda+2\mu}{\rho}}
\omega_0=\frac{1}{2L}\sqrt{\frac{Y}{\rho} \epsilon}
,
\end{equation}
where $Y$ denotes the Young modulus of graphene.
Eq.~\eqref{eq:fundfrequency} takes into account only the effect of tension on the
fundamental frequency. A realistic model of a graphene resonator
%describing the experimental setting analyzed above
must also further consider the electrostatic force between
graphene and the substrate, the elastic restoring force and possible
dissipative effects\cite{LJK12,AIK08}.

Read-out of GNEMS can be performed through electrical or optical
means\cite{CH13}.
%The important role played by strain in electronic transport
%properties of graphene also leads to important consequences of transport
%properties in GNEMS.
It has been shown that, due to the long
wave-length strains induced by vibrations, a GNEM close to
resonance can work as an adiabatic quantum charge pump\cite{LJK12}. If
inversion symmetry is broken by maintaining   different chemical potentials at  graphene contacts
on opposite ends of the trench, one obtains a nonzero pumping current such that the transport of a single
charge per cycle is feasible. Coulomb blockade in such a device would result in
the transport of an integer number of charges per cycle, so that the ratio
between current and frequency will be quantized. This paves the way to the
application of GNEMS as prototypical standards for current and resistance.

\subsection{Other effects of strains on the electronic structure of graphene.}
Strains modify the hoppings between carbon $\pi$ orbitals. As mentioned in the introduction, this 
effect shifts the Dirac point from the corners of the Brillouin Zone, and leads to the appearance of 
effective gauge fields\cite{VKG10}. When the strains are spatially modulated, an effective magnetic 
field s induced. A smoothly increasing strain leads to an effective constant magnetic 
field\cite{GKG10}. This effect was confirmed by observing the local density of states of highly 
strained triangular graphene bubbles on platinum\cite{LBM10}. The combination of large strains, 
$\epsilon \sim 10 \%$, which change over a small scale, $\ell \sim 15$ nm, and a triangular shape 
favour the formation of large effective fields. Numerous experiments have confirmed the existence 
of 
significant effective magnetic fields in strained graphene, see, for 
instance\cite{Yetal11,Ketal12,Detal12}.

%\documentclass[rmp,onecolumn,superscriptaddress]{revtex4-1}
%%\usepackage{savetrees}
%\usepackage{amsmath}
%\usepackage{epsfig}
%\usepackage{graphicx}% Include figure files
%\usepackage{dcolumn}% Align table columns on decimal point
%\usepackage{bm}% bold math
%\usepackage{slashed}
%\unitlength = 1cm
%\usepackage[toc,page]{appendix}
%\usepackage{float}
%\usepackage{graphicx,xspace,units}
%\usepackage{amsmath}
%\usepackage{amssymb}
%%\usepackage[normalem]{ulem}
%%\usepackage{wrapfig}
%\usepackage{epsfig}
%%\usepackage[pdftex]{graphicx}
%\usepackage{hyperref}
%
%
%\usepackage{graphicx}
%\renewcommand{\dag}{^{\dagger}}
%\newcommand{\dl}{\partial_\ell}
%\def\gapp{\lower.35em\hbox{$\stackrel{\textstyle>}{\sim}$}}
%\def\lapp{\lower.35em\hbox{$\stackrel{\textstyle<}{\sim}$}}
%%\newcommand{\bs}[1]{{\boldsymbol{#1}}} 
%%\newcommand{\red}[1]{{\textcolor{red}{#1}}}
%%\newcommand{\blue}[1]{{\textcolor{blue}{#1}}}
%%\newcommand{\magenta}[1]{{\textcolor{magenta}{#1}}}
%\newcommand{\green}[1]{{\textcolor[rgb]{0,0.5,0}{#1}}}
%
%\begin{document}
%\bibliographystyle{apsrev}
%%
%
%%\draft
%\title{Effects of electron-electron interactions in strained graphene}
%
%\author{Nice People} 
%\affiliation{Good places}
%
%\begin{abstract}
%Abstract here
%\end{abstract}
%
%%\maketitle
%
%%\section{Strain and interaction effects in graphene}
%
%
\subsection{Effects of electron-electron interactions in strained graphene}
\label{sec_interactions}
%%%%%%%%%%%%%%%%%%%%%%%%%%%%%%%%%%%%%%%%%%%%%%%%%%%%%%%%%%%%%%%%%%%%%%%%%%%%%%%%
The effect of electron-electron interactions in graphene 
has been a central research area in the field since the early days~\cite{KUP12}. 
At half-filling, and unlike in typical metallic systems, Coulomb interactions have a long-range character due to the 
vanishing density of states at the Dirac point, resulting in
an enhancement of the Fermi velocity at low energies \cite{GGV94,GGV99,EGM11}.
On the other hand interactions in doped graphene are screened, but can nevertheless lead to 
remarkable phenomena. For instance, a particularly appealing scenario occurs when
graphene is doped up to the van Hove singularity, where the divergent density of states
makes it prone to a number of interesting weak coupling instabilities including 
d-wave superconductivity~\cite{G_08,MB_08,VV_08,MGR11,L_12,NLC12}.

The interplay between strain and interaction effects opens a different avenue to explore
and enhance novel interaction related phenomena.  For instance, by applying
uniaxial strain, it is possible to alter the electronic spectrum and bring the
van Hove singularity to lower energies~\cite{PRP10} and rippled
disorder can also enhance interactions~\cite{GHD08}. In what follows, we focus on how electron-electron interactions of different kinds 
can affect strained graphene and the different phenomena they induce.

\subsubsection{Topological phases induced from the interplay between strain and interactions}
\label{Sec:topointeractions}
Unlike symmetry broken phases, topological states of matter are
described and classified by topological invariants~\cite{HK10,QZ11}. This rapidly expanding and 
multidisciplinary field was boosted by and largely benefited from research in graphene 
itself~\cite{H88,KM05}, cf. Sec.~\ref{Sec:topo} .  \\
An exciting alternative to access topological phases is through the interplay between strain and electron-electron interactions.
For instance, an early example was motivated by the magnetic catalysis scenario for 
real magnetic fields in graphene~\cite{GMS06}. This mechanism, identified first in high-energy physics, refers to the enhancement of the dynamical 
symmetry breaking by an external magnetic field~\cite{S13}. In the context of graphene it is responsible for the opening of a gap catalyzed by the strong magnetic field, which
favours the electron-hole pairing formation. For pseudo-magnetic fields, it was shown~\cite{H_08,RH_13,RGG15} that strained graphene is 
similarly unstable towards an interaction induced gap, but in this case the ground state breaks time reversal symmetry and has a finite Hall conductivity~\cite{H88}. 
\\
An experimentally plausible scenario emerges when considering strain configurations that result in strong uniform pseudo-magnetic fields
\cite{GKG10,GGK10}. In this set-up, the uniform pseudo-magnetic field splits the spectrum into pseudo Landau levels (PLL)
\begin{eqnarray}
\label{eq:PL}
E_{n}= \mathrm{sgn}(n)\sqrt{2n}\hbar v_{F}/l_{B} \hspace{1cm} \psi^{\vec{K}}_{n,m} = 
\psi^{\vec{K}'}_{n,m}=
\begin{cases}
		\frac{1}{\sqrt{2}}(\phi_{|n|-1,m}(z), \mathrm{sgn}(n)\phi_{|n|,m}(z)) & \text{if } n\neq0, \\
		(0,\phi_{0,m}) & \text{if} n=0.
	\end{cases} 
\end{eqnarray}
with $l_{B}=\sqrt{\hbar/(eB)}$ being the standard magnetic length corresponding to the 
pseudo-magnetic field strength, $B$, 
and $\phi_{n,m}$ are the wave functions of the $n$-th level of the non-relativistic Landau level 
problem with angular momentum eigenvalue $m$. 
The states are four-fold (spin and valley) degenerate. They enjoy a $\mathbb{Z}_{2}\times SU(2)$ symmetry,
where $\mathbb{Z}_{2}$ represents the time-reversal symmetry while $SU(2)$ are the spin-rotations.
Note that this is in sharp contrast to the $SU(4)$ symmetry in an external real magnetic field where also the wave functions are different for 
different $\vec{K}$ and $\vec{K}'$ points; $\psi^{\vec{K}}_{n,m} \neq 
\psi^{\vec{K}'}_{n,m}$.\\

Long-range electronic interactions can lift the large degeneracy of a PLL defined in Eq. \eqref{eq:PL} and lead to interesting ordered states, including topologically non-trivial phases.
To tackle this many-body problem, it is generally assumed that interactions are not strong
enough to mix different PLL. In this case it is possible to distinguish the two cases in \eqref{eq:PL}, i.e. $n\neq0$ and $n=0$. 
Considering $n \neq 0$, the projected Hamiltonian describing long-range interactions can be written as~\cite{AP_12}
\begin{eqnarray}
\label{eq:int}
\mathcal{H}_{\mathrm{int}}=\sum_{\vec{q}}V_{C}(\vec{q})n(\vec{q})n(-\vec{q}).
\end{eqnarray}
Here $V_{C}(\vec{q})$ is the Fourier transform of the long-range Coulomb interaction and the 
projection is encoded in the definition of the density operator 
$n(\vec{q})=\sum_{\kappa,\sigma}\bar{n}_{\kappa,\sigma}(\vec{q})$, where 
$\bar{n}_{\kappa,\sigma}(\vec{q})$
are density operators projected to the $n$-th PLL that depend on the valley 
$\kappa=\vec{K},\vec{K}'$ and spin $\sigma=\uparrow,\downarrow$ degrees of freedom. In particular, 
the pseudo magnetic field has opposite signs at the $\vec{K}$ and $\vec{K'}$ points, a fact 
that is reflected through the dependence on $\kappa$.\\
Within the Hartree-Fock approximation the leading instability for a 1/4 or 3/4 filled $n\neq0$ PLL 
is a state with broken $\mathbb{Z}_{2}$ valley symmetry. Such a state is not time-reversal symmetric 
and has a quantized $\sigma_{xy}=e^2/h$. In turn, for a 
half-filled $n\neq0$ PLL with spin-orbit coupling a similar analysis shows that interactions drive 
the system to a quantum spin-Hall effect with a quantized spin Hall response~\cite{AP_12}.\\
Interestingly, the possibility of realizing the elusive fractional Chern insulators (FCI) (for a review see~\cite{BL_13}), zero-magnetic field analogues of the fractional quantum Hall effect, has also been proposed within this scenario~\cite{GCS12,AP_12}. Via exact diagonalization it was shown that a partially filled $n=0$ PLL stabilizes a valley polarized 
fractional quantum Hall state at $2/3$ filling through long-range Coulomb interactions that is robust to repulsive nearest neighbour interactions~\cite{GCS12}. By further 
tuning the latter to attractive interactions, a time-reversal symmetric fractional topological insulator state is stable as well as a spin-triplet superconducting state.
Motivated by these findings, the stability of the former against different interactions was also further addressed~\cite{CY_12}.\\

\begin{figure}
\begin{center}
\includegraphics{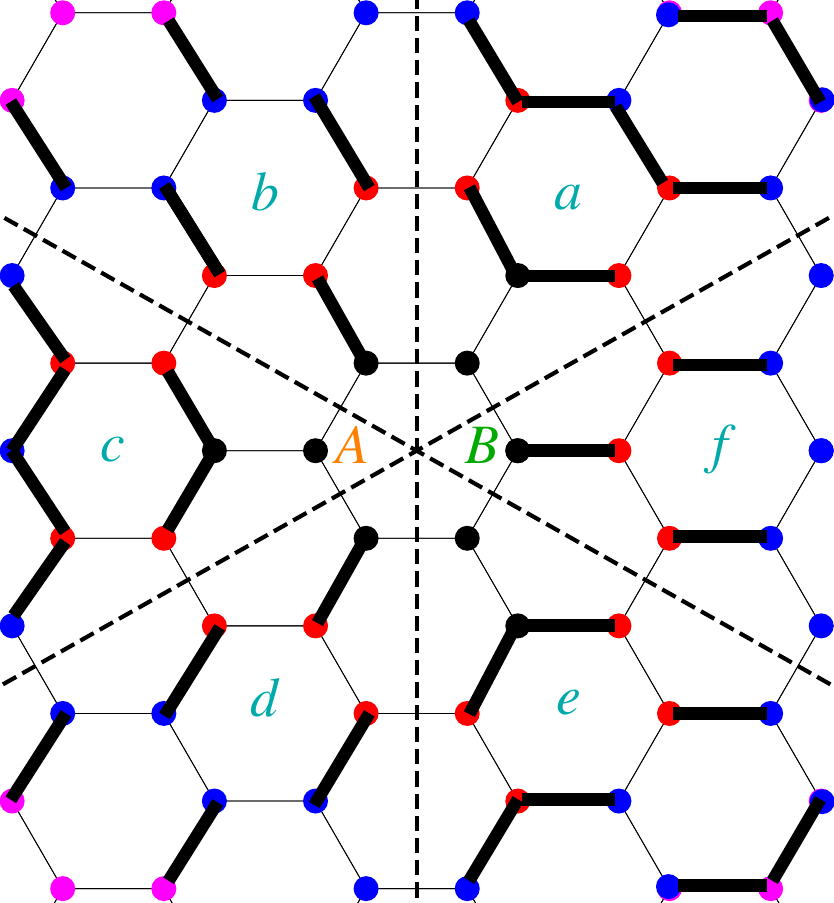}
\caption{Pictorial representations of the hopping amplitudes defined by $\chi(\mathcal{R})$, an increasing of the minimal number of bonds $\mathcal{R}$ to reach a central hexagon.
The bond strength is such that if $\mathcal{R}$ is odd then $\chi(A)>\chi(B)$ else $\chi(A)=\chi(B)$, a strain configuration 
that can interpolate between approximately homogeneous to inhomogeneous magnetic 
fields~\cite{RH_13}. Sections (a,c,e) and (b,d,f) map into each other under a $2\pi$/3 rotation.}
\label{fig:herb}
\end{center}
\end{figure}
Inhomogeneous pseudo-magnetic field configurations can also lead to topological insulating phases. 
For instance, consider a site dependent hopping amplitude of the form~\cite{RH_13}
\begin{eqnarray}
\label{eq:hopingo}
t_{ij}=e^{\chi(i)}te^{-\chi(j)}
\end{eqnarray}
with $i\in A$ and $j\in B$ and $t$ a constant hopping amplitude.
The function $\chi(\mathcal{R})$, with $\mathcal{R}$ defined as a radial coordinate from 
a central hexagon (see Fig.~\ref{fig:herb}) models a lattice gauge potential that can interpolate between an approximately uniform pseudo magnetic field ($\chi(\mathcal{R})\propto\mathcal{R}^2$) to a bell-shaped field ($\chi(\mathcal{R})\propto \ln\mathcal{R}$) while preserving $C_{3}$ rotational symmetry. 
In analogy to the pristine unstrained graphene~\cite{RQH08} fermions at half filling subjected to a strain pattern and repulsive interactions can realize topological Chern and spin Hall insulator phases. The starting point is the Hamiltonian
\begin{eqnarray}
\label{eq:hamint}
\mathcal{H}&=& \sum_{\left\langle i,j\right\rangle} t_{ij}\left( c^{\dagger}_{i}c_{j} 
+\mathrm{h.c.}\right)+U\sum_{i, 
\sigma,\sigma'}n_{i,\sigma}n_{i,\sigma'}+V_{1}\sum_{\left\langle\left\langle 
i,j\right\rangle\right\rangle,\sigma,\sigma'}n_{i,\sigma}n_{j,\sigma'}+V_{2}\sum_{
\left\langle\left\langle i,j\right\rangle\right\rangle,\sigma,\sigma'}n_{i,\sigma}n_{j,\sigma'}
\end{eqnarray}
where $t_{ij}$ encodes the strain pattern and $\left\lbrace U, V_{1}, V_{2}\right\rbrace$ represent the Hubbard on-site, nearest neighbour and next-nearest neighbour repulsion strengths respectively. 
For spinless fermions at half-filling, inducing a topological phase can be achieved by developing a complex bond expectation value
\begin{eqnarray}
\label{eq:hopino}
\eta_{\left\langle\left\langle i,j\right\rangle\right\rangle}=\left\langle c^{\dagger}_{i}c_{j}\right\rangle \in \mathbb{C}
\end{eqnarray}
that breaks time-reversal symmetry and acts like an order parameter of the Chern insulator state \cite{H88,RQH08}. Being an effective 
second-neighbour hopping strength, it is the term proportional to $V_{2}$ in \eqref{eq:hamint} that it is expected to drive the transition.
Indeed, under a mean field decomposition of $V_{2}$ and with $U=V_{1}=0$, it was shown that the strain in Fig.~\ref{fig:herb} and defined by \eqref{eq:hopino} 
develops $\eta_{ij}\neq 0$ for sufficiently strong $V_{2}>V_{c}\sim t$~\cite{RH_13}. 
In the cases with non-uniform strains the order parameter is finite in the vicinity of regions with finite pseudo-magnetic field flux. \\
Spinful fermions allow for other competing orders to emerge, including the spin Hall effect. In this case the on-site 
Hubbard interaction $U$ favours a spin-polarized ferromagnetic state for small $V_{2}$~\cite{GCS12,RS_14} 
that can turn, upon increasing $V_{2}$, into spin-Hall effect~\cite{RH_13,AP_12}.
A nearest neighbour interaction $V_{1}$, on the other hand, favours a charge density wave order which restricts the anomalous Hall phase to appear only when $U$
is sufficiently small and $V_{2}>V_{1}$~\cite{RS_14}, a situation resembling unstrained half-filled graphene~\cite{RQH08,WF_10,GCJ13}. 
An interesting open question is whether these topological phases survive to quantum fluctuations~\cite{GGN13,DH_14,DCH14}.\\

It is worth noting that the effect of including both the pseudo (b) and external (B) uniform magnetic fields on the integer quantum Hall effect in graphene has also been addressed, including the effect of on site $U$ and nearest neighbour $V_{1}$ interactions~\cite{R_11,RHY13}. 
For $B>b$ and at half-filling, the Freeman splitting due to the real magnetic field will lift the 
spin degeneracy while interactions are left with the task of lifting the valley degeneracy 
(either forming a charge density wave or ferrimagnetic order) triggered by the pseudo flux~\cite{H_08}. The physical effect is a standard quantum Hall conductivity 
sequence $\sigma_{xy}=ne^2/h$, in sharp contrast with the non-interacting result $\sigma_{xy}= (4n+2)e^2/h $ with $n \in \mathbb{Z}$ that is a consequence of the valley and spin symmetry of 
non-interacting electrons. In the strain dominated regime ($b>B$) an alternating sequence between $\sigma_{xy}=0$ and $\sigma_{xy}=\mp 2e^2/h$ occurs as a function of electron or hole doping. In this case, and similar to the $B=0$ case discussed above, short range interactions can stabilize quantum spin Hall and anomalous Hall insulators among other ordered phases depending on their relative strengths~\cite{RHY13}.
%
%
%%%%%%%%%%%%%%%%%%%%%%%%%%%%%%%%%%%%%%%%%%%%%%%%%%%%%%%%%%%%%%%%%%%%%%%%%%%%%%%%
%
%
%%%%%%%%%%%%%%%%%%%%%%%%%%%%%%%%%%%%%%%%%%%%%%%%%%%%%%%%%%%%%%%%%%%%%%%%%%%%%%%%
\subsubsection{Interaction induced magnetic and non-magnetic phenomena in strained graphene}
%%%%%%%%%%%%%%%%%%%%%%%%%%%%%%%%%%%%%%%%%%%%%%%%%%%%%%%%%%%%%%%%%%%%%%%%%%%%%%%%
Competing with the topological states described in
Sec.~\ref{Sec:topointeractions}, other important interaction induced orders can
emerge in different strained graphene configurations.\\ 
An important case is that of uniaxial strain~\cite{PCP09}. Moderate uniaxial strain 
is known to deform the low energy graphene spectrum
into an anisotropic Dirac cone resulting in an elliptical Fermi surface~\cite{VKG10}. 
The effect of long-range Coulomb
interactions in anisotropic Dirac models has been studied (prior to the
experimental realization of graphene) in QED$_3$ theories relevant for cuprate
superconductors~\cite{VTF02,LH_02}, and was revisited recently within the
strain context~\cite{SKC12}.
The Dirac anisotropy can be parameterized by the ratio of the Fermi velocities
along the $x$ and $y$ directions 
\begin{eqnarray}
\label{eq:vfanis}
v_{y}/v_{x}=1+\delta.
\end{eqnarray}
Within a renormalization group approach, the anisotropy is
found to be \emph{irrelevant} in the renormalization group sense  if $\delta\ll 1$, i.e. the
isotropic Dirac cone is a stable fixed point~\cite{VTF02,LH_02,SKC12}.  When
$\delta$ is sufficiently strong a quasi one-dimensional excitonic insulator with
\begin{eqnarray}
\label{eq:exci}
\langle\psi^{\dagger}_{A}\psi_{A}-\psi^{\dagger}_{B}\psi_{B}\rangle\neq0
\end{eqnarray}
appears to be the leading instability~\cite{SKC12}.
Furthermore, uniaxial strain can also affect the onset of the ferromagnetic
instability of neutral unstrained graphene under the long range Coulomb
interaction~\cite{PGC05}, driving the instability to smaller couplings within
the range of those expected for suspended graphene~\cite{SKC13}.

Due to its possible experimental relevance, it is important to note that uniaxial strain 
also alters the screening properties of graphene that are encoded in the dielectric function 
$\varepsilon(\vec{q},i\omega)$
at finite frequency ($\omega$) and momentum ($\vec{q}$).
In the random phase approximation it is given by
\begin{eqnarray}
\label{eq:diel}
\varepsilon(\vec{q},i\omega)=1-V_{C}(\vec{q})\Pi^{(1)}(\vec{q},i\omega)
\end{eqnarray}
Here $V_{C}(\mathbf{q})\propto 1/q$ is the bare Coulomb interaction 
and $\Pi^{(1)}(\mathbf{q},i\omega)$ is the one loop polarization function. 
For all $(\mathbf{q},i\omega)$, the latter quantity can be read off from the result in massless quantum electrodynamics (c.f. Eq. (2.14) in ~\cite{KUP12}).
The effect of uniaxial strain is apparent by restoring the dependence on the anisotropic Fermi velocities 
 \begin{eqnarray}
\label{eq:diel2}
\Pi^{(1)}(\vec{q},i\omega)=-\dfrac{N}{16v_xv_y}\dfrac{v^2_xq^2_{x}+v^2_yq^2_y}{\sqrt{v^2_xq_{x}
+v^2_yq_y+\omega^2}}
\end{eqnarray}
where $N=4$ is the number of fermion flavours (one for each spin and valley). The effect of such 
strain on $\varepsilon(\vec{q},i\omega)$ 
permeates into the behaviour of the van der Waals forces
between strained graphene sheets, that can be shown to increase
with increasing strain anisotropy~\cite{SHS14}.  Through a renormalization
group calculation it was shown that, for large separations between undoped strained
graphene sheets, the inclusion of electron--electron interactions reduces the van der
Waals interaction between graphene sheets~\cite{SHS14}.\\

For interacting graphene in a uniform pseudo-magnetic field,
the different ordered phases resulting from Coulomb long-range interaction, 
on-site $U$, nearest neighbour $V_{1}$ and next nearest neighbour interaction $V_{2}$ 
have been extensively studied~\cite{GCS12,RS_14,RH_13,AP_12,RAH14} and discussed in 
Sec.~\ref{Sec:topointeractions}.  
From first principles however,~\cite{WSF11} it is estimated that $U$ is the largest energy scale, which, if strong enough would 
in principle favour a spin-ferromagnet~\cite{GCS12,AP_12,RS_14}. However,
with mean-field theory and Monte-Carlo simulations it was found that, for finite samples,
the system develops a global anti-ferromagnet with zero magnetization~\cite{RAH14}. Although locally 
this state is ferromagnetically ordered, it displays a magnetization that changes sign between the bulk and 
the edges precisely to deliver a zero total magnetization. This is ultimately tied to the fact that although in an infinite
system the support of the zero-modes is restricted to one sub-lattice (say A), in a finite system there is support
for these also in sub-lattice B but only at the edges. The magnetization, being tied to the zero modes, changes sign
as the support of the modes changes from A (bulk) to B (edge) keeping the total magnetization zero.\\
To conclude this part we note that long-range Coulomb interactions can reduce the stability of edge states resulting from strain~\cite{GGR13}. 
On the other hand an on-site repulsion together with uniaxial strain can lead to edge-ferromagnetism~\cite{CYM14}.
\subsubsection{Superconductivity in interacting strained graphene}
\label{Sec:intSC}
We finish this section devoted to interactions with a note regarding superconducting states in strained
graphene induced by attractive interactions. As mentioned above, a spin-triplet
superconducting state was shown to exist via exact diagonalization in strained graphene~\cite{GCS12}
with a finite superfluid density $$n_{s}=\frac{1}{2}\frac{\partial^2E_{g}}{\partial \theta^2},$$
calculated from the variation of the energy of the ground-state $E_{g}$ upon twisting the boundary conditions an angle $\theta$~\cite{MPB05}.
Moreover, strain can lead to interesting superconducting states in graphene~\cite{UB_13,RJ_14}.
In particular, an exotic $f+is$ superconducting state that spontaneously breaks time-reversal symmetry is expected to 
occur when attractive on site $U$ and second nearest neighbours interactions $V_{2}$ are comparable ($U \sim V_{2}$)~\cite{RJ_14}.
This state ultimately originates from the competition between the $U$ favoured singlet $s$-wave 
pairing and the $V_{2}$ favoured triplet $f$-wave pairing.
%
%
%
%\bibliography{interactions}
%
%\end{document}

%---------------------------------
% Sat Mar 14 18:29:35 CET 2015
%
% New \cite{LC_13} added.  Refers to
%@article{LC_13,
%  title = {Effect of dynamical screening on single-particle spectral features of uniaxially strained graphene: Tuning the plasmaron ring},
%  author = {LeBlanc, J. P. F. and Carbotte, J. P.},
%  journal = {Phys. Rev. B},
%  volume = {87},
%  issue = {20},
%  pages = {205407},
%  numpages = {6},
%  year = {2013},
%  month = {May},
%  publisher = {American Physical Society},
%  doi = {10.1103/PhysRevB.87.205407}
%}
%---------------------------------
% Thu Feb 19 20:58:29 CET 2015
%
%@article{WSS06,
%  author={B. Wunsch and T. Stauber and F. Sols and F. Guinea},
%  title={Dynamical polarization of graphene at finite doping},
%  journal={New Journal of Physics},
%  volume={8},
%  pages={318},
%  doi={10.1088/1367-2630/8/12/318},
%  year={2006}
%}
%This is called WSSG06 in anharmonic. I kept that and changed the \cite in optics.tex
%
%
%
%%%%%%%%%%%%%%%%%%%%%%%%%%%%%%%%%%%%%%%%%%%%%%%%%%%%%%%%%%%%%%%%%%%%%%%%%%%%%%%%%%%%%%%%%%%%%%%%%%%%
\subsection{Optical properties of strained graphene}
\label{sec:optics}
%%%%%%%%%%%%%%%%%%%%%%%%%%%%%%%%%%%%%%%%%%%%%%%%%%%%%%%%%%%%%%%%%%%%%%%%%%%%%%%%%%%%%%%%%%%%%%%%%%%%
%
%
%
%
%
The analysis of light scattered inelastically by graphene by means of Raman
spectroscopy has long since become one of the standard characterization
techniques in graphene research \cite{MLN09}.  We give a brief account of the
effect of strain on graphene's Raman spectrum in
Sec.~\ref{subsec:optics_raman}.  However -- and not surprisingly for an
atomically thin structure -- the interaction between graphene and light is weak
within a broad band of frequencies, and elastic scattering is the by far
dominant process  underlying most of the potential applications of graphene in
optical devices \cite{BSH10,SYY13,LYU11,TFM14}.  Thus the optical properties of
graphene are mainly determined by its electronic structure, which is sensitive
to strain.  In particular, it has been proposed to use uniaxial strain to break
the optical anisotropy of two-dimensional graphene, thus converting it into a
dichroic material (see Sec.~\ref{subsec:optics_uniaxial}).  Finally, graphene's
magneto-optical properties, of which, in particular,  a large Faraday effect
has attracted attention recently \cite{CLW11,SYY13}, are determined by the
nonlinear Landau level spectrum of massless Dirac Fermions.  As a zero-field
quantum Hall effect and Landau level formation in graphene can be realized with
strain-induced pseudomagnetic fields \cite{GKG10,LBM10},  the interplay between
real and pseudomagnetic fields in graphene based optical devices  seems to
offer interesting possibilities for future applications, on which we comment in
Sec.~\ref{subsec:optics_magneto}.
%
%
%%%%%%%%%%%%%%%%%%%%%%%%%%%%%%%%%%%%%%%%%%%%%%%%%%%%%%%%%%%%%%%%%%%%%%%%%%%%%%%%%%%%%%%%%%%%%%%%%%%%
\subsubsection{Strain monitoring with Raman spectroscopy}
\label{subsec:optics_raman}
%%%%%%%%%%%%%%%%%%%%%%%%%%%%%%%%%%%%%%%%%%%%%%%%%%%%%%%%%%%%%%%%%%%%%%%%%%%%%%%%%%%%%%%%%%%%%%%%%%%%
Raman spectroscopy typically uses light in the infrared (IR) or near
ultraviolet (UV) spectral range.  Because the laser energy is large compared to
the phonon energy, the scattering mechanism involves electronic excitations in
intermediate states, rather than direct photon-phonon coupling. Thus, apart
from providing information about the phonon spectrum, Raman spectroscopy can
also shed light on the behaviour of electrons \cite{HYH10} and complement
transport measurements.  As tensile strain usually results in phonon softening
(and the opposite for compressive strain), Raman spectroscopy provides a tool
for strain monitoring.

The Raman spectrum of defect-free single layer graphene is characterized by two
peaks: The G peak, which corresponds to the excitation of a phonon in the
doubly degenerate in-plane optical mode $E_{2}$ (see Table~\ref{Tab:phononsIN}), 
and the 2D peak, due to a two-phonon process which involves scattering of electron-hole pairs between
neighbouring Dirac cones \cite{FB_13}.

Under uniaxial strain, graphene's $E_{2}$ mode splits into two components
$E_{2}^\pm$, which are perpendicular and parallel to the strain axis,
respectively.  As the $E_{2}^+$ mode experiences less softening due to tensile
strain, the Raman G  peak is seen to split into a G$^+$ and a G$^-$ band
\cite{HYC09, MLN09}.  Both redshift with increasing strain, and their splitting
increases.  Polarized measurements of the G$^\pm$ intensities allow the
determination of the crystallographic orientation with respect to a known strain
axis.  If the uniaxial strain is applied along zigzag or armchair directions,
the 2D peak likewise splits into two distinct submodes \cite{HYH10}.  Magnitude
and polarization dependence of the mode splitting was found to be consistent
with the strain-induced motion of the Dirac cones predicted by tight binding
calculations \cite{PCP09}.

In surface enhanced Raman spectroscopy, photonic structures supporting plasmon
resonances (such as arrays of metallic nanoparticles),  are deposited on
graphene.  The strong electromagnetic nearfield associated with the plasmon can
increase the intensities of graphene's Raman spectrum by several orders of
magnitude \cite{SLL10}.  In turn, the Raman spectrum of graphene serves as a
detection channel for plasmonic field enhancement.  With graphene placed on top
of a photonic cavity structure, the slight sagging in the suspended areas
locally strains the graphene membrane.  The resulting phonon mode softening
allows  to distinguish between enhanced Raman signals originating from strained
and unstrained areas.  Thus the strained graphene membrane can be used as a
probe that locally resolves the plasmonic field enhancement of photonic
structures \cite{HFO13}.
%
%
%%%%%%%%%%%%%%%%%%%%%%%%%%%%%%%%%%%%%%%%%%%%%%%%%%%%%%%%%%%%%%%%%%%%%%%%%%%%%%%%%%%%%%%%%%%%%%%%%%%%
\subsubsection{Optical properties of uniaxially strained graphene}
\label{subsec:optics_uniaxial}
%%%%%%%%%%%%%%%%%%%%%%%%%%%%%%%%%%%%%%%%%%%%%%%%%%%%%%%%%%%%%%%%%%%%%%%%%%%%%%%%%%%%%%%%%%%%%%%%%%%%
Elastic (Rayleigh) scattering of light by graphene can be described by the
Fresnel equations for a thin film with an appropriate sheet conductivity
$\sigma(\omega)$ \cite{NH_12,NGG11}.  For normal incident light of frequency
$\omega$, this yields 
\begin{align}
A(\omega)
	&=
	1 - |t(\omega)|^2 - |r(\omega)|^2
	=
	\frac{4 \sigma(\omega)/(\varepsilon_0 c)}{[2 + \sigma(\omega)/(\varepsilon_0 c)]^2}
\label{eqn:optics_absorbance}
\end{align}
for the absorbance $A(\omega)$ of free standing graphene, where $r$ and $t$
denote the Fresnel coefficients for reflection and transmission, respectively, $c$ is the speed 
of light and $\varepsilon_0$ the vacuum permittivity.

Modelling the graphene charge carriers as non-interacting, massless fermions in
the Dirac-cone approximation yields a universal conductivity of
$\sigma=e^2/(4\hbar)$ for all $\omega>2 E_F/\hbar$ \cite{WSSG06,HD_07}.  When
inserted into Eq.~\eqref{eqn:optics_absorbance}, this results in $A = \pi\alpha +
\mathcal{O}(\alpha^2)\simeq2.3$ (where $\alpha = e^2/(4 \pi \varepsilon_0 \hbar c)$ is the 
fine-structure constant) which corresponds to the experimental
value observed at wavelengths between 450\,nm and 4.2\,$\mu$m
\cite{MSW08,NBG08}.  At energies below $2 E_F$, graphene supports collective
plasmon oscillations, which can be described by treating electron-electron
interactions within the random phase approximation (RPA) \cite{JBS09}.  In the
long wavelength limit $\hbar\omega \ll 2 E_F$, the plasmon dispersion reads 
\begin{align}
\omega(\vec{k})
	&=
	\bigl(
	2\alpha c  k E_F / \hbar
	\bigr)^{1/2}
\label{eqn:optics_plasmon_disp}
\end{align}
\cite{LA_14,KCG11,GPN12}.  For typical graphene-on-substrate samples,  $E_F$ is
of the order of $0.5$\,eV or lower.  This fixes the application range of
graphene plasmonics to the THz and mid-IR band (wavelengths $\geq1\,\mu$m), while
graphene based optical broadband applications can operate in in the near-IR and
most of the visible spectrum (wavelengths between $450$ and $1\,\mu$m) \cite{BSH10}.

The effect of uniaxial strain on graphene's optical conductivity can be
described by introducing strain dependent hopping parameters into the standard
tight binding Hamiltonian \cite{PAP10,PCP09}, which deforms the Fermi surface
into an ellipse (see also Sec.~\ref{Sec:electron-phonon_coupling}).  This defines a fast and a slow optical axis, with the latter
(and, correspondingly, a lowered conductivity) oriented closely along the
direction of strain.  The frequency independence of absorption in the near-IR
and visible range stays intact under uniaxial strain, but the absorbance  of
linearly polarized light now depends on the polarization direction $\phi_I$
with respect to the slow optical axis \cite{PRP10},
\begin{align}
A
	&\approx
	\pi\alpha[1 - 2 a |\delta k_D| \operatorname{cos} 2 \phi_I]
,
\label{eqn:optics_trans_uni}
\end{align}
and the degree of polarization dependence is determined by the strain-induced
shift $\delta k_D=\epsilon\lambda_s(1+\nu)/2$ of the Dirac points from their
position at $\vec{K},\vec{K}'$ in the Brillouin zone.  Here $\epsilon$
denotes the magnitude of applied strain, $\nu\approx0.16$ is graphene's Poisson
ratio, and $a \lambda_s  \approx 3$-4 \cite{PAP10}.  The periodic modulation
(\ref{eqn:optics_trans_uni}), measured in the transmission of visible light
through bended graphene \cite{NYJ14}, allows for an optical detection of
magnitude  and orientation of uniaxial strain, with possible applications in
passive, atomically thin, and flexible strain sensors.  A reduced optical
absorption along the direction of strain was likewise observed in the Drude
response of uniaxially strained graphene at far-IR wavelengths of
$\lambda=$\,30-300\,$\mu$m \cite{KLB12}.  Importantly, the conductivity
modulations resulting from tensile strain up to nearly $20\%$ have been shown
to be  reversible \cite{KZJ09,NYJ14}.

The square root character of the plasmon branch [Eq.~(\ref{eqn:optics_plasmon_disp})]
as well as its $n^{1/4}$ dependence on the charge density are not changed by
uniaxial strain.  However, corresponding  to the strain dependence obtained for
the optical conductivity and the resulting absorption in
Eq.~(\ref{eqn:optics_trans_uni}), the plasmon dispersion gets steeper for
$\vec{k}$ perpendicular to the direction of strain and is flattened for
wavevectors along the strain direction, with the amplitude of the modification
proportional to $\epsilon$ \cite{PAP10b}.  In a similar fashion, the effect of
scalar and vector potentials with one-dimensional periodicity [that is,
$V_s=V_s(x)$, $\vec{\mathcal{A}}^{el}=\vec{\mathcal{A}}^{el}(x)$], induced by periodic strain waves, 
was
found to render the low-energy part of the plasmon dispersion anisotropic
\cite{DK_12}.
Further, uniaxial strain (via its effect on the electron-electron interaction)
is expected to modify the electron-plasmon scattering structures which are
visible near the Dirac points in ARPES studies~\cite{LC_13}.

For realistic values of $\epsilon \leq 20$\,\%,
and in contrast to non-uniform strain configurations (as discussed in
Secs.~\ref{Sec:electron-phonon_coupling} and \ref{Sec:superlattices}), uniaxial
strain is not expected 
to open a gap  (resulting in a vanishing optical conductivity) in extended
two-dimensional graphene \cite{PCP09}.  However, tight binding calculations for
graphene nanoribbons predict that realistically small uniaxial strain results
in a periodic modulation of the bandgap in armchair ribbons,  while in zigzag
ribbons it creates a band gap proportional to the magnitude of applied tensile
strain \cite{LG_10b}.  As the band-structure is modified, the optical
transition energies between maxima of occupied bands and minima of unoccupied
bands change, resulting in a shift of the maxima in the optical loss-function
$Z(\omega)=[\operatorname{Im} \varepsilon(\omega)]^{-1}$ \cite{JG_14}.
%
%
%%%%%%%%%%%%%%%%%%%%%%%%%%%%%%%%%%%%%%%%%%%%%%%%%%%%%%%%%%%%%%%%%%%%%%%%%%%%%%%%%%%%%%%%%%%%%%%%%%%%
\subsubsection{Magneto-optics with strained graphene}
\label{subsec:optics_magneto}
%%%%%%%%%%%%%%%%%%%%%%%%%%%%%%%%%%%%%%%%%%%%%%%%%%%%%%%%%%%%%%%%%%%%%%%%%%%%%%%%%%%%%%%%%%%%%%%%%%%%
As in the magnetically unbiased case, the transmission of light through a
graphene layer in the presence of a uniform external magnetic field can be
described within the Fresnel formalism \cite{SC_11}, with the Fresnel
coefficients for reflected and transmitted amplitudes now being functions of
both graphene's diagonal and Hall conductivity.  Analytical expressions for
$\sigma_{xy}$ and $\sigma_{xx}$ as functions of frequency and magnetic field
have been derived in Refs.~\cite{GS_06,GSC06}.  Due to the action of the
magnetic field on the conduction electrons, an incident  beam of linear
polarization along the $x$-direction will, after passing the graphene layer,
consist of both $x$ and $y$ components.  The total transmission $T$ through
biased graphene is a sum of the intensities in both polarizations, and for an
incident beam oriented normal to the surface of suspended graphene, we arrive
at \cite{FV_12,SYY13}
\begin{align}
T(\omega)
	&=
	|t_{xx}|^2+|t_{xy}|^2
	=
	1-
	\frac{\operatorname{Re}\sigma_{xx}(\omega)}{\varepsilon_0 c}
	+
	\mathcal{O}(\alpha^2).
\label{eqn:optics_trans_B}
\end{align}

Due to the non-equidistant spacing of graphene's Landau levels, the transmission
spectrum Eq.~(\ref{eqn:optics_trans_B}), as observed by IR spectroscopy
\cite{SMP06,JHT07}, displays resonant dips at frequencies where the incident
light can excite electronic transitions between Landau levels $E_n=\pm
v_F\sqrt{2 e \hbar B n}$ (where $n=0,1,2,\dots$).  Calculations using
density-functional theory based tight-binding theory (DFTB) predict that the
same Landau level spectrum can be obtained at zero external field by twisting
graphene nanoribbons with several hundreds of nanometres width \cite{ZSC14}.
The strain produced by twisting the ribbon around its axis results in a
pseudomagnetic field $\vec{B}_s^K = - \vec{B}_s^{K'}\simeq \gamma^2 x$
(where $\gamma$ denotes the twist rate, and $|x| \leq W/2$ the coordinate
perpendicular to the ribbon axis) that changes sign when going from one ribbon
edge to the other.  The resulting electronic DOS for zig-zag as well as
armchair ribbons is predicted to mimic Landau quantization of Dirac fermions
in magnetic fields up to 160\,T.

For graphene, the Faraday effect (and the related Kerr effect), in which
linearly polarized light transmitted through (or reflected by) a material in
the presence of a magnetic field has its polarization plane rotated, is
surprisingly strong:  monolayer graphene in modest fields has been found to turn
the polarization  by several degrees \cite{CLW11,SYY13,FV_12,SC_11}.  With the
incident beam as well as the external magnetic field oriented perpendicular to
the  surface of suspended graphene, we arrive at a Faraday rotation angle
\cite{CLW11,FV_12}
\begin{align}
\theta_F(\omega)
	&=
	- \frac 1 2 \frac{\operatorname{Re}\sigma_{xy}(\omega)}{\varepsilon_0 c}
	+
	\mathcal{O}(\alpha^2)
.
\label{eqn:optics_theta_F}
\end{align}
$\theta_F$ is large for frequencies where $\sigma_{xy}$ shows resonances.  In
the classical cyclotron resonance regime, this occurs for $\omega\simeq v_F^2
|e B/ E_F|$, while in the quantum Hall regime,  laser frequencies matching the
transition energies between two Landau levels lead to large Faraday angles
\cite{CLW11}.

The Faraday and Kerr effects are of technological importance in the
development of optical diodes and other non-reciprocal optical elements
\cite{BHJ11}.  For these magneto-optical applications, graphene offers a great
potential because Landau-level formation and a nondiagonal optical conductivity
in graphene can be obtained either by applying an external magnetic field or
through strain-induced pseudomagnetic fields \cite{LG_10,RHY13}.  With an
additional  strain-induced pseudomagnetic field $B_s$, electrons in the
$\vec{K}$ and $\vec{K}'$ valleys are exposed  to an effective field
$B_{\rm eff}=B \pm B_s$, respectively \cite{LG_10,RHY13}.  An effect of
pseudomagnetic fields on the Faraday rotation is thus possible at frequencies
where the contributions to $\sigma_{xy}$ resulting from the two inequivalent
valleys do not cancel out in Eq.~(\ref{eqn:optics_theta_F}).  For $\hbar \omega \ll
E_F$, the overall dependence of  $\sigma_{xy}(\omega)$ on the effective field
scales as $B_{\rm eff}^{-1}$ \cite{GS_06,GSC06}, which opens possibilities for
a strain-induced Faraday rotation in the THz band.

\section{Superlattices}
\label{Sec:superlattices}
% Needs `superlattices.bib'

The properties of misaligned graphene multilayers and graphene deposited on other 
two-dimensional crystals have been the subject of intense research recently.
The simplest of these, twisted bilayer graphene, exhibits different stackings (local atomic 
alignments between layers) when the two layers are rotated relative to each other by a finite angle 
$\theta$ \cite{ATH13,RK_93,PAM08}. Such a rotation gives rise to a periodic spatial modulation of 
the stacking, known as a Moir\'e pattern or stacking superlattice, with a period that goes as $\sim 
a/\theta$ at small $\theta$. All sorts of interesting electronic reconstructions arise from the 
stacking modulation, including low-energy van-Hove singularities, secondary Dirac points away from 
charge neutrality and Hofstadter butterflies \cite{LLR11,BMG12,ALD12,KCT12,PGY13,DWM13}. These 
reconstructions can be understood as the consequence of non-Abelian gauge fields that emerge in the 
low energy sector of the system \cite{SGG12}, similarly to the Abelian pseudogauge fields associated 
to strains in a graphene monolayer.

Together with the modulated stacking, a rather strong modulation of the local inter-layer 
adhesion energy density develops in the twisted bilayer. The inhomogeneous adhesion creates a 
surprisingly strong strain modulation in the system, whereby the areas with the preferred stackings 
(AB, or Bernal stacking, with higher adhesion) become expanded at the expense of other regions. This 
transforms the Moir\'e patterns into a characteristic hexagonal mesh of stacking solitons (linear 
stacking faults, around $\sim 10$ nm wide), usually with the same period as the original (unrelaxed) 
Moir\'e \cite{ATH13}. Once more, a range of intriguing electronic effects emerge from the soliton 
strain fields, e.g. the development of a helical network of states carrying valley currents along 
the solitons when the bilayer is placed under a perpendicular electric field \cite{SP_13}, or the 
suppression of conductance close to neutrality due to scattering on the solitons 
\cite{SGG14,LSP13}.

More complicated Moir\'e structures and inhomogeneous adhesion-induced strains are possible, 
and even ubiquitous in more complex multilayers \cite{BMG12,LGP14}. Graphene trilayers, for example, 
can exhibit ABA and ABC stacking domains, visible with Raman spectroscopy \cite{LYZ15}, and 
separated by compressed stacking solitons. Unlike in the bilayer, these two types of stackings, 
despite being very close in energy, are electronically very different. In particular, ABA domains 
remain metallic under large out-of-plane electric fields, while ABC domains develop a sizeable gap. 
This enables all-electric manipulation of stacking domains in the trilayer, and of their elastic 
soliton boundaries, using a biased STM tip \cite{YWB14}. These examples clearly show that the 
spatial modulation of stacking alignment in graphene multilayers creates a fascinating playground 
where elasticity and electronic structure become strongly coupled, and where quite a few problems 
remain open \cite{HSS11,YLD12}.

But stacking Moir\'e superlattices are not restricted to graphene multilayers. A simple 
graphene monolayer grown or deposited over different insulating crystals such as SiC and hBN also 
exhibits Moir\'e patterns, stacking solitons and a remarkable electronic structure.
High quality samples with outstanding transport properties are provided by these setups (cf. 
Sec.~\ref{Sec:general}), and other heterostructures, with \cite{GG_13,BGJ12} or without 
\cite{ZKC14} graphene as one of their building blocks. Due to their current interest, we will now 
review in more detail the physics of graphene/hBN structures, as a representative of this class of 
hybrid Moir\'e superlattices.

Superlattices arise in this context due to the superposition of 2D crystals with different periodicities, stackings and/or alignments.
To study their electronic band structure, at least two aspects must be considered.
First, the deformation of the layers' pristine periodicity that may take place spontaneously due to Van der Waals adhesion potentials, which drives the multilayer into a non-trivial equilibrium configuration.
Second, the electronic coupling between layers in this new scenario, usually regarding the derivation of an effective description.

The Frenkel-Kontorova model addresses the first issue, namely to study the ground state of a system 
where a substrate potential and the elastic energy of a 1D or 2D lattice 
compete.\cite{A__83,AL_83,BK_04,CL_00}
It exhibits three possible ground state phases, depending on the relative periodicities of the 
lattices and the ratio of adhesion versus elastic energy:
(i) a generalized floating phase, where there are only local deformations
due to a weak interaction with the substrate, the
Moir\'e pattern preserving its undeformed periodicity;
(ii) a commensurate phase, where there is a modified overall periodicity
of the Moir\'e pattern, resulting from a finite global deformation;
(iii) an incommensurate phase, characterized by the presence of aperiodic solitons, and separating two commensurate phases as a function of lattice mismatch.

\begin{figure}
\begin{center}
\includegraphics[width=0.44\textwidth]{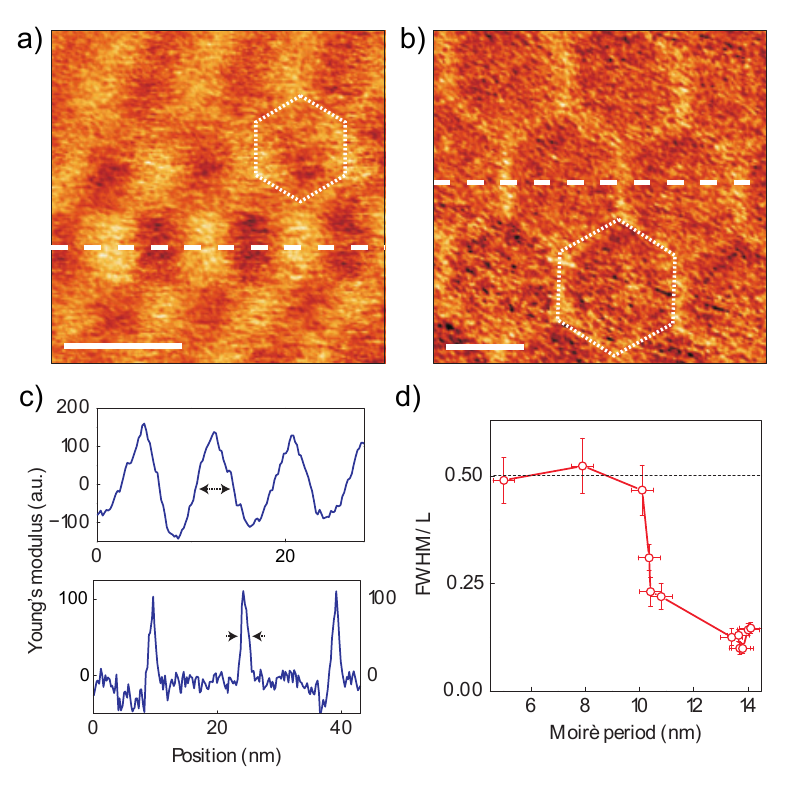}
\end{center}
\caption{(a) and (b): the Young modulus distribution, measured by AFM, for structures with 8 and 
14nm Moir\'e patterns, respectively. Scale bars are 
$10\,\text{nm}$ long. (c): Cross-sections of the Young modulus distribution taken along the dashed lines in (a) (top) and (b) (bottom). (d): Ratio 
between the full width half maximum (FWHM) of the peak in the Young modulus distribution (as 
marked by arrows in (c)) and the period of the Moir\'e 
structure $L$, as a function of the period of the Moir\'e structure for several samples. Reproduced 
from Ref. \cite{WBE14} with permission of the 
authors.}
\label{Fig:young}
\end{figure}

These three phases have been experimentally encountered in a variety of heterostructures containing graphene.
For a twisted bilayer, Ref.~\cite{ATH13} observed, in different regions of the sample, both 
{commensurate structures with hexagonal symmetry and the 
structural solitons}.
For graphene over hBN, Ref.~\cite{WBE14} reported a phase transition between a {commensurate} 
structure and a floating phase as the Moir\'e  
period exceeds $\sim 10\,\text{nm}$.\footnote{The notation in Ref.~\protect \cite{WBE14} differs from the customary denominations in the 
Frenkel-Kontorova framework: they refer to the floating phase as incommensurate, in the sense that the top layer does not conform to the substrate. 
They disregard what in this review we call incommensurate phases, since no aperiodic solitons are observed.
}
Fig. \ref{Fig:young} is a reproduction of this experimental evidence, with the Young modulus FWHM 
over the Moir\'e wavelength as the order parameter.

Determining the strain field that carries the system to its ground state is the first issue to be addressed.
The application of the Frenkel-Kontorova model to two dimensions is not trivial, since it demands the use of a complex mathematical machinery,\cite{AL_83,A__83} as well as an accurate knowledge of the shape of the adhesion potential.
To predict theoretically the deformation in this framework is, therefore, a difficult task.

\begin{figure}
\begin{center}
\includegraphics[width=0.35\textwidth]{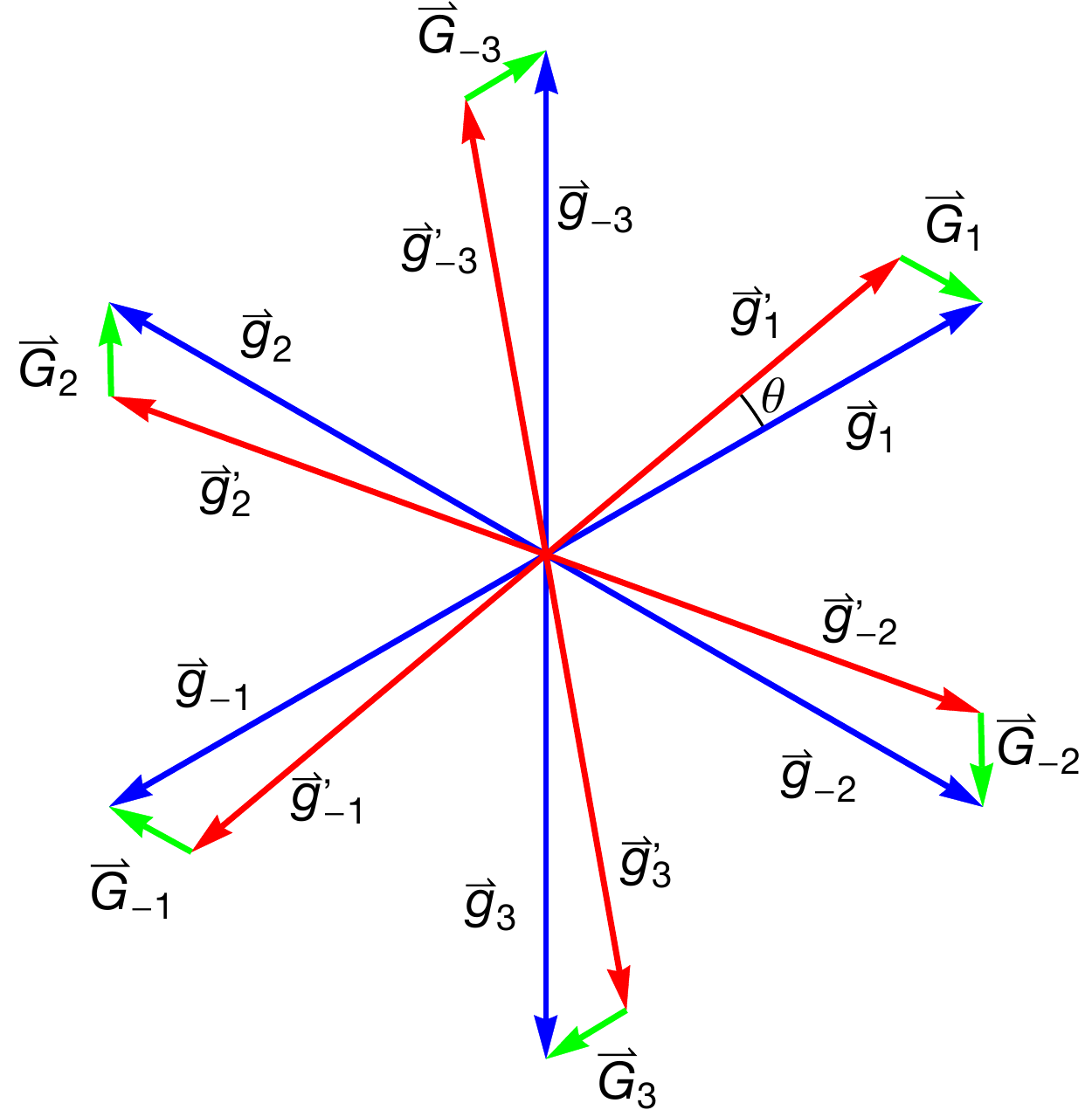}
\end{center}
\caption{Shortest reciprocal lattice generating vectors of two misaligned honeycomb lattices (red and blue) and of the resulting superlattice (green).}
\label{Fig:momenta}
\end{figure}

A first strategy to tackle this problem consists on focusing solely on floating-phase deformations, 
namely strain superlattices with the same periodicity as the Moir\'e pattern. We consider here the 
simple case of a bilayer. Then, such deformations conform minimal commensurate configurations with 
reciprocal superlattice generating vectors\cite{BM_10,MWF13,WPM13,YXC12}
\begin{align}
\vec{G}_j=\vec{g}_j-\vec{g}^{\prime}_j. \label{Eq:G_j}
\end{align}
Vectors $\vec{g}_j$ and $\vec{g}^{\prime}_j$ are the reciprocal lattice generating vectors of each of the two crystals, chosen such that the $|\vec G_j|$ are minimal.
%shortest reciprocal lattice generating vectors of individual layers such that the angle between them is minimal.
A sketch of this construction for graphene on hBN (honeycomb lattices with a finite mismatch and a relative rotation) is shown in Fig. \ref{Fig:momenta}.
$\vec{G}_j$ are often referred to as ``first star'' vectors when dealing with hexagonal symmetry. 

The reciprocal vectors $\vec{G}_j$ and the corresponding superlattice vectors $\vec{A}_j$ define a superlattice that may or may not be commensurate with each of the two layers. In the incommensurate case, no true periodicity of the bilayer exists, and the basic tenet of Bloch's theorem cannot be invoked when considering electronic structure.
It has been shown, however, that for a small value of the graphene lattice constant $a$ over the interlayer distance $d$, incommensurability and exact commensurability yield the same observable electronic properties.\cite{DPN12}
It is then legitimate to approximate any system with $a/d\to 0$ as a commensurate structure with a 
reciprocal superlattice generated by Eq. (\ref{Eq:G_j}). This is known as the continuum 
approximation, and allows one to consider the concept of a low energy band structure assuming an 
effectively periodic superlattice, even though Bloch's theorem does not apply at the smallest 
length scales. 
Considering in-plane deformations $\vec{u}(\vec r)$ with the same periodicity (floating phase 
approximation, with purely local deformations), one arrives at the simplest nontrivial description 
of the coupled electronic and elastic properties of these Moir\'e crystals. In subsequent sections 
we focus on the continuum approximation approach, which turns out to be enough to describe a number 
of systems, such as graphene on hBN.  

%Focusing on this approach, that dodges the aperiodicity characterizing every incommensurate system, the electronic structure has been studied so far for different setups described in subsequent sections.

\subsection{Undeformed Moir\'e superlattices}
\label{Sec:electronic_properties_undeformed}

\begin{figure}
\begin{center}
\includegraphics[width=0.44\textwidth]{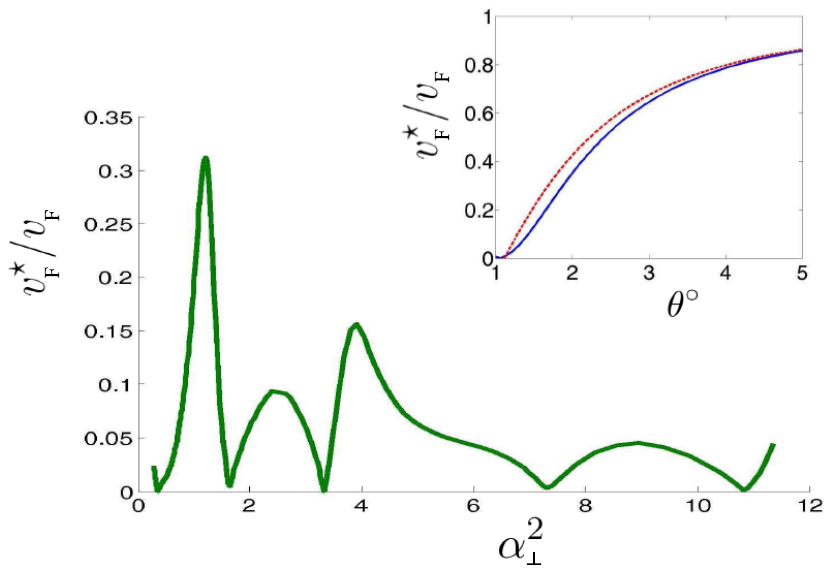}
\end{center}
\caption{Renormalization of the Fermi velocity in bilayer graphene for small twist angles $\theta$ 
(strong coupling). $\alpha_\perp$ is given in the main 
text. The inset depicts $v_F^*/v_F$ in the weak coupling regime. Red is respective to Eq. (\ref{Eq:vF^*}), whereas the blue curve corresponds to more 
accurate numerical calculations. Reproduced from Ref. \cite{BM_10} with permission of the authors.}
\label{Fig:vFrenormalized}
\end{figure}

The first studies of graphene superlattices were restricted to floating phases with zero deformations, very particularly for the case of twisted graphene bilayers.
An effective continuum model for unstrained twisted bilayer graphene can be found in Refs. \cite{BM_10,M__11,LPN07}.
The Dirac cone structure is preserved, although with a renormalized Fermi velocity $v_F^*$. For large enough twist angles $\theta$, and at low enough energies, a Dirac model remains valid for carriers in each of the two monolayers, which behave as if decoupled,
\begin{align} 
\mathcal{H}_{\text{eff}}=v_F^*\sum_{\vec{k}}\psi^\dagger_{\vec{k}}\,\vec{\sigma}\cdot\vec{k}\,\psi_{
\vec { k } }
, \label{Eq:Heffbeginning}
\end{align}
Here $\psi^\dagger_{\vec{k}}=(a^\dagger_{\vec{k}},b^\dagger_{\vec{k}})$, while $a^\dagger_{\vec{k}}$ ($b^\dagger_{\vec{k}}$) are the creation operator of the $\vec{k}$ Bloch state in the A (B) sublattice.
The correction to the Fermi velocity $v_F^*/v_F$ as a function of the twist angle is plotted in Fig. \ref{Fig:vFrenormalized}.
In the weak coupling regime (large relative rotation angles $\theta\gtrsim 1^{\circ}$),
\begin{align}
v^*_F\simeq v_F\frac{1-3\alpha_{\perp}^2}{1+6\alpha_{\perp}^2}; \label{Eq:vF^*}
\end{align}
where $\alpha_{\perp}=\frac{t_{\perp}}{3v_Fk_\theta}$, $t_\perp$ is the nearest neighbour 
interlayer 
hopping ($t_\perp\simeq 330\,\text{meV}$), $k_\theta=2K\sin(\theta/2)$ and $K=4\pi/3a$ is the 
modulus of the $K$-point wavevector.\cite{BM_10}
At energies around $v_F^*k_\theta/2$, the decoupled monolayer description breaks down. Carriers from different layers hybridize, giving rise to a $\theta$-dependent van-Hove singularity in the density of states. In contrast with unrotated bilayer graphene, a gap does not open in its twisted counterpart when applying a potential difference between the layers.\cite{LPN07}

As the relative rotation angle $\theta$ falls below $1^\circ$, the system enters a more complicated 
strong coupling regime. When neglecting deformations, Dirac cones survive in this regime. However, 
the corresponding renormalized Fermi velocity becomes much smaller and is even completely suppressed 
for a set of `magic' rotation angles, implying a strong enhancement of the density of states at the 
neutrality point.
Ref. \cite{M__11} also points out the possibility of annihilation and regeneration of Dirac nodes based on the ratios of the Slonczewski-Weiss-McClure 
parameters, as well as encountering a quadratic dispersion relation near the Fermi energy.

\begin{figure}
\begin{center}
\includegraphics[width=0.44\textwidth]{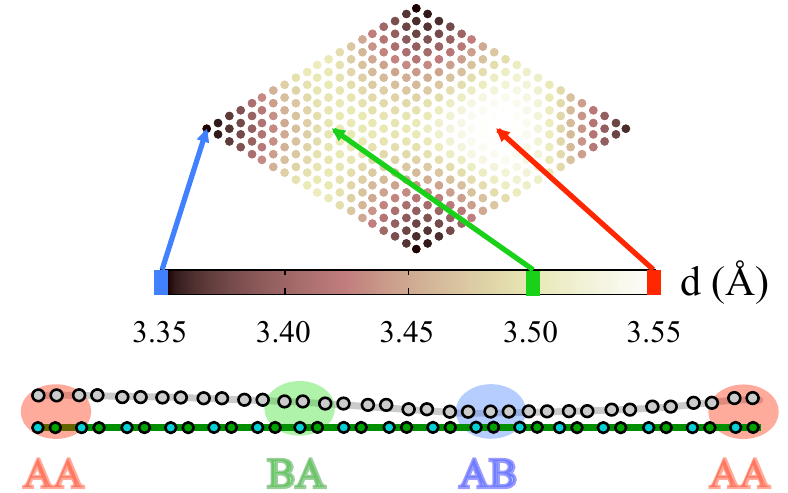}
\end{center}
\caption{Modulation of the graphene distance to hBN as a function of the stacking according to \emph{ab initio} calculations. With AB (BA) stacking, 
we refer to one sublattice of graphene overlapping exactly with boron (nitrogen) atoms. Reproduced 
from Ref. \cite{BAB14} with permission of the 
authors.}
\label{Fig:outofplane}
\end{figure}

As for graphene over hBN, a floating phase analysis was conducted in Refs. \cite{BAB14,GKB07}.
Not including in plane deformations, \emph{ab initio} calculations suggest a modulation in the distance between layer and substrate (Fig. \ref{Fig:outofplane}), which in turn results in a gap opening.
Interaction between electrons has been shown to enhance this gap to the extent of experimentally measured values ($\sim 30\,\text{meV}$).\cite{BAB14,JDM14,SSL13}
If, on the other hand, the interlayer distance is assumed constant throughout the system and interactions are neglected, the gap averages to zero without deformations.
%{\color{red} Comments about the Fermi velocity: I would not include them. We do not have an effective model giving a value for the renormalized Fermi velocity.}

\subsection{Spontaneous deformations in the continuum approximation}
\label{Sec:spontaneous_deformations}

Strains, such as the spontaneous deformations expected from adhesion interactions with a crystalline 
substrate, induce significant changes in the electronic structure of graphene.\cite{LBM10,GKG10} Let 
us assume a certain purely in-plane deformation, described at $\vec r$ by an in-plane displacement 
$\vec u(\vec r)=\left( u_x(\vec r),u_y(\vec r) \right)$ with respect to its unstrained 
configuration.
The change in hopping integrals arising from said deformations can be recast into the scalar ($V_s$) 
and pseudogauge ($\vec{\mathcal{A}}^{el}$) potentials\cite{SA_02} discussed in Sec. 
\ref{Sec:electron-phonon_coupling}, with the symmetry described by Eq. (\ref{Eq:strain_sym}) and 
yielding the Hamiltonian of Eq. (\ref{HTBcomplete}).

In the continuum approximation, the influence of the substrate on the electronic structure acts via two mechanisms.
First, as  pseudopotentials $(V_s,\vec{\mathcal{A}}^{el})$ associated to strains, induced in turn 
by the adhesion potential.
Second, through the electronic coupling with the substrate. To obtain an effective model out of 
these ingredients one must add the pseudopotentials of Eqs. \eqref{Eq:deformation_potential}, 
\eqref{Eq:gauge_field} to a pristine layer, and also integrate out the electronic degrees of freedom 
of its substrate, which creates a self-energy in the form of a local potential in graphene for 
strongly insulating substrates.

We will focus on the case of small rotation angles between graphene and a hBN substrate, as it has caught the attention of most part of the recent literature due to the high quality of the ensemble.
Since in the continuum approximation we may consider the superlattice to be minimally commensurate 
for any rotation angle, the deformation field $\vec{u}(\vec{r})$ will be assumed to have the same 
periodicity as the graphene/hBN Moir\'e, encoded in $\vec{G}_j$.
Furthermore, the bulk substrate will be treated henceforth as undeformable (hBN is assumed to be a thick and rigid crystal).
The smoothest possible deformation field $\vec{u}(\vec{r})$ in the superlattice unitary cell thus reads
\begin{align}
\vec{u}(\vec{r})=\sum_{j}\vec{u}_{\vec{G}_j}\,e^{i\vec{G}_j\cdot\vec{r}}.\label{Eq:u}
\end{align}
Since $\vec{u}(\vec{r})$ is real, its harmonics satisfy $\vec{u}_{\vec{G}_j}=\vec{u}^*_{-\vec{G}_j}$.
Moreover, the symmetry point group $S$ of the superlattice imposes some more constraints, namely $\mathcal{S}\vec{u}_{\vec{G}_j}=\vec{u}_{\mathcal{S}^{-1}\vec{G}_j}\,\forall \mathcal{S}\in S$.
For example, when dealing with $S=C_3$, only the vector $\vec{u}_{\vec{G}_1}$ ($\vec{G}_1$ belonging 
to the superlattice basis) will be independent in Eq.~(\ref{Eq:u}).
%Finally, we will also assume in the following that deformations $\vec{u}(\vec{r})$ are of the order of lattice parameters,\cite{SGS14,WBE14} and much smaller than the moir\'e unit cell length.

\emph{Symmetry-based analysis.---}
\label{Sec:symmetry-based_analysis}
Under the foregoing considerations, effective theories dependent on the deformation $\vec{u}(\vec{r})$ have been formulated.
A first foray into the computation of the Moir\'e band structure in the continuum approximation, 
including deformations, was based on symmetry principles.\cite{WPM13} 
% analyzing the band structure consists on its calculation taking $\vec{u}_{\vec{G}_j}$ as free parameters.\cite{WPM13}

\begin{figure}
\begin{center}
\includegraphics[width=0.44\textwidth]{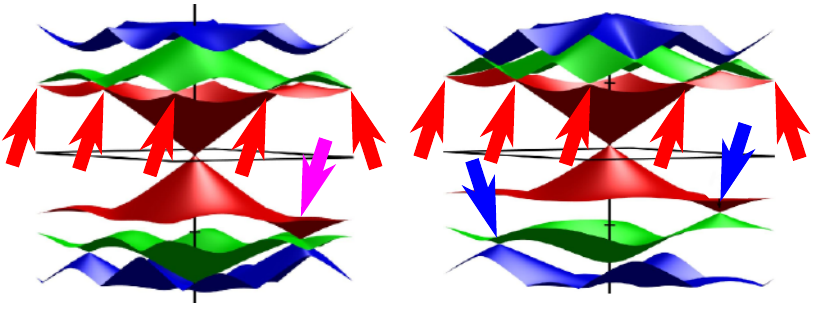}
\end{center}
\caption{Band structure for different choices of the parameters calculated in Ref. \cite{WPM13}.
Red, blue and magenta arrows point to multiple, single mDPs and triple degenerate band crossings, respectively. Reproduced with permission of the authors.
}
\label{Fig:mDPs}
\end{figure}

For monolayer graphene over a hBN substrate, the Dirac cone at the Fermi energy is preserved.
Moreover, due to the folding of the single layer bands over the superlattice 1st Brillouin zone, minibands were predicted to exhibit a combination of three possible features:
% for different values of the parameters:
(i) three mini Dirac points (mDPs) in the conduction or valence band, located at the superlattice 1st Brillouin zone edge with momentum $\hbar v_F G_j/2$;
(ii) a single mDP at the $\pm \kappa_\text{super}$ point (which is analogous to graphene's $K$ 
point 
but respective to the superlattice);
(iii) a triple degenerate band crossing.
Examples of band structures where (i)-(iii) are encountered appear in Fig. \ref{Fig:mDPs}.

\begin{figure}
\begin{center}
\includegraphics[width=0.44\textwidth]{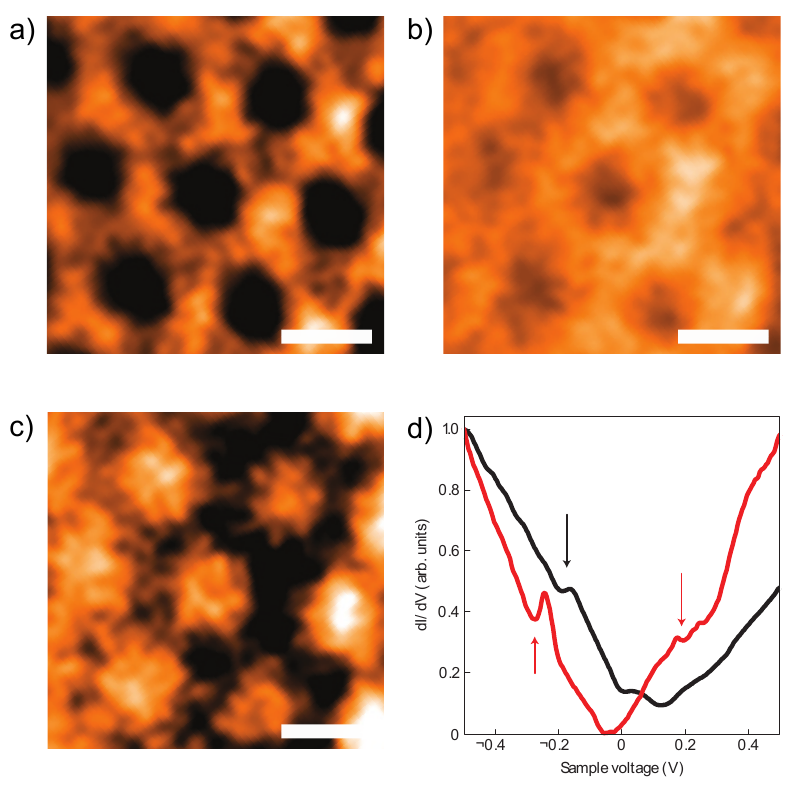}
\end{center}
\caption{Experimental LDOS for energies close to valence mDPs (a), to the main Dirac point (b) and to conduction mDPs (c) in a sample with a $13.4\,\text{nm}$ period.
The scale bars in all images are $10\,\text{nm}$ long.
(d): Measured $dI/dV$ (proportional to the DOS) with arrows pointing to the dips respective to mDPs.
Red (black) corresponds to a $13.4\,\text{nm}$ ($9.0\,\text{nm}$) Moir\'e wavelength.
Plots were extracted from Ref. \cite{YXC12} with permission of the authors.
}
\label{Fig:DOSLDOS}
\end{figure}

In agreement with these calculations, mDPs were observed in Ref. \cite{YXC12} at energies between $0.2$ and $1\,\text{eV}$ by means of STM 
measurements.
Dips in $dI/dV$ curves like the ones reproduced in Fig. \ref{Fig:DOSLDOS} are their main signature.
The predicted electron hole asymmetry comes from the interaction with the substrate, although 
next-nearest neighbour hopping could also contribute to it.
%{\color{red} Level repulsion implies the flattening of the minibands with a subsequent enhancement of the density of states.}
Further experimental evidence of mDPs could be extracted from (far) infrared spectra, where dissimilarities with respect to suspended graphene's constant conductivity would show up; or by Hall coefficient measurements, which would vary non monotonically upon doping with electrons or holes.

As for the dispersion relation pinned to mDPs described in (i), further analysis shows that it is anisotropic with a new Fermi velocity\cite{YXC12}
\begin{align}
v_F(\theta)=
\sqrt{\left(v_F\cos\delta\theta\right)^2+\left[V\sin\delta\theta/(2\hbar G_j)\right]^2}.
\label{Eq:vFtheta}
\end{align}
Here, $\delta\theta$ is the angle between $\vec{G}_j/2$ and $\delta\vec{k}=\vec{k}-\vec{G}_j/2$, and $V$ is the first star amplitude of the adhesion potential Fourier expansion.

Bilayer graphene over hBN was featured in Ref. \cite{MWF13} using a similar approach. 
%from the same perspective, namely taking $\vec{u}_{\vec{G}_j}$ as free parameters to explore the influence of that substrate on a different band structure.
mDPs also appear in this system, although the presence of a single cone is rare.
%, restricted to a unique choice of $\vec{u}_{\vec{G}_j}$.
As a new feature, gaps between the first Moir\'e miniband and the rest of the spectrum are common 
in this case.
The dependence of the spectra on the rotation angle is now enhanced with respect to single layer graphene due to higher trigonal warping corrections.

\emph{Microscopic derivations of spontaneous deformations.---}
\label{Sec:microscopic_derivations_deformations}
A step beyond the symmetry-based approaches described above was recently taken,\cite{SGSG14} and 
specific values for the spontaneous deformations $\vec{u}_{\vec{G}_j}$ were predicted as a function 
of the relative rotation angle between layers. From this, the specific form of the low energy 
electronic model was also computed.\cite{SGS214}
This program requires the knowledge of the adhesion potential $V_S$, whose spatial periodicity is 
given by that of the substrate, and its amplitude modulation is parameterized by adhesion energy 
differences $\Delta\epsilon_{AB}=\epsilon_{AB}-\epsilon_{AA}$ and 
$\Delta\epsilon_{BA}=\epsilon_{BA}-\epsilon_{AA}$. Here $\epsilon_{ij}$ denotes the adhesion energy 
per graphene unit cell for local $ij$ stacking. The stacking notation is $AA$ for perfectly aligned 
hexagonal lattices, $AB$ for Carbon-on-Boron and $BA$ for Carbon-on-Nitrogen Bernal stackings.  
The values of the different adhesion energies $\epsilon_{ij}$ can be extracted from \emph{ab 
initio} calculations,\cite{BAB14,BF_11,GKB07,PLK11} which almost invariably predict a stronger 
adhesion for $AB$-stacking. Ref.~\cite{SGSG14} constructs a first star adhesion potential 
corresponding to a given $\Delta\epsilon_{BA}$ and $\Delta\epsilon_{AB}$
\begin{align}
&
V_S(\vec{r})=\sum_{j=\pm 1}^{\pm 3}
v_j\,e^{i\vec{g}^{\prime}_j\cdot\vec{r}}+v_0;
\label{Eq:Vsr}
\\
&
v_{j>0}=v_{j<0}^*=
-\frac{\Delta\epsilon_{AB}+\Delta\epsilon_{BA}}{18}+i\frac{\Delta\epsilon_{AB}-\Delta\epsilon_{BA}}{6\sqrt{3}};
\label{Eq:vjgl0}
\end{align}
$\vec{g}'$ belonging to the reciprocal lattice of hBN. The uniform offset $v_0=(\epsilon_{AB}+\epsilon_{BA}+\epsilon_{AA})/3$ may be disregarded in the following.
In the continuum limit, the total adhesion energy can be written as
\begin{align}
&
U_S=\frac{1}{\Omega_S}\int_{\Omega_S} d^2\vec{r}\, \tilde{V}_S[\vec{r},\vec{u}(\vec{r})];
\label{Eq:US}
\\
&
\tilde{V}_S[\vec{r},\vec{u}(\vec{r})]=\sum_{j=\pm 1}^{\pm 3}v_j\,
e^{-i\vec{G}_j\cdot\vec{r}}e^{i\vec{g}_j^{\prime}\cdot\vec{u}(\vec{r})}.
\label{Eq:tildeVS}
\end{align}
$\Omega_S$ is the area of the superlattice unit cell.
The linear distortion regime may be considered,
\begin{align}
\tilde{V}_S[\vec{r},\vec{u}(\vec{r})]\simeq \tilde{V}_S[\vec{r},0]
+
\vec{u}(\vec{r})
\cdot
\partial_{\vec{u}}\tilde{V}_S[\vec{r},\vec{u}(\vec{r})]\big|_{\vec{u}=0}, \label{Eq:V_S}
\end{align}
This approximation allows us to compute analytically an equilibrium deformation by minimizing the 
adhesion energy $U_S$ plus the elastic energy due to in-plane distortions 
$F^{\text{in-plane}}_{\text{st}}$ (see below), and 
can be shown to be legitimate, since the inclusion in Eq. (\ref{Eq:V_S}) of higher order terms in $\vec{u}(\vec{r})$ does not significantly modify the result.\cite{SGSG14}
 
The elastic energy $F^{\text{in-plane}}_{\text{st}}$ prevents the layer from conforming completely 
to $V_S$.
According to continuum elasticity theory, the elastic energy due to in-plane 
stretching, $F^{\text{in-plane}}_{\text{st}}$, for an isotropic medium is obtained from Eq. 
(\ref{eq:stretching_energy}) by replacing the strain tensor $u_{ij}$ by the linear strain tensor 
$u_{(i,j)}$,
\begin{align}
F^{\text{in-plane}}_{\text{st}}
&=\frac{1}{2} \int d^2 \vec{r} \left( \lambda u_{(i,i)}^2 + 2 \mu u_{(i,j)} u_{(i,j)} 
\right)
\notag
\\
&=\frac{1}{2}\sum_{\vec{q}}\vec{u}_{-\vec{q}} \cdot \underline{W}_{\vec{q}} \cdot 
\vec{u}_{\vec{q}},
\label{Eq:UE}
\end{align}
with
\begin{align}
&
\underline{W}_{\vec{q}}
=(\lambda+2\mu)\Omega_g\underline{W}_{\vec{q}}^\parallel
+\mu\,\Omega_g \underline{W}_{\vec{q}}^\perp, \label{Eq:Wq}
\\
&
\underline{W}_{\vec{q}}^\parallel
=\left(
\begin{array}{cc}
q_x^2 & q_xq_y \\
q_xq_y & q_y^2
\end{array}
\right),\ 
\underline{W}_{\vec{q}}^\perp
=\left(
\begin{array}{cc}
q_x^2 & -q_xq_y \\
-q_xq_y & q_y^2
\end{array}
\right).\label{Eq:Wqparperp}
% \\
%&
%B=\lambda+2\mu. \label{Eq:Blambda}
\end{align}
$\Omega_g$, analogous to $\Omega_S$, corresponds to the area of the graphene unit cell. $\lambda$ and $\mu$ are the Lam\'e constants.

\begin{figure}
\begin{center}
\includegraphics{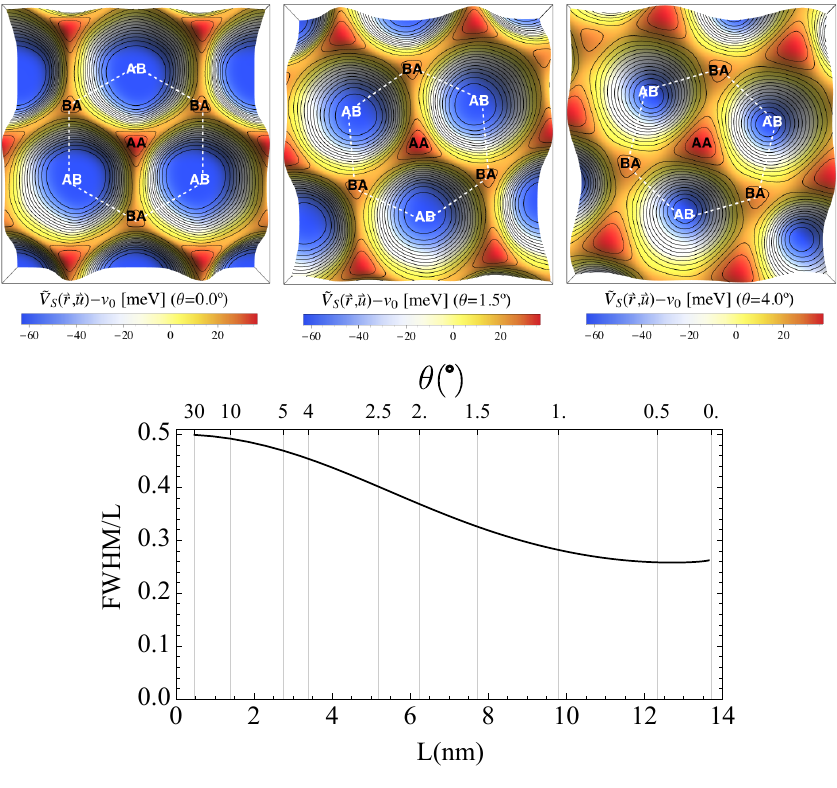}
\end{center}
\caption{TOP: Adhesion energy density $\tilde{V}_S[\vec{r},\vec{u}(\vec{r})]$ in real space, 
relative to the average adhesion $\nu_0$.
%First star deformations $\vec{u}(\vec{r})$ calculated by inserting Eq. (\ref{Eq:uG}) in Eq. (\ref{Eq:u}).
The rotation angles are $\theta =0$ (left), $\theta=1.5^{\circ}$ (middle) and $\theta=4^{\circ}$ (right).
BOTTOM: Theoretical FWHM of a section of the deformation profile as a function of the rotation angle 
or Moir\'e period.}
\label{Fig:deformations}
\end{figure}

The minimization of the total energy of the lattice, $U=F^{\text{in-plane}}_{\text{st}}+U_S$, with 
respect to the parameters $\vec{u}_{\vec{G}_j}$ yields the equilibrium configuration,
\begin{align}
\vec{u}_{\vec{G}_j}=iv_j^*\underline{W}_{\vec{G}_j}^{-1} \cdot \vec{g}_j^{\prime}.
\label{Eq:uG}
\end{align}
Fig. (\ref{Fig:deformations}) plots the resulting adhesion energy 
$\tilde{V}_S[\vec{r},\vec{u}(\vec{r})]$ for different relative rotation angles.
It is expected to be representative of the Young modulus measured in Ref. \cite{WBE14}.
A good qualitative agreement is achieved with experiments (see Fig. \ref{Fig:young}): greater angles 
give rise to configurations close to an undeformed floating phase, whereas a small twist drives the 
system to sharper deformations without a global expansion or contraction (a `generalized' floating 
phase). The inclusion of e.g. global deformations as additional degrees of freedom could be 
considered, but remain unexplored. Other works have studied also the role of out of plane 
instabilities.\cite{JDM14} These could account for quantitative differences with experimental 
results, noticeable in particular in the normalized width of stacking solitons $\text{FWHM}/L$ for 
long Moir\'e periods. 
In fact, an abrupt jump at $L\sim 10\,\text{nm}$ was reported therein that is not reproduced by the 
linear deformation model described here. 
Interestingly, however, some conflicting results exist in this regard, with e.g. Ref. \cite{HSY13} 
reporting no evidence of such discontinuity in 
electronic observables.
%{\color{red}This shortcoming is due to having departed from a continuum approximation. A model mimicking such a non analytic behavior is still to be constructed, which demands overcoming the aforementioned difficulties that arise in a 2D Frenkel-Kontorova model.}

\emph{Microscopic derivations of effective electronic model.---}
\label{Sec:microscopic_derivations_electronic}
Once an equilibrium deformation solution such as Eq. (\ref{Eq:uG}) has been established, and bearing 
also Eqs. \eqref{Eq:deformation_potential},\eqref{Eq:gauge_field} in mind, an effective low energy 
Hamiltonian may be derived microscopically for graphene on the crystalline substrate, which has the 
advantage over symmetry-based approaches of predicting specific values of all model parameters. The 
first step involves writing a low energy model including graphene and hBN degrees of freedom,
\begin{align}
\bm H=\left(\begin{array}{cc}
\bm H_\mathrm{gr}\left[\vec{k}-\frac{1}{2}\Delta 
\vec{K}-\frac{e}{\hbar}\vec{\mathcal{A}}^{el}(\vec{r})\right]+\bm V_s(\vec{r}) & \bm \tilde{\bm 
T}^\dagger(\vec{r}) \\
\bm \tilde{\bm T}(\vec{r}) & \bm H_\mathrm{hBN}\left(\vec{k}+\frac{1}{2}\Delta \vec{K}\right)\end{array}\right);\ \ 
\bm H_{\mathrm{hBN}}\simeq \bm \Delta_{\mathrm{hBN}}=
\left(
\begin{array}{cc}
\epsilon_v & 0\\
0 & \epsilon_c
\end{array}
\right). \label{Eq:H}
\end{align}
$\bm H_{\mathrm{gr}}$ denotes the Hamiltonian of monolayer graphene and $\bm H_{\mathrm{hBN}}$ 
corresponds to a single layer of hBN, that is 
approximated by a perfect insulator.
$\epsilon_v$ ($\epsilon_c$) are the energies of its valence (conduction) band.
$\tilde{\bm T}(\vec{r})$ is the interlayer hopping taking deformations in the linear regime into account,\cite{SGS214} and $\Delta\vec{K}$ joins the Dirac point of pristine graphene with the analogous in hBN.
hBN orbitals can then be integrated out of Eq. (\ref{Eq:H}) exactly, which results in a local self energy,
\begin{align}
\bm\Sigma(\vec{r})\approx -\tilde{\bm T}^\dagger(\vec{r})\bm \Delta^{-1}_\mathrm{hBN}\tilde{\bm T}(\vec{r})
\label{Eq:Sigmar}
\end{align}
that must be added to $\bm H_{\mathrm{gr}}$.
Together with the induced pseudopotentials $(V_s,\vec{\mathcal{A}}^{el})$, the final effective 
Hamiltonian reads\cite{SGS214}
\begin{align}
&
\bm H_\mathrm{eff}=\bm 
H_\mathrm{gr}\left[\vec{k}-\frac{e}{\hbar}\vec{\mathcal{A}}^{el}(\vec{r})\right] +\bm V_s 
(\vec{r})+\bm\Sigma(\vec{r}). \label{Eq:Heff}
\end{align}

\begin{figure*}
   \centering
     \begin{tabular}{ccc}
     \fbox{\includegraphics[width=0.44\textwidth]{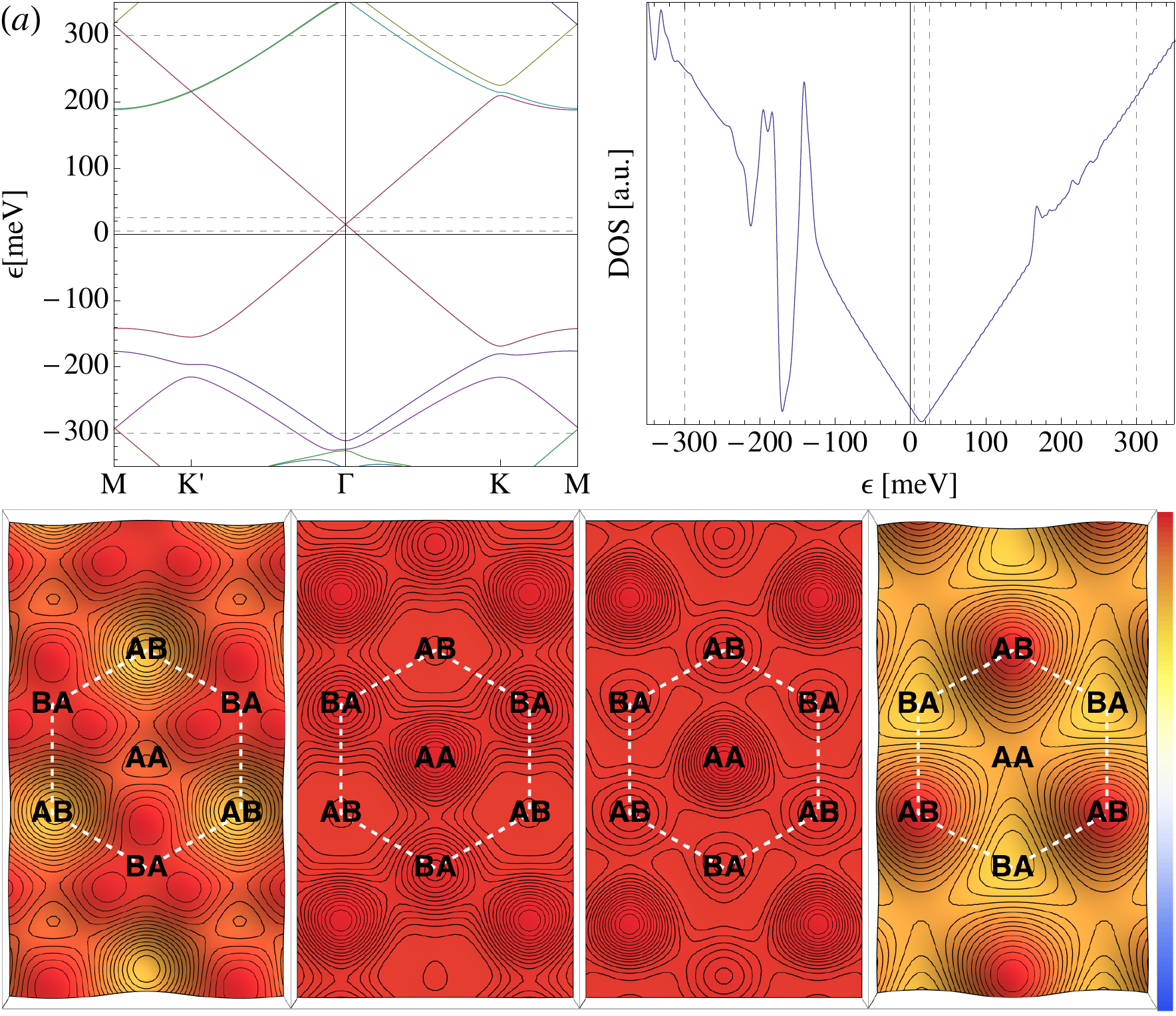}} & &
     \fbox{\includegraphics[width=0.44\textwidth]{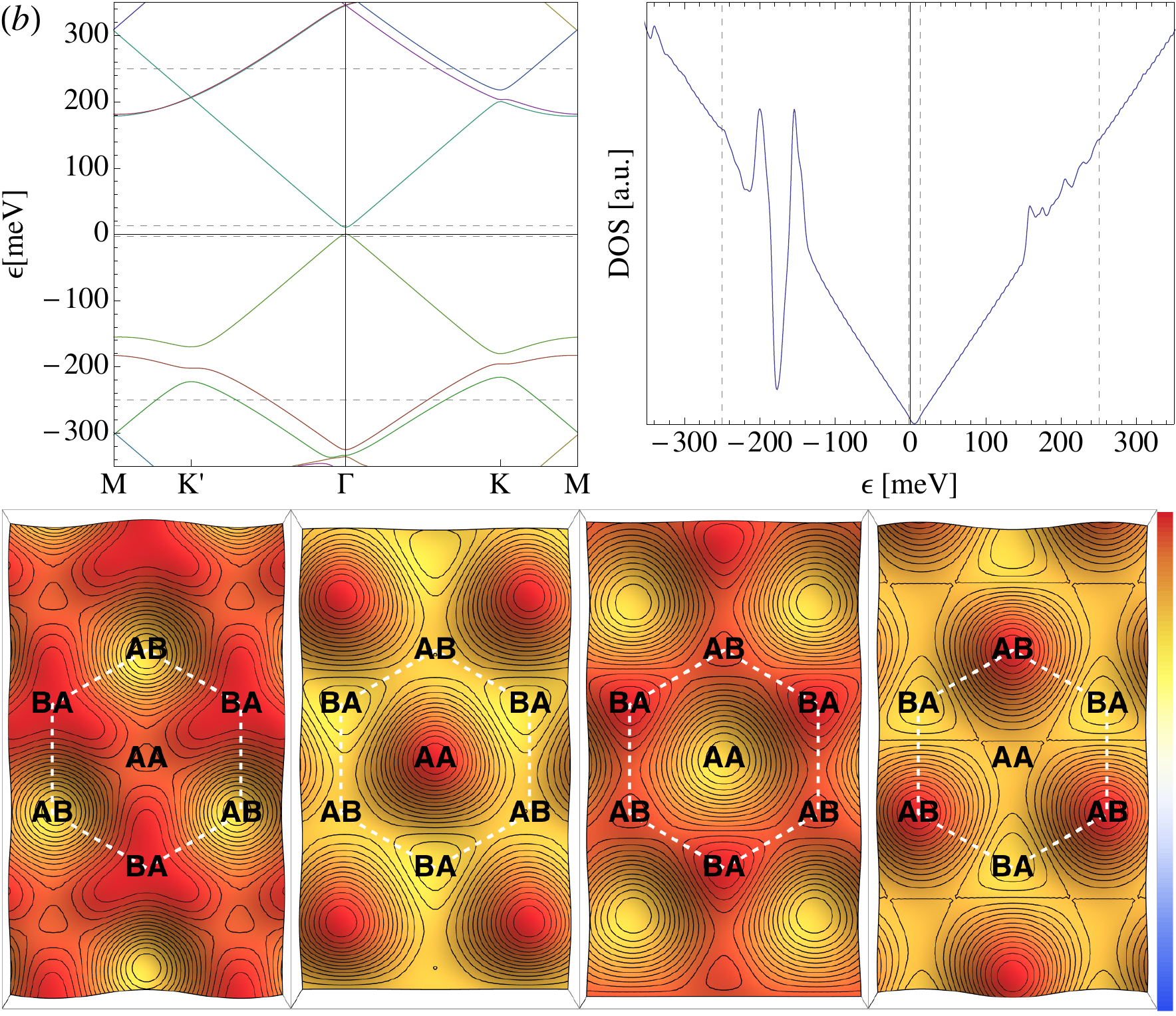}} \\
     $\left(z_u=0, t_\perp\neq 0, g_2= 0\right)  \rightarrow \Delta\approx 0.14\,\mathrm{meV}$ & & 
     $\left(z_u\neq 0, t_\perp\neq 0, g_2=0\right) \rightarrow \Delta\approx 9.8\,\mathrm{meV}$
     \\ \\
     \fbox{\includegraphics[width=0.44\textwidth]{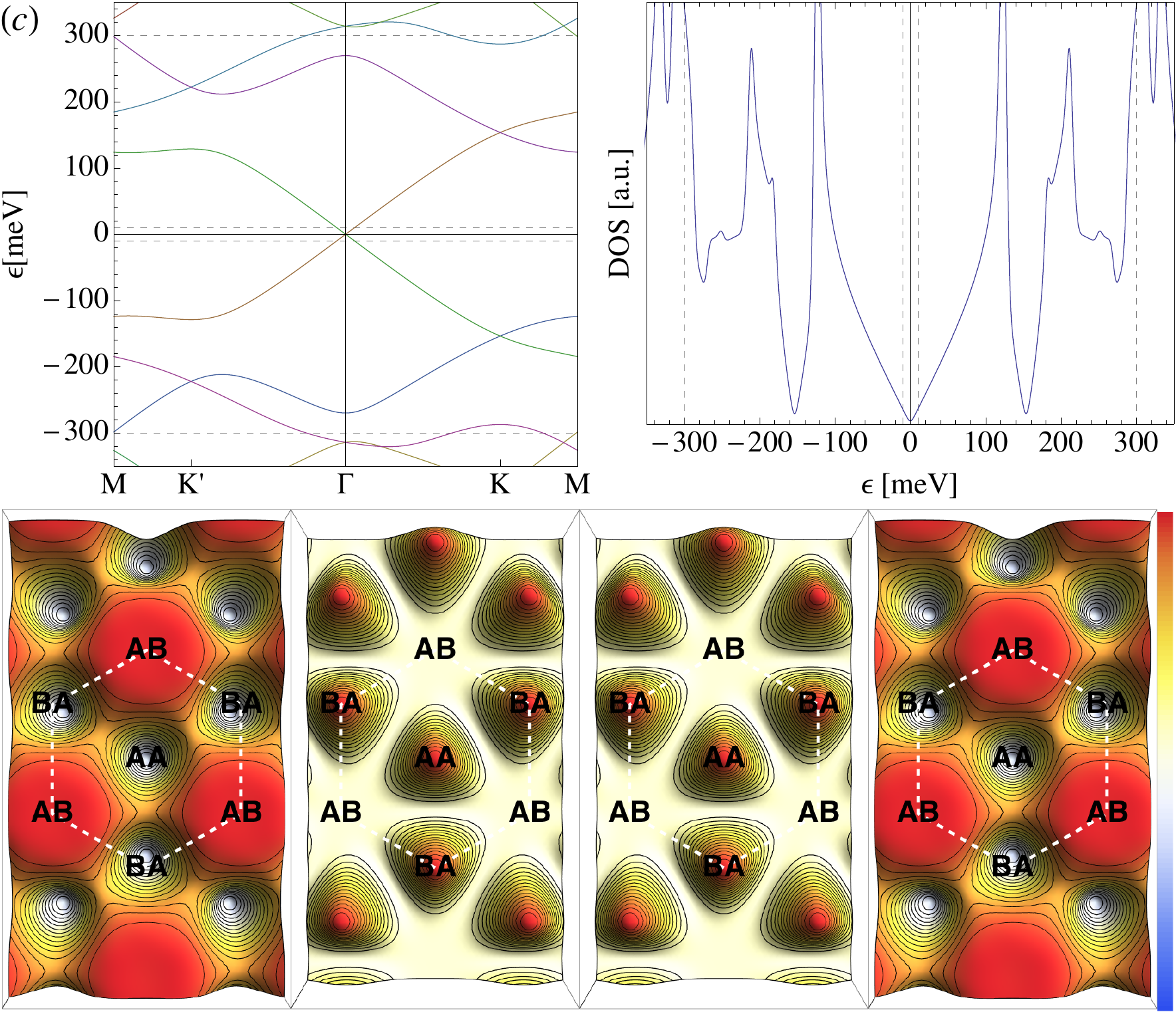}} & &
     \fbox{\includegraphics[width=0.44\textwidth]{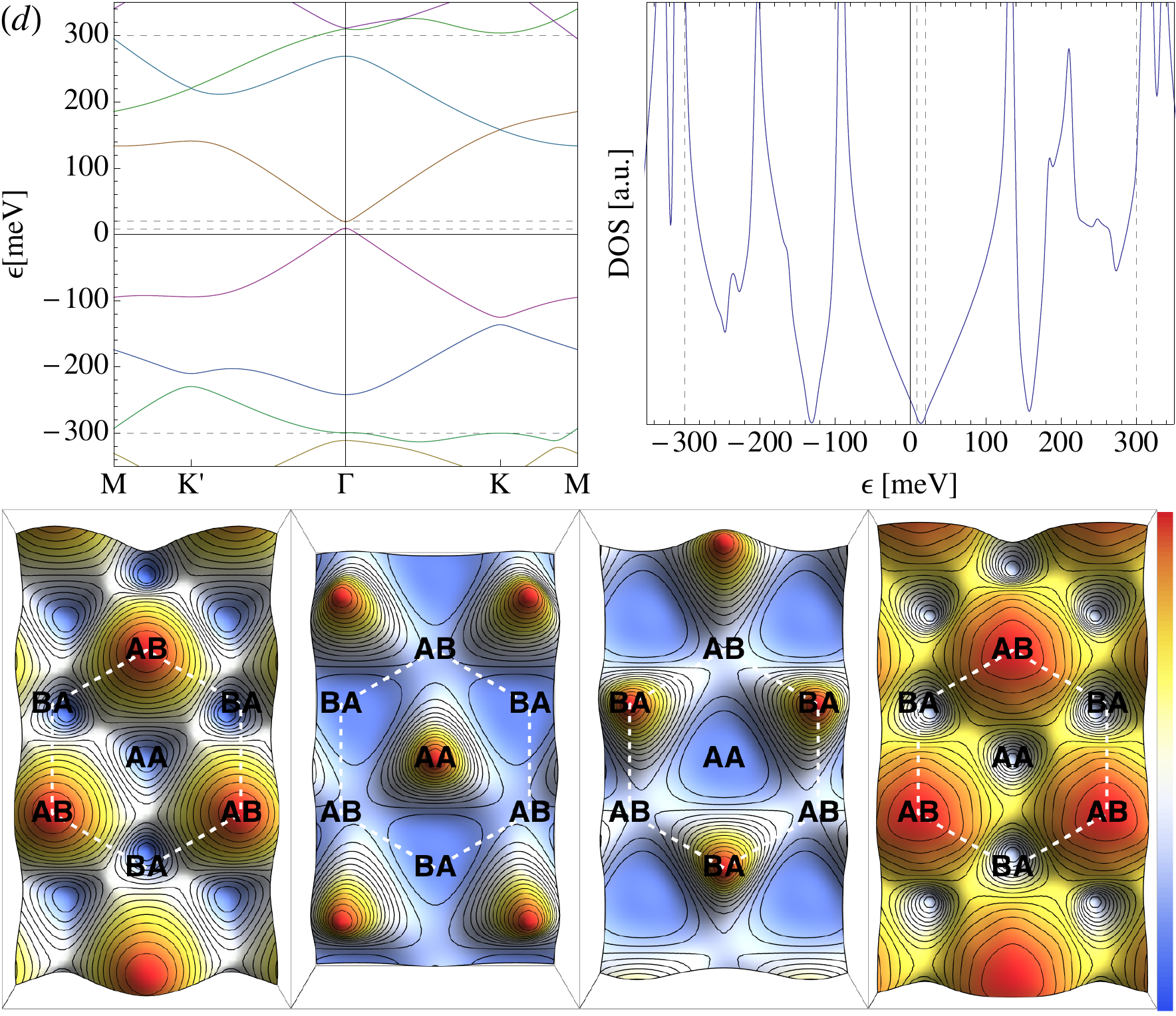}} \\ 
     $\left(z_u\neq0, t_\perp=0, g_2\neq 0\right)  \rightarrow \Delta=0$ & & 
     $\left(z_u\neq0, t_\perp\neq0, g_2\neq 0\right)  \rightarrow \Delta\approx 9.6\,\mathrm{meV}$
     \end{tabular}
     \caption{Electronic structure given by the model of Eq. (\ref{Eq:Heff}) for unrotated graphene 
over hBN. Different combinations of physical ingredients have been considered: unstrained graphene 
on hBN (a), strained graphene on hBN without pseudogauge potential (b), strained graphene 
electronically decoupled from hBN (c), and the complete model of a strained graphene on hBN 
including pseudogauge field (d). Represented are the low-energy band structure, the corresponding 
total DOS, and the LDOS in real space for four energies shown as dashed lines in the band structure 
and DOS. In the LDOS the colouring corresponds to blue for zero, and red for the maximum. We have 
included the value for the (non-perturbative) gap $\Delta$ at the primary Dirac point in each case.}
   \label{Fig:ES2}
\end{figure*}

Ref. \cite{SGS214} focuses on unrotated graphene over hBN, explicitly  spelling out Eq. (\ref{Eq:Heff}) after decomposing all contributions in terms 
of $\vec{\sigma}$ matrices, and into spatially symmetric and antisymmetric components.
Every harmonic beyond the first star may be neglected in the model. 
The solution depends solely on dimensionless parameters $z_u$ (maximum difference in local relative expansion between different regions, see Ref. 
\cite{SGS214}), $m_\pm$ (associated to the hBN gap and the interlayer hopping $t_{\perp}$), $g_1$ [energy scale of the deformation potential, Eq. 
(\ref{Eq:deformation_potential})] and $g_2$ [relating strains to pseudomagnetic fields, Eq. (\ref{Eq:gauge_field})].

The electronic spectrum, density of states (DOS) and local density of states (LDOS) were analysed 
as a function of strains, hopping to hBN, $V_s$ and $\vec{\mathcal{A}}^{el}$.
The most notable conclusions are:
(i) $V_s$ and $\vec{\mathcal{A}}^{el}$ do not generate a gap by themselves (i.e., when the 
interlayer hopping $t_\perp\to 0$), and in fact tend to cancel each other's contribution when 
$t_\perp\neq 0$; 
(ii) the spectral gap at the primary Dirac point, that can be calculated analytically, is linear in the distortions $z_u$ to leading order;
(iii) new Dirac Points emerge in the valence and conduction bands, in agreement with the symmetry-based analysis;
(iv) LDOS homogeneity is strongly broken by $(V_s,\vec{\mathcal{A}}^{el})$;
(v) the overall electronic structure at finite energies shows a strong sensitivity to the values of $(z_u,m_\pm,g_1,g_2)$ within their realistic range.
As an example, the spectral gap can range from $0$ to $\sim 10\,\text{meV}$ depending on their choice. Fig. \ref{Fig:ES2} illustrates these results.
Therefore, a direct measurement of electronic structure properties, such as the DOS, could allow to obtain accurate values of $(z_u,m_\pm,g_1,g_2)$, to be contrasted with the theoretical predictions.

Ref. \cite{SGS214} uses the values $z_u=-0.18$ (which corresponds to $u_{kk}/2=2.8\%$, 
with a graphene lattice parameter of $a=2.4$ \AA,  lattice mismatch between hBN ($a_\text{hBN}$) 
and graphene of $\delta=a_{\text{hBN}}/a-1=0.018$, 
$\lambda+2\mu=19.1$ eV/\AA${}^2$ and $\Delta\epsilon_{AB}=-100$ meV/unit cell), $t_\perp=0.3$ eV and $g_2=4.8 \mathrm{nm}^{-1}$.
Moreover, $t=3.16$ eV, $\epsilon_c=3.34$ eV and $\epsilon_v=-1.4$ eV.
As recent works have shown that electronic screening makes graphene's deformation potential $g_1$ negligible in practice,\cite{PBS14,SCP14} $g_1=0$ was assumed here.
Intriguingly, this choice of parameters achieves the best fit to LDOS experimental data\cite{YXC12} 
when switching off $\vec{\mathcal{A}}^{el}$ but taking into account in plane deformations and 
hoppings to hBN (see Figs. \ref{Fig:DOSLDOS}b and \ref{Fig:ES2}d).
%A qualitative agreement of the DOS is also achieved between Figs. \ref{Fig:ES2} and \ref{Fig:DOSLDOS}d concerning the dips that unravel the presence of mDPs. 

\section{Strain in the new families of 2D crystals beyond graphene}
\label{sec_others}

The progresses in the fabrication and characterization of graphene have been successfully applied to
the isolation of single layers of other 2D materials with properties complementary to graphene 
\cite{NJS05}. Atomically thin 2D crystals of boron nitride, transition metal dichalcogenides (TMDs) 
like MoS$_2$ or WS$_2$, black phosphorus, etc, have been recently obtained by means of mechanical 
exfoliation \cite{MLH10,DYM10,KDS14}, chemical vapour deposition (CVD) \cite{SHJ10,LZZ12} or laser 
thinning \cite{CBG12}. Similar to graphene, these other families of 2D crystals are extremely strong 
materials with high elasticity and Young modulus, and they can withstand non-hydrostatic (e.g., 
tensile or shear) stresses up to a significant fraction of their ideal strength without inelastic 
relaxation by plasticity or fracture\cite{CSZ14}. Furthermore, the techniques to apply strain in 
graphene can also be applied to other 2D materials, and strain engineering has been proposed as  an 
effective approach to continuously tune their electronic and optical properties \cite{LSM14}. In 
this section, we provide a brief overview of the effect of strain in other layered materials 
different from graphene, with special focus on TMDs and black phosphorus. For a comprehensive 
review on strain engineering in semiconducting 2D crystals, we refer the reader to 
Ref.~\cite{RCC15}.

\subsection{Transition Metal Dichalcogenides}

\begin{figure*}[t]
\begin{center}
\includegraphics[scale=0.8,clip=]{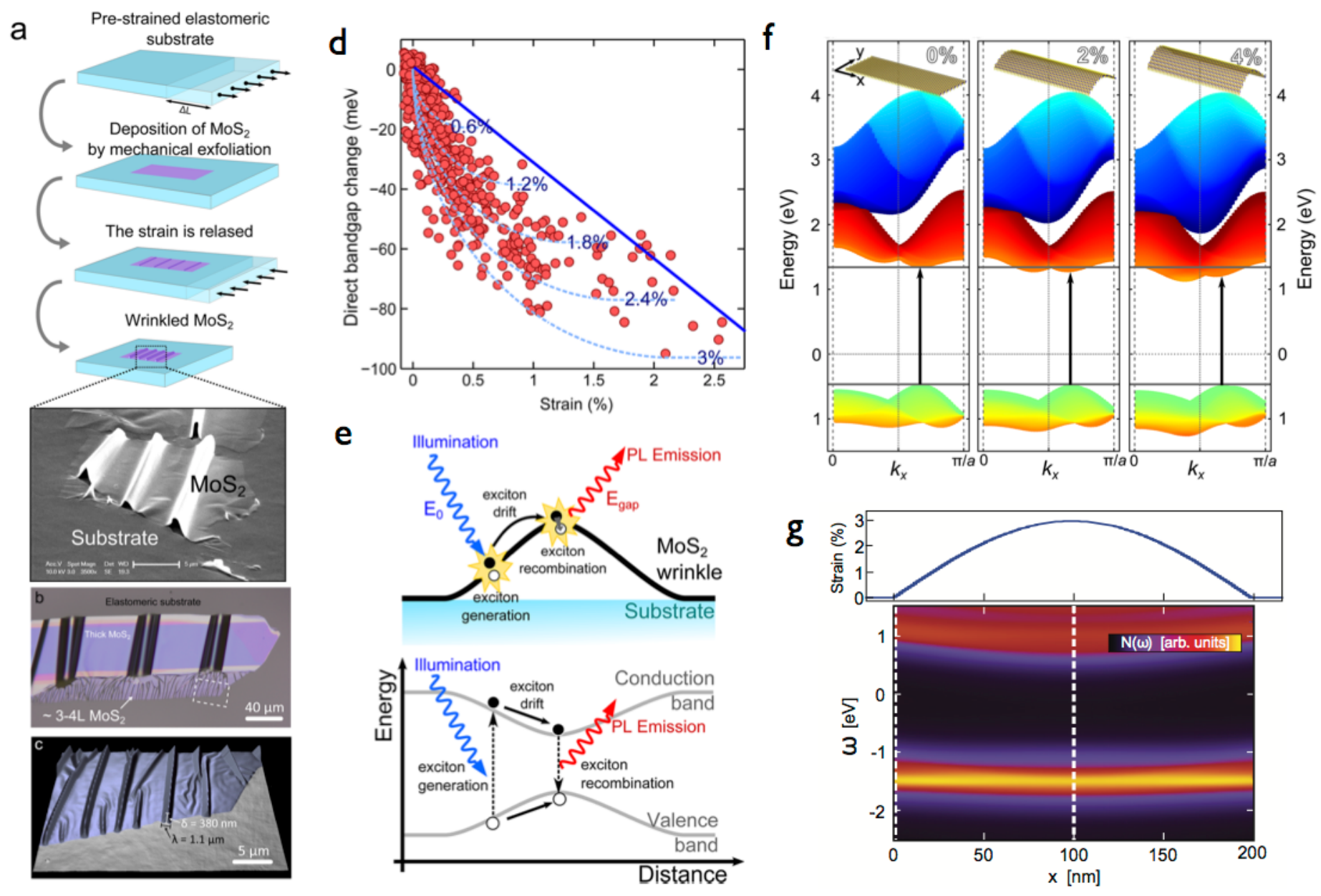}
\end{center}
\caption{ (a) Schematic diagram of the fabrication process of wrinkled MoS$_2$ nanolayers. An elastomeric substrate is stretched prior depositing 
MoS$_2$ by mechanical exfoliation. The strain is released afterwards, producing buckling-induced 
delamination of the MoS$_2$ flakes. (b) Optical 
microscopy image of a wrinkled MoS$_2$ flake fabricated by buckling-induced delamination. (c) Atomic force microscopy (AFM) topography image of the 
region marked by the dashed rectangle in (b).  (d) Change in the energy of the direct bandgap transition as a function of the strain. The dashed lines 
show the expected bandgap vs strain relationship after accounting for the effect of the finite laser spot size and the funnel effect. (e) Schematic 
diagram explaining the funnel effect due to the nonhomogeneous strain in the wrinkled MoS$_2$. (f) Theoretical band structure for nonuniform strained 
MoS$_2$ calculated for a zigzag ribbon. The different panels correspond, from left to right, to the band structure for the case of an unstrained 
ribbon, and for wrinkled ribbons with a 2\% and 4\% maximum tensile strain, respectively. In the photoluminescence experiments, the wavelength of the 
emitted light is determined by the direct bandgap transition energy, indicated in each panel by a 
vertical arrow. (g) Colour map of the local density 
of states as a function of the position along the wrinkle profile, calculated theoretically for a wrinkle under non-uniform uniaxial strain. (Figures 
with permission from Ref. \cite{CRC13})}
\label{Fig:MoS2}
\end{figure*}

 Among the new families of 2D layered materials, TMDs are recently object of special attention. They are semiconductors with a gap in the visible range of the electromagnetic spectrum, present a strong spin-orbit coupling, and offer the possibility to control quantum degrees of freedom as the electron spin, and the valley and layer pseudospin, among other interesting features\cite{WKK12,JSL14,XYX14,RSL14}. Semiconducting TMDs have the form $MX_2$, where $M={\rm Mo,W}$ is a transition metal  and $X={\rm S, Se}$ is a chalcogen atom. In their bulk configuration, they are composed of two-dimensional $X-M-X$ layers stacked on top of each other, coupled by weak van der Waals forces. The metal atoms are ordered in a triangular lattice, each of them bonded to six chalcogen atoms located in the top and bottom layers. The electronic band structure of $MX_2$ changes from an indirect band gap for multilayer samples to a direct gap semiconductor for single layers \cite{MLH10}.

TMDs have a phonon structure highly sensitive to strain. This is revealed by the shift of the Raman 
peaks corresponding to the A$_{1g}$ out-of-plane  mode (where the top and bottom $X$ atoms are 
moving out of plane in opposite directions while $M$ is fixed), and the $E_{2g}^1$ in-plane mode 
(where the $M$ and $X$ atoms are moving in-plane in opposite directions) \cite{LYB10}. In 
particular,  applying uniaxial strain lifts the degeneracy of the $E_{2g}^1$ mode, leading to red 
shifting and splitting into two distinct peaks for strain of just 1\% \cite{CRC13,CWZ13,HLJ13}. 
Strain in the samples can be induced by the application of forces across specific axes of the 
crystal structure. This effect can be used to modify the band structure of TMDs, and in fact a 
reduction of the band gap can be achieved under uniaxial compressive strain across the $c$-axis of 
the crystal structure of $MX_2$.\cite{PV12}  On the other hand, tensile strain is expected to lower 
the electron effective mass\cite{SPZ13} and consequently improve electron mobility. A semiconductor 
to metal transition in single layer MoS$_2$ has been predicted for compressive (tensile) biaxial 
strain of about $15\%$ ($ 8\%$) \cite{SHP12}, which are strengths achievable for these kind of 2D 
crystals \cite{CSZ14}. First principle calculations have predicted that strain can be induced by 
depositing MoS$_2$ on hexagonal boron nitride, which can lead to a direct-to-indirect gap 
transition \cite{HHQ14}. See Fig. \ref{Fig:deformations} and  Sec. V for a discussion on how 
superlattices may also induce local variations of the LDOS that is seen experimentally by STM.

Within the framework of the effective mass theory, the effect of strain in the low energy electronic spectrum can be easily incorporated by means of 
group theory. Similarly to the case of graphene, Eq.~\eqref{Eq:strain_sym}, strain tensor components can be arranged in a scalar ($A_1'$) and 
vectorial ($E'$) irreducible representations of $D_{3h}=D_3\otimes\sigma_h$, the point group of TMDs monolayers of the hexagonal polytype. This vector 
may be interpreted as a pseudo-gauge field that, together with the large spin-orbit coupling 
provided by $M$ atoms and the lack of a centre of 
inversion in the unit cell (note that $D_{3h}$ cannot be decomposed into a direct product of the inversion group and the group of rotations of the 
crystal) can lead to a quantum spin Hall effect and time-reversal invariant topological phases in these single-layer materials. For low carrier 
densities, as suggested in Ref. \cite{COG14}, semiconducting TMDs under shear strain will develop spin-polarized Landau levels residing in different 
valleys. For the case of MoS$_2$, gaps between Landau levels have been estimated to scale as $\hbar\omega_c/ k_B \sim 2.7 B_0[\rm T]$ K, where 
$\omega_c=eB_0/m^*$ is the cyclotron frequency in terms of the effective mass $m^*$, and $B_0[\rm T]$ is the strength of the pseudomagnetic field in 
Tesla. Considering that, in the case of graphene, pseudomagnetic fields of $B_0[{\rm T}]\sim 10-10^2$ T have been experimentally demonstrated 
\cite{LBM10}, and taking into account the intrinsic limitations imposed by the value of the shear modulus and the maximum tensile strength of these 
materials, Landau level gaps of up to $\approx 20$ K are within experimental reach. Other proposals for different polytypes have appeared in the 
recent literature \cite{QLFL14}. In Ref.~\cite{RRC15} it has been calculated a low energy 
${\bf k}\cdot {\bf p}$ Hamiltonian that goes beyond the simple Dirac-like model and captures some of 
the main features of the electronic band dispersion of TMDs under strain, such as the shift of the 
conduction and valence band edges of $MX_2$ in the presence of homogeneous strain, with the 
corresponding transition from direct to indirect gap, as well as the coupling between spin degrees 
of freedom and strain.

Some of the aforementioned theoretical results have been confirmed experimentally, and a change in 
the direct band gap up to $\sim 45$ meV per 1\% applied strain has been measured 
\cite{CWZ13,HLJ13,DDJ14,YCZ14}. Strain engineering has been proposed as a powerful tool to create a 
broad-band optical funnel in layered 2D crystalline semiconductors \cite{FL12}. Recent experimental 
observations have shown that, a continuous change in strain across an atomically thin sheet of 
MoS$_2$,  leads to a continuous variation of the optical band gap \cite{CRC13}. This could allow not 
only to capture photons across a wide range of the solar spectrum, but also can guide  the resulting 
generated excitons towards the contacts. As sketched in Fig. \ref{Fig:MoS2} (a)-(c), buckling induced 
delamination can be used to obtain samples of TMDs with a non-uniform profile of strain, up to 2.5\% 
tensile strain. The reduction of the direct band gap with strain, shown in Fig. \ref{Fig:MoS2} (d), 
makes that the  photo-induced excitons, in order to minimize their energy, move to lower bandgap 
regions before recombining. The nonuniform bandgap profile induced by the local strain of the 
wrinkles therefore generates a trap for photogenerated excitons ({\it funnel effect}) with a depth 
of up to 90 meV, as  sketched in Fig. \ref{Fig:MoS2} (e).  The effect of non-uniform strain in the 
band structure and in the local density of states of MoS$_2$, shown in Fig. \ref{Fig:MoS2} (f) and 
(g), has been theoretically studied with a tight-binding model that properly accounts for both, 
$d$-orbital contribution from the metal as well as $p$-orbital contribution from the chalcogen, 
which are relevant for the valence and conduction bands \cite{CRS13,CRC13,RLG14}. Recently it 
has been demonstrated that an optoelectronic crystal consisting of a superlattice of {\it artificial 
atoms} can be made from biaxially strained single layer MoS$_2$ deposited on a patterned nanocone 
substrate \cite{LCQ15}.

The large mechanical stretchability and flexibility of single layer TMDs together with the absence 
of inversion symmetry makes them good candidates for 
high-performance piezoelectric materials. This has been recently demonstrated experimentally by Wu {et al.}, Ref. \cite{WWL14}, who have shown that 
cyclic stretching and releasing of thin MoS$_2$ flakes with an odd number of atomic layers produces oscillating piezoelectric voltage and current 
outputs, converting mechanical energy into electricity. On the other hand, no output is observed for flakes with an even number of layers, which 
possess inversion symmetry \cite{WWL14}. The strain-induced polarization charges in single layer MoS$_2$ can modulate charge carrier transport at a 
MoS$_2$-metal barrier and enable enhanced strain sensing. This demonstrates the potential of 2D crystals in powering nanodevices and 
tunable/stretchable electronics/optoelectronics.

\subsection{Black Phosphorus}

\begin{figure}[t]
\begin{center}
\includegraphics[scale=0.7,clip=]{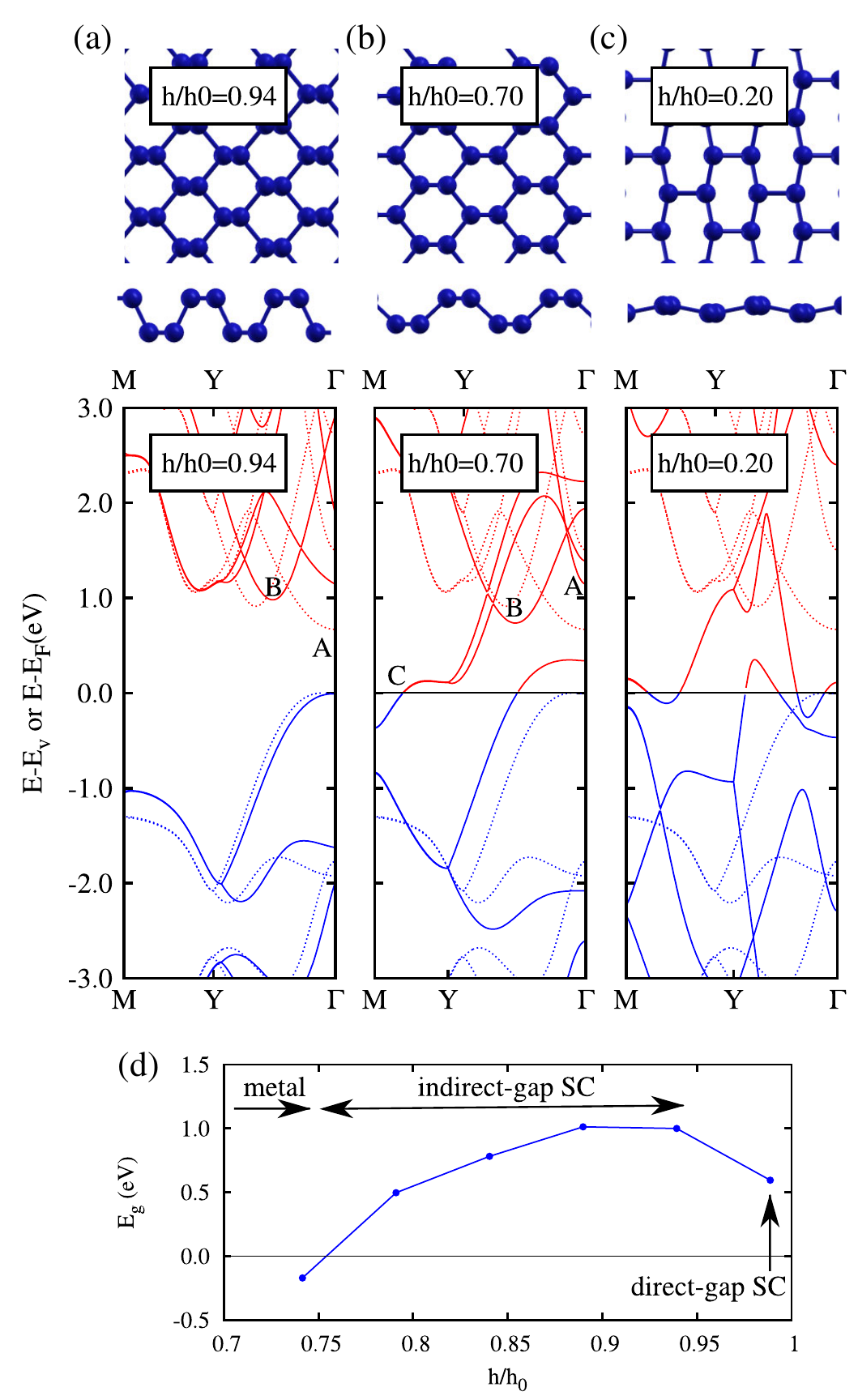}
\end{center}
\caption{Single layer BP under uniaxial compression along the out-of-plane direction. Results corresponding to three values of $h=h_0$ are shown 
(solid line), along with the electronic band dispersion of the unstrained material (dotted lines). Top and side views represent the respective relaxed 
structure. (d) Change of band gap with height. The original layer thickness used in the calculations is $2 h_0$ . (Reproduced with permission from 
Ref. \cite{RCC14})}
\label{Fig:BP}
\end{figure}

Black phosphorus (BP) is another layered material that has been recently synthesized in its single layer form, also known as {\it phosphorene} \cite{LNZ14}. BP is a stable allotrope of phosphorus and an elemental semiconductor. At ambient conditions, orthorhombic BP forms puckered layers which are stacked together by weak Van der Waals forces (see Fig. \ref{Fig:BP}). Similarly to graphite and TMDs, the layered structure allows for mechanical exfoliation to prepare thin layers on a substrate. The band gap is direct, located at the $\Gamma$ point of the rectangular Brillouin zone, and its size highly depends on the number of layers, suggesting that multi-layer BP optical devices could allow photo-detection over a broad spectral range. Furthermore, BP is another promising material in nanoelectronics applications since the first field-effect transistors fabricated with few layer samples show drain current modulation on the order of $10^5$ (four orders of magnitude larger than that in graphene) and high charge-carrier mobilities \cite{LYY14}. 

Before the isolation of single layer BP, the effect of pressure on bulk samples was studied experimentally \cite{CSR79} and theoretically 
\cite{ASM82}. The stability of BP nanotubes was studied with density function tight-binding methods \cite{SH00}, and a Young modulus of about 300 
GPa was estimated. Recent {\it ab initio} calculations have shown that single layer  phosphorene can withstand  strain up to 30\%, and stress up to 18 
GPa and 8 GPa in the zigzag and armchair directions, respectively \cite{PWC14,WP14}. Furthermore, uniaxial stress along the direction perpendicular to 
the layer can be used to change the gap size in the system, eventually driving this semiconducting material into a 2D metal \cite{RCC14}. DFT 
calculations of Ref. \cite{RCC14} considered the effect of strain by including in the monolayer unit cell and corresponding atomic positions the 
constraint $z=\pm h$ for all the atoms (see Fig. \ref{Fig:BP}). Therefore an in-plane expansion of the unit cell appears under compressive strain 
($h<h_0$, where $2h_0$ is the thickness of the free layer). The structure of the BP bonds remains until $h/h_0\sim 0.4$, leading to a lattice 
structure similar to that of a puckered graphene layer. The electronic band structure in BP is highly sensitive to strain, suffering a direct to 
indirect gap transition as we increase the compression, and a transition from indirect gap semiconductor to metal is expected for $h/h_0\sim 0.75$, as 
shown in Fig. \ref{Fig:BP}. The application of normal compression on bilayer samples of BP can lead to a direct-to-indirect gap transition for $\sim 
3\%$ of strain, and to a fully reversible semiconductor to metal transition at $\sim 13 \%$ of applied compression \cite{MSP14}.

The peculiar puckered structure of BP layers leads to highly anisotropic in-plane optical and 
electronic properties \cite{XWJ14,LRW14}. By applying appropriate uniaxial or biaxial strain in 
single layer and few-layer BP, it has been shown that the anisotropy of the electron effective mass 
and corresponding mobility direction can be rotated by $90^{\circ}$ while the anisotropy of holes is 
not affected by the strain \cite{FY14}. External strain has been shown to drive a Dirac-like 
dispersion in BP, which differs from the previously reported for graphene, silicene, and germanene, 
in the fact that the anisotropic dispersion of BP allows carriers to behave as either massless Dirac 
or massive carriers, depending on the transport direction along the armchair or zigzag axes, 
respectively \cite{EKT15}.  This strain-engineered anisotropic conductance makes BP a promising 
material for mechanical and electronic applications, such as stretchable electrical devices and 
mechanically controlled logical devices. A huge modulation of the gap, up to $\sim 1$ eV, has 
been reported in rippled BP samples under inhomogeneous strain, suggesting that this material is an 
excellent candidate to create exciton funnels. In this work it has been theoretically predicted that 
strong confinement of carriers is possible into narrow one-dimensional $\sim 150$ nm wide quantum 
wires formed along the valleys of the sample ripples, where the sample is under compression and the 
gap is minimum \cite{QPS15}.

 \subsection{Other 2D crystals}

Whereas graphene, hBN, TMDs and BP are the most studied 2D crystals due to their demonstrated 
stability, there are other 2D crystals which present interesting features that deserve to be 
mentioned.  {\it Silicene}, a single layer of Silicon whose atoms are ordered in a low-buckled 
crystalline structure, is a 2D material with non-trivial topological properties. It has been 
suggested that strain can be used to experimentally observe quantum spin Hall effect in this 
compound \cite{LFY11}. A single layer of Germanium, {\it Germanane}, is a 2D semiconductor with a 
(irrespective of stacking pattern and thickness) direct band gap and strain can be used to drive the 
membrane into its metallic phase \cite{LC14}. Two-dimensional layers of Arsenic or {\it arsenene} 
have been predicted to be stable in two types of honeycomb structures: buckled and puckered 
\cite{KE14}. Both forms present an indirect gap, that can be tuned into a direct gap by applying 
strain. Strain induced tuning of the band gap has been proven in the transition metal 
trichalcogenide TiS$_3$ \cite{BFA15}, where it was found that single layer and bilayer samples 
undergo a direct-to-indirect band gap transition when compressive strain is induced along the 
preferred transport axis.

In summary, we have seen that all the experimental and theoretical techniques developed to study and control the effect of strain in graphene are being applied and adapted to other 2D crystals which have electronic and optical properties highly sensitive to mechanical deformations. This is broadening a field of research which is interesting not only from a fundamental point of view, but also because of potential applications in flexible and stretchable electronics and optoelectronics nanodevices.

\section*{Acknowledgements}
We acknowledge the financial support of the following institutions:
Funda\c{c}\~{a}o para a Ci\^encia e a Tecnologia, Portugal, through Grant No. SFRH/BD/78987/2011 (BA); European Union through ESF funded-JAE doc 
program (AC); JAE-Pre (CSIC, Spain) (AGR); Juan de la Cierva Program (MINECO, Spain) and FCT-Portugal through grant no. EXPL/FIS-NAN/1720/2013 (RR); 
ERC Advanced Grant 290846 (FG, VP);
Spanish Ministry of Economy (MINECO) through Grant No. FIS2011-23713 (PSJ) and FIS2014-58445-JIN (RR); CONICET (PIP 0747) and ANPCyT (PICT 2012-1724) (MS); Spanish MECD grant 
PIB2010BZ-00512,
and  the European Union Seventh Framework Programme under Grant Agreement No. 604391 Graphene Flagship (MV, JS).

\bibliography{Intro,groups,Fernando,Alberto,topology,interactions,anharmonic,transport,optics,superlattices,Others,mechanical_resonators}
\end{document}